\newif\ifcras
\crasfalse
\ifdefined\CRAS\crastrue\fi    
\makeatletter\def\input@path{{cras/}}\makeatother

\ifcras
  \documentclass[CRMECA,Unicode,manuscript,biblatex]{cedram}
  \usepackage{mathtools}
  \usepackage{stmaryrd}
  \usepackage{booktabs}
  \usepackage{algorithm}
  \usepackage{algpseudocode}
  \usepackage{lineno}   
  \linenumbers
  \usepackage{ccicons}  
  \makeatletter\cdr@showtranslationtrue\makeatother
\else
  \documentclass[11pt]{article}
  \usepackage[T1]{fontenc}
  \usepackage[utf8]{inputenc}
  \usepackage[english]{babel}
  \usepackage{amsmath,amsfonts,amssymb,amsthm}
  \usepackage{mathtools}
  \usepackage{hyperref}
  \usepackage[]{graphicx}
  \usepackage[a4paper, margin=2cm]{geometry}
  \usepackage{algorithm}
  \usepackage{algpseudocode}
  \usepackage{color}
  \usepackage{booktabs}
  \usepackage[backend=biber,style=authoryear,giveninits=true,
  url=false, doi=true, eprint=true,date=year,isbn=false,natbib=true]{biblatex}
  \usepackage{stmaryrd}
  \usepackage{paralist}
\fi

\title{Strength-degradation phase-field regularization of cohesive fracture: the antiplane case}

\date{\today}

\newcommand{\keywordlist}{Phase-field fracture, Cohesive fracture, Variational
  fracture mechanics, Antiplane shear, Crack nucleation, Strength, Conic
  programming}

\newcommand{\ackstext}{BB acknowledges the support of the Natural Sciences and
Engineering Research Council of Canada (NSERC), RGPIN-2022-04536 and the Canada
Research Chair program.
CM received funding from the European Union's Horizon 2020 research and
innovation program under the Marie Sk\l{}odowska-Curie grant agreement
No.~861061 --- NEWFRAC Project.}

\ifcras
  \author{\firstname{Blaise} \lastname{Bourdin}}
  \address{Department of Mathematics \& Statistics, McMaster University,
    Hamilton, Ontario L8S-4K1, Canada}
  \email[B. Bourdin]{bourdin@mcmaster.ca}
  \thanks{BB acknowledges the support of the Natural Sciences and Engineering
    Research Council of Canada (NSERC), RGPIN-2022-04536 and the Canada
    Research Chair program.}
  \author{\firstname{Corrado} \lastname{Maurini}\IsCorresp}
  \address{Institut Jean Le Rond d'Alembert, Sorbonne Universit\'e, CNRS,
    F-75005 Paris, France}
  \email[C. Maurini]{corrado.maurini@sorbonne-universite.fr}
  \thanks{CM received funding from the European Union's Horizon 2020 research
    and innovation program under the Marie Sk\l{}odowska-Curie grant agreement
    No.~861061 --- NEWFRAC Project.}
  \keywords{\keywordlist}
  \COI{The authors do not work for, advise, own shares in, or receive funds
    from any organization that could benefit from this article, and have
    declared no affiliations other than their research organizations.}
\else
  \author{Blaise Bourdin, Corrado Maurini}
\fi

\newcommand{\tensd}[1]{\underline{\underline{#1}}}
\newcommand{\jump}[1]{\left \llbracket #1 \right\rrbracket}
\newcommand{\avg}[1]{\left\{ #1 \right\}}
\newcommand{\uline}[1]{{\underline{{#1}}}}
\newcommand{\uuline}[1]{\underline{\underline{{#1}}}}
\newcommand{\ssig}{\uuline{\sigma}}
\newcommand{\veps}{{\varepsilon}}
\newcommand{\R}{\mathbb{R}}

\newcommand{\e}{\boldsymbol{\veps}}
\newcommand{\vveps}{\uuline{\veps}}

\newcommand{\uu}{\uline{u}}
\newcommand{\p}{\boldsymbol{p}}
\newcommand{\q}{\boldsymbol{q}}
\newcommand{\pp}{\uuline{p}}
\renewcommand{\t}{\boldsymbol{\tau}}

\renewcommand{\k}{\mathsf k}
\newcommand{\cw}{\mathsf{c_w}}
\newcommand{\w}{\mathsf w}
\newcommand{\K}{\mathbb K}

\newcommand{\dV}{\,\mathrm{d}V}
\newcommand{\dA}{\,\mathrm{d}A}
\newcommand{\Gc}{\mathsf{G}_\mathrm{c}}
\newcommand{\Gceff}{\Gc^{\mathrm{eff}}}
\newcommand{\dDda}{\mathrm{d}\mathcal{D}/\mathrm{d}a}
\newcommand{\lch}{\ell_{\mathrm{ch}}}
\DeclareMathOperator*{\argmin}{arg\,min}

\ifcras
  \makeatletter
  \AtBeginDocument{\theoremstyle{plain}%
    \newtheorem{hypothesis}[\cdr@thm]{Hypothesis}}
  \makeatother
\else
  \newtheorem{remark}{Remark}
  
  \newtheorem{hypothesis}{Hypothesis}
\fi

\ifcras
  \graphicspath{{./Figures/}{./cras/}}
\else
  \graphicspath{{./Figures/}}
\fi

\newcommand{\abstracttext}{%
Phase-field approaches to fracture, initially designed as regularization of the Griffith model of brittle fracture, are now commonly viewed as gradient-damage models whose regularization length becomes a material property driving crack nucleation.
One weakness of this approach is that the strength surface cannot be arbitrary: its shape is dictated by the elastic energy, and its magnitude by the regularization length.
We focus on the antiplane version of the model introduced by Bourdin, Marigo, Maurini and Zolesi (arXiv:2506.22558), which handles crack propagation along unknown paths and nucleation governed by an arbitrary convex strength surface by degrading the strength instead of the stiffness.
It can be interpreted as a regularization of softening plasticity in which localization bands obey an equivalent cohesive law set by the strength domain and the toughness, while the role of the regularization length, when small compared to the elasto-cohesive length, is purely numerical.
Strength, stiffness, and toughness thus become independent material data, and limit analysis, perfect plasticity, cohesive fracture, and brittle fracture merge into a single variational framework.
We derive closed-form solutions for a simple shear problem, propose a numerical scheme combining alternate minimization and conic programming, and numerically verify the equivalent cohesive law, its independence of the regularization, and the size effect governed by the elasto-cohesive length.
A ``surfing'' simulation highlights the structure of the propagating crack while a re-entrant V-notch is used to show how the model bridges small-scale yielding, cohesive fracture, and brittle fracture without \emph{a priori} hypotheses.
}

\newcommand{\altabstracttext}{%
Les modèles de rupture de type champ de phase ont été initialement conçus comme des régularisations du modèle de Griffith pour la rupture fragile.
Ils sont souvent interprétés comme des lois à gradient d'endommagement où la longueur de régularisation est identifiée à une propriété matérielle pilotant la nucléation des fissures.
Une faiblesse de cette approche est que la surface de résistance ne peut être arbitraire : sa forme est dictée par les propriétés élastiques, et son amplitude par la longueur de régularisation.
Nous nous concentrons sur la version antiplane du modèle introduit par Bourdin, Marigo, Maurini et Zolesi (arXiv:2506.22558) et qui permet de rendre compte de la propagation de fissures le long de trajets inconnus ainsi que d'une nucléation gouvernée par une surface de résistance convexe arbitraire, en dégradant la résistance plutôt que la rigidité.
Ce modèle peut s'interpréter comme de la plasticité adoucissante régularisée, où les bandes de localisation obéissent à une loi cohésive équivalente fixée par le domaine de résistance et la ténacité, alors que le rôle de la longueur de régularisation, lorsqu'elle est petite devant la longueur élasto-cohésive, est purement numérique.
La résistance, la rigidité et la ténacité deviennent ainsi des données matériau indépendantes, et l'analyse limite, la plasticité parfaite, la rupture cohésive et la rupture fragile sont liées dans un cadre variationnel unique.
Nous obtenons des solutions analytiques pour un problème de cisaillement simple, proposons un schéma numérique combinant minimisation alternée et programmation conique, et vérifions numériquement la loi cohésive équivalente, son indépendance vis-à-vis de la régularisation, ainsi que l'effet d'échelle gouverné par la longueur élasto-cohésive.
Une simulation de type «surfing» met en évidence la structure des fissures durant leur propagation, tandis qu'une entaille en V rentrante permet de montrer comment le modèle fait le lien entre plasticité confinée (small-scale yielding), rupture cohésive et rupture fragile, sans hypothèse \emph{a priori}.
}

\ifcras
  \begin{abstract}\abstracttext\end{abstract}
  \begin{altabstract}\altabstracttext\end{altabstract}
\fi

\begin{document}

\ifcras
  \makeatletter
  \begingroup
    \let\Hy@raisedlink\@gobble
    \let\@makefnmark\relax\let\@thefnmark\relax
    \@footnotetext{Manuscript version of \today.}%
  \endgroup
  \makeatother
\fi

\maketitle

\ifcras\else
  \begin{abstract}\abstracttext\end{abstract}

  \medskip
  \noindent\textbf{Keywords:} \keywordlist
\fi

\section{Introduction}
The variational approach to fracture~\autocite{francfort1998Revisiting,bourdin2008VariationalApproach} and its phase-field regularizations~\autocite{BFM00} have become a standard framework to predict crack nucleation and propagation in brittle materials.
Mechanically, phase-field models can be interpreted as gradient-damage models~\autocite{Pham-Marigo-2010a,pham2011GradientDamageModels}, where a scalar variable $\alpha$ degrades the elastic stiffness and a regularization length $\ell$ controls the width of the localization bands.
In this framework, the material strength is not an independent constitutive ingredient: it emerges from the combination of the elastic stiffness, the fracture toughness, and the regularization length, which must therefore be tuned as a material parameter to fit the observed nucleation loads~\autocite{tanne2018CrackNucleation}.
This approach becomes impractical for stiff materials of modest strength, for which the required regularization length far exceeds the relevant structural sizes, and for nominally brittle materials at very large scales, where the need to resolve the regularization length with the mesh leads to prohibitively large problem sizes.
Moreover, the shape of the resulting strength surface in a multiaxial stress state is essentially dictated by the form of the elastic energy and cannot be freely prescribed, failing to reproduce the experimentally measured multiaxial strength data of common brittle materials~\autocite{kumar2020RevisitingNucleation}.

  Plasticity theory, in contrast, accommodates an arbitrary convex strength domain~\autocite{hill1950MathematicalTheoryPlasticity,prager1951TheoryPerfectlyPlastic}. Small-scale yielding (SSY) theory applies it to fracture through near-tip plastic deformation~\autocite{mcclintock1965PlasticityAspects,hutchinson1968SingularBehaviour,rice1968PlaneStrainDeformation}.
  In SSY, cracks and stress concentrators generate confined plastic zones near the tip~\autocite{hult1956ElasticPlastic,rice1966ContainedPlastic}. Plastic deformations can also concentrate on a curve in two dimensions or on a surface in three, acting as cohesive cracks, as in the classical slit problem~\autocite{dugdale1960YieldingSteelSheets,bilby1963SpreadPlasticYield}.
  Perfect plasticity alone, however, sustains a constant stress across such a band, so that the dissipated energy grows without bound with the displacement jump: it endows the material with a strength but not with a finite toughness, since the traction across the band can vanish only if the strength itself degrades.
  Degrading the strength in the bulk leads to softening plasticity, whose pathologies are well documented --- deformations localize on sets of vanishing measure and numerical solutions suffer a spurious mesh dependence --- and are classically mitigated by nonlocal, gradient, or micromorphic regularizations~\autocite{pijaudier-cabot1987NonlocalDamageTheory,muhlhaus1991VariationalPrincipleGradient,leblond1994BifurcationEffectsDuctile,bazant2002NonlocalIntegralFormulations,jirasek2003ComparisonIntegraltypeNonlocal,forest2004LocalizationPhenomenaRegularization,lorentz2008NumericalSimulationDuctile,bacquaert2025RegularizationSofteningPlasticity}.
  These regularizations, however, are conceived as localization limiters: they leave the limit properties of the localization bands and their link with sharp-interface cohesive models unidentified.
  Cohesive models of fracture~\autocite{barenblatt1962MathematicalTheoryEquilibrium,hillerborg1976AnalysisCrackFormation,needleman1987ContinuumModelVoid,delpiero1999OneDimensionalDuctileBrittle,delpiero2009ElasticBarsCohesive,marigo2023ModellingFractureCohesive} take this step at the interface level, letting strength and toughness enter as independent material data through a traction-separation law acting on surfaces of displacement discontinuity.
  Unfortunately, they raise difficulties of their own.
  First, their direct numerical implementation through interface elements is delicate~\autocite{ortiz1999ClassCohesiveFinite,klein2001PhysicsbasedModelingBrittle,rimoli2015MeshingStrategiesAlleviation}: the crack is confined to the mesh facets, and the interface stiffness introduces a spurious compliance.
  Second, as recently shown by~\textcite{rodella2026SharpInterface}, sharp cohesive interfaces are incompatible with a linearly elastic bulk: no solution exists in the form of a regular crack with a simple tip, the stress exceeding the cohesive strength over a finite region around the tip.
  The associated variational problem then lacks lower semi-continuity~\autocite{bouchitte1995RelaxationResults}, calling for bulk energies whose growth is consistent with the interfacial strength.
  Finally, no cohesive model is to date widely accepted for crack propagation along unknown paths, and the geometric regularity of the resulting cracks remains poorly understood: the existence theory for such energies does not rule out minimizers whose ``fracture'' is not a set of codimension one --- a curve in two dimensions, a surface in three --- but a Cantor-like set of intermediate dimension.

The formulation of regularized counterparts of cohesive models, enjoying the same flexibility as the phase-field models of brittle fracture, is thus the object of active research.
Gradient-damage models recovering a cohesive response with an $\ell$-independent strength have been obtained by a suitable tuning of the stiffness-degradation and dissipation functions~\autocite{lorentz2011ConvergenceGradientDamage,lorentz2012ModellingLargeCrack}, or through directional energy decompositions implementing a prescribed mixed-mode cohesive law~\autocite{feng2023UnifiedRegularizedVariational}, but in both cases the strength remains tied to the stiffness degradation and does not enter as a constitutive strength domain independent of the elastic properties.

We recently introduced in~\autocite{bmmz25}\footnote{Closely related models were independently introduced around the same time in~\autocite{vicentini2026TunableStrength,fengLi2026ConvergenceCZM}.} a variational phase-field fracture model that incorporates arbitrary convex strength domains. Given a three-dimensional domain ${\Omega_\mathrm{3D}}$, a vector-valued displacement field $\underline{u}:\underline{x}\in {\Omega_\mathrm{3D}}\to \underline{u}(\underline{x})\in\R^3$, and a phase field $\alpha:\underline{x}\in {\Omega_\mathrm{3D}}\to {\alpha}(\underline{x})\in[0,1]$, this model is characterized by an energy functional of the form
\begin{equation}
    \mathcal E_\ell(\underline{u},\alpha):=\int_{\Omega_\mathrm{3D}}\psi_{\mathrm{3D}}(\tensd{\varepsilon}(\underline{u}),\alpha)+D_\ell(\alpha,\nabla \alpha) \dV,
    \label{eq:plastic-damage-energyIntro}
\end{equation}
where $\tensd{\varepsilon}(\underline{u})=\mathrm{sym}({\tensd{\nabla}}\,\underline{u})$ denotes the linearized strain tensor.

The first term in~\eqref{eq:plastic-damage-energyIntro} is the elastic energy density, defined as
\begin{equation}
    \label{eq:phialpha}
    \psi_{\mathrm{3D}}(\tensd{\varepsilon},\alpha)=
    \min_{\tensd{p}\in\mathbb{M}^3_s}\varphi_{\mathrm{3D}}(\tensd{\varepsilon},\tensd{p},\alpha),
    \quad
    \varphi_{\mathrm{3D}}(\tensd{\varepsilon},\tensd{p},\alpha)=\dfrac{1}{2}\mathsf{A}_0(\tensd{\varepsilon}-\tensd{p})\cdot(\tensd{\varepsilon}-\tensd{p})+\mathsf{k}(\alpha)\mathsf{H}_{\mathbb{K}_0}(\tensd{p}),
\end{equation}
where  $\mathsf{A}_0$ is the fourth-order elasticity tensor, $\mathbb{M}^3_s$ is the space of symmetric $3\times 3$ tensors, $\mathsf{k}(\alpha)$ is a decreasing function such that $\mathsf{k}(0)=1$ and $\mathsf{k}(1)=0$, the dot ``$\cdot$'' standing for the usual scalar product.
The single-valued function
\begin{equation}
    \mathsf{H}_{\mathbb{K}_0}(\tensd{p}):=\sup_{\tensd{\sigma}\in\mathbb{K}_0}\tensd{\sigma}\cdot\tensd{p}
    \label{eq:support-function-3d}
\end{equation}
denotes the support function of the convex domain $\mathbb{K}_0\subset \mathbb{M}^3_s$ representing the material strength, \emph{i.e.}~the domain of admissible Cauchy stress tensors in the undamaged material.
The symmetric tensor $\tensd{p}$ can be regarded as the nonlinear contribution to the geometric deformation $\tensd{\varepsilon}(\underline{u})$.

The second term in~\eqref{eq:plastic-damage-energyIntro} represents a fracture-type dissipation, which, as in classical phase-field fracture models, includes a gradient-type regularization and is defined as
\[
    D_\ell(\alpha,\nabla \alpha):=
    \dfrac{\Gc}{4\cw}\left(\dfrac{\mathsf{w}(\alpha)}{\ell}+\ell\,\nabla\alpha\cdot\nabla\alpha\right),
\]
where $\Gc$ is the fracture energy, $\ell$ is a regularization length scale, and $\mathsf{w}(\alpha)$ is an increasing function satisfying $\mathsf{w}(0)=0$ and $\mathsf{w}(1)=1$.
The constant $\cw:=\int_0^1\sqrt{\mathsf{w}(s)}\,\mathrm{d}s$ is a normalization factor ensuring that the energy dissipated to create a fully developed crack of unit area is equal to $\Gc$.

The nonlinear dependence of $\psi_{\mathrm{3D}}$ on $\tensd{\varepsilon}$ endows the model with a strength domain $\mathbb{K}_0$ in the undamaged state.
The elastic energy potential is akin to the energy density of perfect plasticity~\autocite{DalMaso-DeSimone-EtAl-2006a}, with the damage variable $\alpha$ playing the role of a softening variable acting on the strength domain.
The model can thus be interpreted as a phase-field regularization of a softening plasticity model \emph{à la} Hencky (\emph{i.e.} without the irreversibility condition on the plastic strain), in contrast with classical phase-field models, where the damage variable induces stiffness degradation.

Considering quasi-static evolution governed by minimization of $\mathcal{E}_\ell(\underline u,\alpha)$ under an irreversibility constraint on $\alpha$, \textcite{bmmz25} report analytical solutions of a three-dimensional model problem indicating that in the limit $\ell\to 0$ the model recovers a cohesive fracture behavior~\autocite{dugdale1960YieldingSteelSheets,barenblatt1962MathematicalTheoryEquilibrium,marigo2023ModellingFractureCohesive} with a traction-separation law directly related to the strength domain $\mathbb{K}_0$.
The conjectured limiting cohesive model is characterized by an energy functional of the form
\begin{equation}
    \label{eq:limit-energy}
    \mathcal{E}_0(\underline{u},\hat\alpha)=
    \int_{{\Omega_\mathrm{3D}}\setminus J_{\underline{u}}}
\psi_\mathrm{3D}(\tensd{\varepsilon},0) \,\mathrm{d} V +  \int_{J_{\underline{u}}}\phi(\jump{\underline{u}},\hat\alpha) \,\mathrm{d}S,\end{equation}
where $\psi_\mathrm{3D}(\tensd{\varepsilon},0)$ is the bulk elastic energy density of the undamaged material, $J_{\underline{u}}$ is the set of displacement discontinuities, $\jump{\underline{u}}$ is the displacement jump across $J_{\underline{u}}$, and $\hat\alpha:J_{\underline{u}}\to[0,1]$ is a local damage variable supported on $J_{\underline{u}}$.
The surface energy density $\phi$ takes the form
\begin{equation}
    \phi(\jump{\underline{u}},\hat\alpha) =
    \hat{\mathsf{k}}(\hat\alpha)
    \mathsf{H}_{\mathbb{K}_0}(\underline{n} \odot \jump{\underline{u}})
    + \Gc\,\hat{\alpha},
    \label{eq:surface-energy-densities-limit-model}
\end{equation}
where $\underline{n}$ is the normal to the jump set $J_{\underline{u}}$, $\odot$ denotes the symmetric tensor product, and
\begin{equation*}
    {\hat{\alpha}} (\alpha):= \frac{\int_0^{\alpha} \sqrt{\mathsf{w}(\beta)}\, \mathrm{d}\beta}{\int_0^1 \sqrt{\mathsf{w}(\beta)}\, \mathrm{d}\beta},\qquad
    \hat{\mathsf{k}}({\hat{\alpha}}):=\mathsf{k}(\alpha({\hat{\alpha}})).
\end{equation*}
The $\Gamma$-convergence towards the sharp cohesive energy has been proven, for the choice $\mathsf{k}(\alpha)=(1-\alpha)^2$ and $\mathsf{w}(\alpha)=\alpha^2$, in the spatially discrete antiplane setting by~\textcite{maggiorelli2025GammaConvergence}; a similar result had previously been established for gradient-damage models coupled with plasticity, also in the antiplane setting, by~\textcite{dalmaso2016FractureModelsElastoplastic}.

The present work is related to several lines of research.
Sharp-interface cohesive models with bulk energies consistent with the traction-separation law have been investigated numerically in~\autocite{rodella2026SharpInterface}.
Phase-field approximations by $\Gamma$-convergence of specific cohesive fracture energies have been established in the scalar and, more recently, vector-valued settings~\autocite{conti2016PhaseFieldApproximation,conti2024PhaseFieldApproximationVectorial}.
A unified framework spanning the $\Gamma$-convergence analysis, the reconstruction of the phase-field model realizing a prescribed cohesive law, and its mechanical assessment has recently been developed in the three-part series~\autocite{alessi2026CohesivePartI,alessi2025CohesivePartII,alessi2025CohesivePartIII}.
A family of gradient models coupling damage and plasticity and where the strength is accounted for through a convex domain inherited from the variational formulation of perfect plasticity~\autocite{duvaut1976InequalitiesMechanicsPhysics,temam1980FunctionsBoundedDeformation,anzellotti1980ExistenceDisplacementsField,suquet1981EquationsPlasticiteExistence}, was proposed in~\autocite{alessi2014GradientDamageModels,Alessi-Marigo-EtAl-2015a,alessi2017CouplingDamagePlasticity} and studied in~\autocite{dalmaso2016FractureModelsElastoplastic,dalmaso2020JerkyCrackGrowth,dalmaso2022PureJumpNature}.
The model that we study here differs from these in that the damage variable degrades only the strength domain, leaving the elastic stiffness unaffected, and that no irreversibility condition is imposed on the nonlinear deformation: this minimal structure makes the limit properties of the localization bands explicit, with an equivalent cohesive law independent of both the regularization length and the elastic stiffness (Section~\ref{sec:localizedSolutions}).

The goal of this work is to discuss in detail the properties of this model in the simplified setting of antiplane shear and to present a first set of numerical simulations of quasi-static crack growth exhibiting a cohesive-like behavior.
The antiplane shear setting greatly simplifies the analysis since the displacement field reduces to a scalar function $u:\mathbf{x}\in \Omega\to u(\mathbf{x})\in\R$, the strain tensor reduces to the gradient vector $\nabla u$, and the stress tensor is represented by a shear-stress vector in $\R^2$.
Moreover, the nonlinear deformation is a vector, hence always representable as $\p=\jump{u}\,\mathbf{n}$ for a suitable unit normal $\mathbf{n}$, and its support function $\tau_c\Vert\p\Vert$ is finite for every jump.
The compatibility condition between the strength domain and the admissible jumps identified in~\autocite{bmmz25} is therefore always satisfied, and the limit cohesive model is well-defined.
In three dimensions, by contrast, a symmetric tensor $\tensd{p}$ is in general not of the form $\underline{n}\odot\jump{\underline{u}}$, and $\mathsf{H}_{\mathbb{K}_0}(\underline{n}\odot\jump{\underline{u}})$ is finite only for certain relative orientations of $\jump{\underline{u}}$ and $\underline{n}$: a strength domain unbounded along the hydrostatic axis, for instance, forces $\jump{\underline{u}}\cdot\underline{n}=0$ and thus admits no opening.

Accordingly, for an isotropic material the strength domain $\mathbb{K}_0$ reduces to the disk of radius equal to the shear strength $\tau_c$, its support function reduces to $\mathsf{H}_{\mathbb{K}_0}(\p)=\tau_c\|\p\|$ with $\p\in\R^2$, and the energy functional~\eqref{eq:plastic-damage-energyIntro}, with the nonlinear deformation $\p$ kept as an explicit unknown, specializes to
\begin{equation}
    \label{eq:energy-antiplane-intro}
    \mathcal{E}_\ell(u,\p,\alpha)=
    \int_{\Omega}\Big(\tfrac{\mu}{2}\|\nabla u-\p\|^2+\mathsf{k}(\alpha)\,\tau_c\|\p\|\Big)\dA
    +\frac{\Gc}{4\cw}\int_{\Omega}\Big(\frac{\mathsf{w}(\alpha)}{\ell}+\ell\,\|\nabla\alpha\|^2\Big)\dA,
\end{equation}
where $\mu>0$ is the shear modulus.
Eliminating $\p$ by pointwise minimization yields a nonlinear elastic energy density that is quadratic in $\nabla u$ below the degraded strength $\mathsf{k}(\alpha)\,\tau_c$ and grows linearly above it: the model is a phase-field regularization of an antiplane softening plasticity model in which the damage variable $\alpha$ degrades the strength while leaving the elastic stiffness $\mu$ unchanged.
Consistently, the surface energy density~\eqref{eq:surface-energy-densities-limit-model} of the limit cohesive model reduces to $\phi(\jump{u},\hat\alpha)=\hat{\mathsf{k}}(\hat\alpha)\,\tau_c\,\left|\jump{u}\right|+\Gc\,\hat\alpha$, a mode-III cohesive law that is set by two independent material data, the strength $\tau_c$ and the toughness $\Gc$, and whose precise shape can be fine-tuned through the constitutive functions $\mathsf{k}$ and $\mathsf{w}$.

In this setting the model problem of Section~\ref{sec:modelProblem} can be solved in closed form: the homogeneous solutions fix the equivalent material \emph{strength} and may display a constant-stress plateau at $\tau=\tau_c$ before softening; the localized solutions yield an explicit equivalent \emph{cohesive law} set by the toughness $\Gc$ and the strength $\tau_c$, both independent of the regularization length $\ell$; and the corresponding no-snap-back conditions are governed by $\ell/\ell_{\mathrm{ch}}$ for the material response and by the size ratio $L/\ell_{\mathrm{ch}}$ for the structural response of a bar, where $\ell_{\mathrm{ch}}=\mu\Gc/\tau_c^2$ is the elasto-cohesive length.
This analysis is the antiplane specialization of the three-dimensional construction of~\autocite{bmmz25}; we report it in full because the scalar setting makes every step explicit and easy to follow, giving the complete picture while factoring out the effects of multiaxiality and of the tensorial nature of stress and strain, and because it provides the exact reference solutions against which the numerical results are assessed.

The main original contributions of this paper are the following.
First, we devise a numerical scheme that exploits the separately convex, non-smooth structure of the energy: the alternate-minimization subproblems are recast as second-order cone programs and solved by conic optimization~\autocite{krabbenhoft2007FormulationSolutionPlasticity,bleyer-convex-optim,lobo1998ApplicationsSecondOrderCone,aps2024mosek}, without any smoothing or penalization of the strength term.
Second, we verify the numerical solution against the closed-form solutions: the simulations recover the equivalent cohesive law and its independence of $\ell$, quantify the mesh-induced toughening, and reproduce the size effect in $L/\ell_{\mathrm{ch}}$, the antiplane counterpart of the Griffith--Barenblatt transition analyzed in~\autocite{marigo2004InitiationPropagationFracture}; in the linearly rigid limit $\mu\to\infty$, where the structural response coincides with the cohesive law, we measure the latter directly.
Third, we extract the effective toughness $\Gceff$ from a ``surfing'' experiment: it recovers the prescribed $\Gc$ when $\p$ is unconstrained, but exceeds it when $\p$ is irreversible, because of the residual plastic wake left behind the crack tip.
Finally, we describe how, within one and the same model, the nonlinear fields at the crack tip evolve from the small-scale-yielding zone of perfect plasticity~\autocite{hult1956ElasticPlastic,rice1966ContainedPlastic} to a cohesive crack~\autocite{barenblatt1962MathematicalTheoryEquilibrium} and then to a brittle crack.
Taken together, these results present limit analysis~\autocite{salencon2013YieldDesign}, perfect plasticity, cohesive fracture, and brittle fracture not as separate constitutive descriptions, but as regimes of a single variational model with one set of material data $(\mu,\tau_c,\Gc)$; we return to this point in Section~\ref{sec:conclusions}.

The article is organized as follows.
Section~\ref{sec:phaseFieldModelAntiplane} casts the model in the antiplane shear setting and derives the constitutive law and the first-order optimality conditions.
Section~\ref{sec:modelProblem} constructs the homogeneous and localized solutions of a one-dimensional model problem, the associated equivalent cohesive law, and the corresponding no-snap-back conditions.
Section~\ref{sec:M1} introduces the one-parameter family of constitutive functions used in all the numerical simulations, for which the results of Section~\ref{sec:modelProblem} take a fully explicit form.
Section~\ref{sec:numericalImplementation} presents the dimensionless formulation and the alternate-minimization scheme; the conic-programming solution of the two subproblems is detailed in Appendix~\ref{sec:numericalAppendix}.
Section~\ref{sec:numericalSimulations} compares analytical predictions and numerical simulations for three problems: a rectangular domain under simple shear, a ``surfing'' simulation, and a V-notch.
Section~\ref{sec:conclusions} draws conclusions and discusses the extension to the multiaxial case treated in the companion paper.

\section{The phase-field model with strength degradation in the antiplane setting}
\label{sec:phaseFieldModelAntiplane}
\subsection{The antiplane shear setting}
\label{sec:antiplaneShearSetting}
Consider an isotropic homogeneous linearly elastic material occupying a cylindrical domain $\Omega_\mathrm{3D} =\Omega\times (-H,H)$ in the reference configuration with $\Omega\subset \R^2$ and $H\gg \mathrm{diam}(\Omega)$.
The position vector can be decomposed as $\underline{x}=\mathbf{x}+z\,\underline{e}_3$ with $\mathbf{x}=x\,\underline{e}_1+y\,\underline{e}_2 \in \Omega$ and $z\in(-H,H)$.
In the antiplane setting, one assumes that the displacement field is of the form
$$\underline{u}(x,y,z) = u(x,y) \,\underline{e}_3$$
with $u: (x,y)\in\Omega \mapsto u(x,y)\in\mathbb{R}$.
Under these hypotheses, one has that
$$
\vveps(\uu)  = \partial_x u \,(\underline{e}_1 \odot \underline{e}_3) + \partial_y u\,(\underline{e}_2 \odot \underline{e}_3),
$$
where $\underline{a} \odot \underline{b} := \frac12\left(\underline{a} \otimes \underline{b} + \underline{b} \otimes \underline{a}\right)$.
The strain tensor $\vveps(\uu)$ can be represented by the vector-valued function $\e = \nabla u:\Omega \to \R^2$, whose components are twice the tensor ones: $\e = 2\varepsilon_{13}\,\underline{e}_1+ 2\varepsilon_{23}\,\underline{e}_2$.
Moreover, all the components of the stress $\ssig$ and the nonlinear deformation $\pp$ vanish except for $(\sigma_{13},\sigma_{23})$ and $(p_{13},p_{23})$, which can be collected in the vectors $\t := (\tau_1, \tau_2) = \sigma_{13}\,\underline{e}_1+ \sigma_{23}\,\underline{e}_2$ and $\p := (p_1, p_2) = 2p_{13}\,\underline{e}_1+ 2p_{23}\,\underline{e}_2$ in $\R^2$.
With this convention --- strain-like vectors collect twice the tensor components, the stress vector the bare ones --- the duality pairings coincide: $\ssig \cdot \vveps = \t \cdot \e$ and $\ssig \cdot \pp = \t \cdot \p$.

\subsection{Strain energy density and stress-strain constitutive relation at fixed $\alpha$}
\begin{figure}[h]
    \centering
    \includegraphics[width=.9\textwidth]{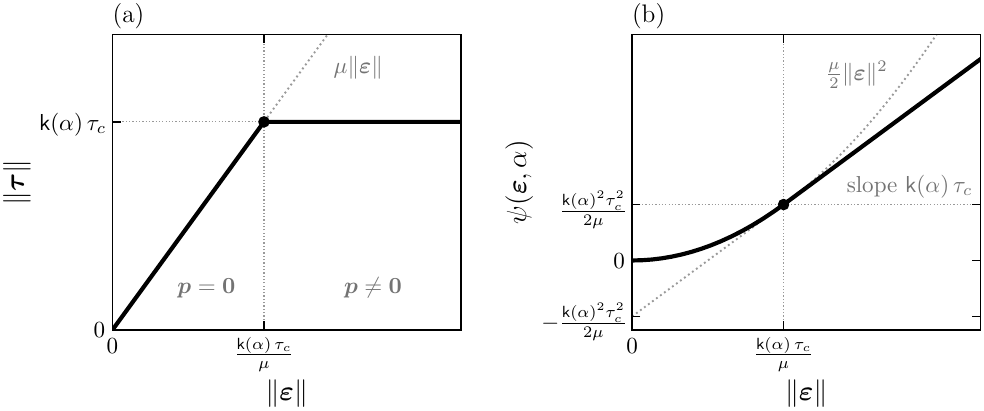}
    \caption{Equivalent nonlinear constitutive relation (a) and elastic energy density (b) at fixed $\alpha$.
    }
    \label{fig:nonlinear_elasticity}
\end{figure}

Under the same hypotheses, one has that $\frac{1}{2}\mathsf{A}_0 \vveps \cdot \vveps = \frac{\mu}{2} \Vert \e\Vert^2$, where $\mu>0$ is the \emph{shear modulus}.
Moreover, for an isotropic strength domain $\mathbb{K}_0$ and for $\pp$ of the antiplane form above, the support function~\eqref{eq:support-function-3d} reduces to
$$
\mathsf{H}_{\mathbb{K}_0}(\pp) = \sup_{\ssig \in \mathbb{K}_0}  \ssig \cdot\pp = \sup_{\ssig \in \mathbb{K}_0} \t \cdot\p =\tau_c \|\p\|,
$$
where $\tau_c=\sup_{\ssig \in \mathbb{K}_0} \Vert \t\Vert$ is the \emph{shear strength} of the undamaged material.
Hence, the energy density~\eqref{eq:phialpha} simplifies to
$$
\psi(\e,\alpha) = \min_{\p \in \R^2} \varphi(\e, \p, \alpha),\quad
\varphi(\e, \p, \alpha) = \frac{\mu}{2} \|\e - \p\|^2 + \mathsf{k}(\alpha)\,\tau_c \|\p\|.
$$
Using the fact that the minimum over $\p$ is attained by taking $\p$ collinear to $\e$ and pointing in the same direction, in which case $\|\e - \p\|^2= (\|\e\| - \|\p\|)^2$, we can also compute the energy density explicitly:
\begin{equation}
    \label{eq:elastic-energy-density-antiplane}
    \begin{split}
        \psi(\e,\alpha) &= \min_{\|\p\| \geq 0} \frac{\mu}{2} (\|\e\| - \|\p\|)^2 + \mathsf{k}(\alpha)\,\tau_c \|\p\|\\
        &=
        \begin{cases}
            \dfrac{\mu}{2} \|\e\|^2 & \text{ if } \|\e\| \le \mathsf{k}(\alpha)\,\dfrac{\tau_c}{\mu},\\
            \mathsf{k}(\alpha)\,\tau_c \left(\|\e\| - \mathsf{k}(\alpha)\,\dfrac{\tau_c}{2\mu} \right)& \text{ otherwise.}
        \end{cases}
    \end{split}
\end{equation}
The constitutive relation for the antiplane stress vector $\t$ is obtained by differentiating $\psi$ with respect to $\e$:
\begin{equation}
    \label{eq:constitutiveRelationAntiplane}
    \t(\e,\alpha) = \begin{cases}
        \mu \e & \text{ if } \|\e\| \le \mathsf{k}(\alpha)\,\dfrac{\tau_c}{\mu},\\
        \mathsf{k}(\alpha)\,\tau_c \dfrac{\e}{\|\e\|} & \text{ otherwise.}
    \end{cases}
\end{equation}
The equivalent nonlinear constitutive relation and elastic energy density at fixed $\alpha$ are illustrated in Figure~\ref{fig:nonlinear_elasticity}.

\subsection{Quasi-static variational formulation}

In all that follows, we consider loadings in the form of applied displacement boundary conditions $u=\bar{u}(t)$ on a part of the boundary  $\partial_u \Omega$.
Assimilating the loading parameter to a time variable, the quasi-static evolution problem consists in finding ($u(t),\alpha(t)$) for each value of $t\in[0,T]$.
This can be formulated in the framework of rate-independent processes~\autocite{MielkeRoubicek2015}  by assuming that the evolution is quasi-static, \emph{i.e.}~neglecting the effects of inertia and viscosity.
Discretizing the time variable in a sequence of loading steps $0 = t_0 < t_1 <\ldots < t_N=T$ and enforcing a growth condition on  $\alpha$ to account for the irreversible nature of the damage process,  we obtain an incremental formulation for the quasi-static evolution, which is amenable to a numerical implementation.
Namely, at loading step $i$, the state of the system is obtained as a unilateral minimizer of the energy functional~\eqref{eq:plastic-damage-energyIntro} which reduces to
\begin{equation}
    \label{eq:model}
    \mathcal{E}_\ell(u,\p,\alpha) =
    \int_{\Omega} \left( \frac{\mu}{2} \|\nabla u - \p\|^2 + \mathsf{k}(\alpha)\,\tau_c \|\p\| \right) \dA + \frac{\Gc}{4\cw} \int_{\Omega} \left( \frac{\mathsf{w}(\alpha)}{\ell} + \ell \|\nabla\alpha\|^2 \right) \dA.
\end{equation}

At loading step $i$, denoting by $\bar u_i$ and $\bar{\alpha}$ prescribed displacements and damage on parts $\partial_u \Omega \subset \partial \Omega$ and $\partial_\alpha \Omega\subset \partial \Omega$ respectively, we solve
\begin{equation}
    \label{eq:incremental-minimization}
    (u_i, \p_i, \alpha_i) \in \argmin_{\substack{u \in \mathcal{C}_i,\, \alpha \in \mathcal{D}_i},\, \p \in \mathcal{P}} \mathcal{E}_\ell(u, \p, \alpha),
\end{equation}
where the spaces of the admissible fields are:
\begin{eqnarray}
    \mathcal{C}_i &=& \{u \in BV(\Omega): u = \bar{u}_i \text{ on } \partial_u \Omega\},\\
    \mathcal{D}_i &=& \{\alpha \in H^1(\Omega): \alpha_{i-1} \le \alpha \le 1 \text{ in } \Omega,\alpha=\bar\alpha\,\text{on}\,\partial_\alpha\Omega\},\\
    \mathcal{P} &=& \mathcal{M}(\Omega;\R^2),
\end{eqnarray}
where $BV(\Omega)$ and $H^1(\Omega)$ denote respectively the space of scalar-valued functions with bounded variation and the Sobolev space of square-integrable functions with square-integrable gradient and $\mathcal{M}(\Omega;\R^2)$ that of $\R^2$-valued Radon measures on $\Omega$.

Formally, because the energy functional exhibits linear growth at infinity, the nonlinear deformation $\p$ may concentrate along regions of dimension strictly less than $n = 2$.
Moreover, since the elastic strain $(\nabla u - \p)$ must be square-integrable, the singular parts of the distributional derivative and the nonlinear deformation $\p$ must coincide.
Consequently, $u$ may develop jump discontinuities along the set $J_u$ where $\p$ concentrates.
It is therefore natural to introduce an additive decomposition of the strain fields into regular ($\mathrm{R}$) and singular ($\mathrm{S}$) parts
\begin{equation}
  \label{eq:strain-SR-decomposition}
  \nabla u = \nabla^{\mathrm{R}} u + \nabla^{\mathrm{S}} u, \quad \p = \p^{\mathrm{R}} + \p^{\mathrm{S}},
\end{equation}
where the singular parts are supported on the jump set $J_u$.
The singular parts are such that
\begin{equation}
  \nabla^{\mathrm{S}} u = \p^{\mathrm{S}} = \jump{u} \mathbf{n} \,\delta_{J_u},
  \label{eq:singular-strain-parts}
\end{equation}
where $\jump{u} = u^+ - u^-$ is the displacement jump across $J_u$ compatible with a unit normal vector $\mathbf{n}$.
$\delta_{J_u}$ denotes a measure concentrated on $J_u$ such that for any continuous test function $f$, $\int_{\Omega}f \,\delta_{J_u} \dA = \int_{J_u} f \mathrm{d}s$, where $s$ is the arc-length measure on $J_u$, assuming implicitly that $J_u$ is regular enough to admit a normal almost everywhere.

Using~\eqref{eq:strain-SR-decomposition} and decomposing the integral on $\Omega$ into the sum of bulk integrals on $\Omega\setminus J_u$ where the strain fields are smooth and the surface integral on the jump set $J_u$, the energy functional can be equivalently rewritten explicitly as
\begin{multline}
    \mathcal{E}_\ell(u,\p,\alpha) =
    \int_{\Omega\setminus J_u} \left( \frac{\mu}{2} \|\nabla u - \p\|^2 + \mathsf{k}(\alpha)\,\tau_c \|\p\| \right) \dA
    + \int_{J_u} \mathsf{k}(\alpha)\,\tau_c\, \left|\jump{u}\right| \,\mathrm{d}s\\
    + \frac{\Gc}{4\cw} \int_{\Omega} \left( \frac{\mathsf{w}(\alpha)}{\ell} + \ell \nabla \alpha \cdot \nabla \alpha \right) \dA.
    \label{eq:energyFunctionalAntiplaneSR}
\end{multline}

\begin{remark}[Irreversibility of the nonlinear deformation]
\label{rem:irreversibility}
    The formulation above does not include an irreversibility condition on the nonlinear deformation $\p$, which is then regarded as purely elastic and reversible.
    Imposing an irreversibility condition on $\p$ would lead to a more complex model, akin to a phase-field regularization of an elastoplastic model with softening.
    The corresponding incremental energy  minimization problem would read as
    \begin{equation}
        (u_i, \p_i, \alpha_i) \in \argmin_{\substack{u \in \mathcal{C}_i,\, \alpha \in \mathcal{D}_i},\, \p \in L^1(\Omega)} \mathcal{E}_\ell(u, \p, \alpha;\p_{i-1},\bar{p}_{i-1}),
    \end{equation}
    where  the energy functional is extended to
    \begin{multline}
        \mathcal{E}_\ell(u,\p,\alpha;\p_{\mathrm{old}},\bar{p}_{\mathrm{old}}) =
        \int_{\Omega} \left( \frac{\mu}{2} \|\nabla u - \p\|^2 + \mathsf{k}(\alpha)\,\tau_c (\bar{p}_{\mathrm{old}}+\|\p-\p_{\mathrm{old}}\|)\right) \dA\\
        + \frac{\Gc}{4\cw} \int_{\Omega} \left( \frac{\mathsf{w}(\alpha)}{\ell} + \ell \,\nabla \alpha \cdot \nabla \alpha \right) \dA,
    \end{multline}
$\p_{\mathrm{old}}$ and $\bar{p}_{\mathrm{old}}$ being the nonlinear deformation and the cumulated plastic deformation at the previous loading step, as in variational approaches to fracture coupled with plasticity~\autocite{alessi2014GradientDamageModels,Alessi-Marigo-EtAl-2015a,alessi2017CouplingDamagePlasticity}.
    Given an initial state $(\alpha_0,\p_{0},\bar{p}_0)$, the updating rule for the cumulated plastic deformation is $\bar{p}_{i}:=\bar{p}_{i-1} + \|\p_i-\p_{i-1}\| $.
 Accounting for the plastic irreversibility is not necessary to obtain a cohesive fracture behavior in the limit $\ell \to 0$, as shown in~\autocite{bmmz25}, but can be relevant to model ductile fracture.

\end{remark}

\begin{remark}[Two-field formulation] In the absence of an irreversibility condition on $\p$, we can also consider a two-field formulation in terms of the displacement $u$ and the damage $\alpha$ only, by eliminating the nonlinear deformation $\p$ through partial minimization of the energy functional with respect to $\p$.
This leads to the equivalent formulation
    \begin{equation*}
        (u_i, \alpha_i) \in \argmin_{u \in \mathcal{C}_i, \alpha \in \mathcal{D}_i} \mathcal{E}_\ell(u, \alpha),
    \end{equation*}
    where
    \begin{equation}
      \label{eq:totalEnergy}
        \mathcal{E}_\ell(u,\alpha) =
        \int_{\Omega\setminus J_u} \psi(\nabla u, \alpha) \dA
        +
        \int_{J_u} \mathsf{k}(\alpha)\,\tau_c\, \left|\jump{u}\right| \,\mathrm{d}s
        +
        \frac{\Gc}{4\cw} \int_{\Omega} \left( \frac{\mathsf{w}(\alpha)}{\ell} + \ell \nabla \alpha \cdot \nabla \alpha \right) \dA,
    \end{equation}
    with $\psi(\nabla u, \alpha)$ being the elastic energy density defined above.
\end{remark}
\subsection{First order optimality conditions for local minimizers}
\label{sec:firstOrderOptimality}
A \emph{global minimizer} of the energy  at loading step $i$ satisfies the following  condition:
$$
(u,\p,\alpha)\in \mathcal{C}_i\times \mathcal{P} \times \mathcal{D}_i \quad \text{s.t.} \quad
\mathcal{E}_\ell(u,\p,\alpha)\leq \mathcal{E}_\ell(\hat u,\hat \p,\hat \alpha),\quad \forall
(\hat u,\hat \p,\hat \alpha)\in \mathcal{C}_i\times \mathcal{P} \times \mathcal{D}_i.
$$
Global minimization is useful to characterize the problem within the direct method of the calculus of variations~\autocite{Dacorogna1989}.
However, computing global minimizers of non-convex functionals is typically intractable numerically, and physically questionable: it would require the system to overcome arbitrarily large energy barriers to reach the absolute minimum.

The notion   of a \emph{local minimizer} is relevant from the physical point of view.
We therefore state the optimality conditions at the time step $t_i$ through the \emph{directional local minimality} condition~\autocite{pham2011GradientDamageModels,baldellimaurini2020}:
\begin{multline} \text{Find } (u,\p,\alpha)\in \mathcal{C}_i\times \mathcal{P} \times \mathcal{D}_i \text{ such that } \forall  (\hat u,\hat \p,\hat \alpha)\in \mathcal{C}_i\times \mathcal{P} \times \mathcal{D}_i, \exists {\bar{h} >0} :\\
    \forall 0 <  h < \bar{h},\ \mathcal{E}_\ell(u+h(\hat u-u),\p+h(\hat\p-\p),\alpha+h(\hat \alpha-\alpha))-\mathcal{E}_\ell(u,\p,\alpha)\geq 0.
    \label{eq:local-minimizer}
\end{multline}

A set of \emph{first-order necessary optimality conditions} for the minimization of $\mathcal{E}_\ell(u,\p,\alpha)$ at loading step $i$ can be obtained by taking a first order expansion of \eqref{eq:local-minimizer} in $h>0$.
Because of the box constraint on $\alpha\in[\alpha_{i-1},1]$, the non-smoothness of the energy functional with respect to $\p$ and $\jump{u}$, the optimality conditions are expressed as variational inequalities.
To obtain pointwise strong form of the optimality conditions, we distinguish the subset $J_u$ of $\Omega$ where $u$ jumps and its complement $\Omega \setminus J_u$, where $u$ is smooth enough to admit classical derivatives.
For the sake of simplicity, we assume that $J_u\cap\partial_u\Omega=\emptyset$, otherwise the Dirichlet boundary conditions should be relaxed to account for possible jumps on the Dirichlet boundary~\autocite[see \emph{e.g.}][]{MORA2016725}.

\subsubsection{Optimality condition for the displacement field $u$: mechanical equilibrium}
Setting $\hat{\p}=\p$, $\hat{\alpha}=\alpha$, and $\hat{u}=u+\hat{v}$ with $\hat{v}$ in the vector space $\mathcal{C}_0:=\{v \in BV(\Omega): v = 0 \text{ on } \partial_u \Omega,\; J_v\subseteq J_u\}$ of admissible variations, the perturbed displacement in~\eqref{eq:local-minimizer} is $u_h:=u+h\,\hat{v}$, and the first order expansion in $h$ leads to
\begin{equation*}
    \int_{\Omega\setminus J_u} \t \cdot \nabla \hat{v} \dA + \int_{J_u}
        \mathsf{k}(\alpha)\,\tau_c \, \text{sgn}\left(\jump{u}\right)
         \,\jump{\hat{v}}\,\mathrm{d}s \geq  0, \quad \forall \hat{v} \in \mathcal{C}_0.
\end{equation*}
    which, since $\mathcal{C}_0$ is a vector space, must hold as an equality (variations jumping outside $J_u$ only return the strength condition $\vert\t\cdot\mathbf{n}\vert\le\k(\alpha)\,\tau_c$ on the candidate jump surface), that can be rewritten as\footnote{Here we use that, for $J_u\cap\partial\Omega=\emptyset$,
    $$
     \int_{\Omega\setminus J_u} \t \cdot \nabla v \dA
     = -\int_{\Omega\setminus J_u} \mathrm{div}(\t) v \dA + \int_{\partial \Omega} (\t v) \cdot \mathbf{n}  \,\mathrm{d}s- \int_{J_u} \jump{\t v} \cdot \mathbf{n}  \,\mathrm{d}s,
    $$
    and $\jump{\t v}=\jump{\t} \avg{v}+\avg{\t} \jump{v}$.}
    \begin{multline*}
    -\int_{\Omega\setminus J_u} \mathrm{div}(\t) v \dA + \int_{\partial \Omega} (\t \cdot \mathbf{n})v  \,\mathrm{d}s-\int_{J_u}\jump{\t} \cdot \mathbf{n} \avg{v} \,\mathrm{d}s\\
    +\int_{J_u}\left( \mathsf{k}(\alpha)\,\tau_c \, \text{sgn}\left(\jump{u}\right)-\avg{\t} \cdot \mathbf{n} \right)\jump{v} \,\mathrm{d}s= 0,
    \end{multline*}
where $\t=\mu (\nabla u - \p)$ denotes the stress vector, and $\avg{(\cdot)} = ((\cdot)^+ +(\cdot)^-)/2$.

By the arbitrariness of the test function $v\in \mathcal{C}_0$, we obtain the strong form of the mechanical equilibrium equations.
Taking $v$ with $\jump{v}=0$ on $J_u$ gives
\begin{subequations}
    \begin{equation}
        \begin{cases}
    \mathrm{div}(\t) = 0 \text{ in } \Omega\setminus J_u,\\
    \t \cdot \mathbf{n} = 0 \text{ on } \partial \Omega\setminus \partial_u \Omega,\\
    \jump{\t} \cdot \mathbf{n} = 0 \text{ on } J_u,
    \end{cases}
    \label{eq:equilibrium}
    \end{equation}
 while taking $\jump{v}$ arbitrary on $J_u$ gives:
    \begin{equation}
    {\t} \cdot \mathbf{n}=\mathsf{k}(\alpha)\,\tau_c \, \text{sgn}\left(\jump{u}\right) \text{ on } J_u,
    \label{eq:equilibriumJu}
    \end{equation}
\end{subequations}
where $\t=\avg{\t}$  on $J_u$ because of~\eqref{eq:equilibrium}.
\subsubsection{Optimality condition for the nonlinear deformation $\p$}
Setting $\hat{u}=u$, $\hat{\alpha}=\alpha$, and $\hat{\p}=\p + \q$ with $\q\in \mathcal{M}(\Omega;\R^2)$ absolutely continuous with respect to the area measure --- the singular part of $\p$ is tied to $\jump{u}$ by~\eqref{eq:singular-strain-parts} and is varied through $\hat u$ --- the perturbed field in~\eqref{eq:local-minimizer} is $\p_h:=\p+h\,\q$, and the first order expansion in $h$ leads to, for all such $\q$,
\begin{equation}
        \int_{(\Omega\setminus J_u) \cap \{\Vert\p\Vert> 0\}} \left( -\t + \mathsf{k}(\alpha)\,\tau_c \dfrac{\p}{\|\p\|} \right) \cdot \q \dA +
        \int_{(\Omega\setminus J_u) \cap \{\Vert\p\Vert= 0\}} \left( -\t\cdot {\q} + \mathsf{k}(\alpha)\,\tau_c {\|\q\|} \right)  \dA
        \ge 0,
\end{equation}
where we have distinguished the subsets of $\Omega\setminus J_u$ where $\|\p\|$ is strictly positive or vanishing to account for the non-smoothness of the energy with respect to $\p$.
This leads to the following pointwise optimality conditions almost everywhere in $\Omega\setminus J_u$:

\begin{equation*}
\text{ in }\Omega\setminus J_u:    \begin{cases}
         \t =\mathsf{k}(\alpha)\,\tau_c \dfrac{\p}{\|\p\|} & \text{ where } \|\p\| > 0  \\
        \|\t\| \le \mathsf{k}(\alpha)\,\tau_c & \text{ where } \|\p\| = 0
    \end{cases},
\end{equation*}
which using that $\t=\mu(\nabla u-\p)$ can be equivalently rewritten as the \emph{flow rule} for the nonlinear deformation $\p$:
\begin{equation}
\text{ in }\Omega\setminus J_u:  \p=
 \begin{cases}
    \mathbf{0} & \text{ if } \Vert\t\Vert=\mu\|\e\| \le \k(\alpha)\,{\tau_c},\\
    \left(\ \|\e\|- \k(\alpha)\dfrac{\tau_c}{\mu}\right)\dfrac{\e}{\|\e\|}& \text{ otherwise.}
\end{cases}
\label{eq:flow-rule}
\end{equation}

\subsubsection{Optimality condition for the damage field $\alpha$}
Setting $\hat{u}=u$, $\hat{\p}=\p$, and perturbing the damage as $\alpha_h:=\alpha + h (\hat{\alpha}-\alpha)$ with $\hat{\alpha} \in \mathcal{D}_i$, the first order expansion of \eqref{eq:local-minimizer} in $h$ leads to,
$\forall \hat{\alpha}\in \mathcal{D}_i$,
    \begin{multline*}
        \int_{\Omega\setminus J_u}
        \left(
        \left(\k'(\alpha)\,\tau_c \|\p\|+  \frac{\Gc}{4\cw\ell}{\mathsf{w}'(\alpha)}
        \right)\, (\hat{\alpha}- \alpha)
         +
         \left(
            \frac{\Gc\ell}{2\cw}\,\nabla\alpha\cdot\nabla(\hat{\alpha}- \alpha) \right) \right) \dA \\
        + \int_{J_u}\k'(\alpha)\,\tau_c \,\left|\jump{u}\right|\,(\hat{\alpha} - \alpha) \,\mathrm{d}s \ge 0.
    \end{multline*}
Integrating by parts the term involving $\nabla \alpha$ and rearranging, we obtain
\begin{multline*}
        \int_{\Omega\setminus J_u}
        \left(
        \k'(\alpha)\,\tau_c \|\p\| +  \frac{\Gc}{4\cw} \left( \frac{\mathsf{w}'(\alpha)}{\ell} - 2\ell\,\Delta \alpha \right)
        \right)\, (\hat{\alpha}- \alpha) \dA \\
        + \int_{\partial \Omega \setminus \partial_\alpha \Omega} \frac{\Gc\ell}{2\cw}\,\nabla \alpha \cdot \mathbf{n} \,(\hat{\alpha}- \alpha) \,\mathrm{d}s
        + \int_{J_u}\left( \k'(\alpha)\,\tau_c \,\left|\jump{u}\right| - \frac{\Gc\ell}{2\cw}\,\jump{\nabla \alpha}\cdot\mathbf{n} \right)\,(\hat{\alpha} - \alpha) \,\mathrm{d}s \ge 0,
\end{multline*}
where $\jump{\nabla \alpha}\cdot\mathbf{n}$ is the jump of the normal derivative of $\alpha$ across $J_u$.
This leads to the following pointwise optimality conditions  in $\Omega\setminus J_u$ wherever $\alpha<1$.
\begin{subequations}
    \begin{equation}
         \text{ on } \Omega\setminus J_u:
        \begin{cases}
        \alpha-\alpha_{i-1}\ge 0, &\\
    \k'(\alpha)\,\tau_c \|\p\| + \dfrac{\Gc}{4\cw} \left( \dfrac{\mathsf{w}'(\alpha)}{\ell} - 2\ell\,\Delta \alpha \right)\ge 0, &\\
    \left(\alpha-\alpha_{i-1} \right)
    \left(\k'(\alpha)\,\tau_c \|\p\| + \dfrac{\Gc}{4\cw} \left( \dfrac{\mathsf{w}'(\alpha)}{\ell} - 2\ell\,\Delta \alpha \right) \right)=0.
     &
        \end{cases}
        \label{eq:alpha-optimality-bulk}
    \end{equation}
with the boundary condition $\alpha-\alpha_{i-1} \geq 0$, $\nabla \alpha \cdot \mathbf{n} \geq 0$ and $(\alpha-\alpha_{i-1} )(\nabla \alpha \cdot \mathbf{n}) = 0$
    on $\partial \Omega \setminus \partial_\alpha \Omega$.
On the jump set $J_u$, the conditions above must be read in the sense of distributions as
    \begin{equation}
         \text{ on } J_u:
        \begin{cases}
        \alpha-\alpha_{i-1} \ge 0,&\\
    \k'(\alpha)\,\tau_c \,\left|\jump{u}\right| - \dfrac{\Gc\ell}{2\cw}  \,\jump{\nabla \alpha}\cdot\mathbf{n}  \ge 0, &\\
      \left(\alpha-\alpha_{i-1} \right)
    \left(\k'(\alpha)\,\tau_c \left|\jump{u}\right| -
    \dfrac{\Gc\ell}{2\cw}  \,\jump{\nabla \alpha}\cdot\mathbf{n}  \right)=0.
     &
        \end{cases}
        \label{eq:alpha-optimality-jump}
    \end{equation}
\end{subequations}

In the regions where $\alpha=1$, the necessary optimality condition reduces to the reversed inequality $$  \k'(\alpha)\,\tau_c \|\p\| + \dfrac{\Gc}{4\cw} \left( \dfrac{\mathsf{w}'(\alpha)}{\ell} - 2\ell\,\Delta \alpha \right)\leq 0,$$
and similarly on the Neumann boundary and the jump set $J_u$.

\begin{remark}
    In the damage criterion~\eqref{eq:alpha-optimality-bulk}, the nonlinear deformation $\p$ is the only \emph{driving force} for damage, entering through the term $\k'(\alpha)\,\tau_c\,\|\p\|$ with $\k'(\alpha)<0$. In turn, the flow rule~\eqref{eq:flow-rule} yields $\p\neq\mathbf{0}$ only where the norm of the stress $\Vert\t\Vert=\mu\|\e\|$ exceeds the current strength $\k(\alpha)\,\tau_c$. The two conditions combine into a stress criterion for damage nucleation: damage cannot grow until the stress attains the strength $\k(\alpha)\,\tau_c$.
\end{remark}

\subsection{Specific constitutive models}

The properties of the model depend on the strength degradation function $\k$ and the dissipation function $\w$.
We ask that they satisfy the following minimal requirements.

\begin{hypothesis}[Constitutive assumption: strength degradation and dissipation function]
\label{hyp:kw}
We assume that the strength domain degrades homothetically, \emph{i.e.} $\K(\alpha) = \k(\alpha)\, \K_0$, and that the constitutive functions $\k: \alpha \in [0,1] \to [0,1]$ and $\w: \alpha \in [0,1] \to [0,1]$ are continuous on $[0,1]$ and differentiable on $(0,1)$, with
\begin{equation}
  \label{eq:kwassumptions}
 \w'(\alpha)> 0,\quad\k'(\alpha)< 0\qquad\forall{\alpha\in(0,1)},
\end{equation}
with $\k(0)=1$, $\k(1)=0$, $\w(0)=0$, and $\w(1)=1$.
We further assume the monotonicity conditions
\begin{equation}
    \begin{aligned}
    	\textnormal{\textsf{(SH)}}:\quad &
    \alpha\mapsto\dfrac{{\w'(\alpha)}}{\k'(\alpha)}
    \ \text{is strictly decreasing on $(0,1)$},\\
    	\textnormal{\textsf{(SS)}}:\quad &
    \alpha\mapsto\dfrac{\sqrt{\w(\alpha)}}{\k'(\alpha)}
    \ \text{is strictly decreasing on $(0,1)$}.
    \end{aligned}
    \label{eq:slopesfromkw}
\end{equation}
\end{hypothesis}
When $\k$ and $\w$ are twice differentiable, conditions~\eqref{eq:slopesfromkw} take the explicit form
\begin{equation}
    \begin{aligned}
    &\dfrac{\mathrm{d}}{\mathrm{d}\alpha}\dfrac{{\w'(\alpha)}}{\k'(\alpha)}
    =\frac{\w''(\alpha)\k'(\alpha)-\w'(\alpha)\k''(\alpha)}{\k'(\alpha)^2} < 0,\\
    &\dfrac{\mathrm{d}}{\mathrm{d}\alpha}\dfrac{\sqrt{\w(\alpha)}}{\k'(\alpha)}
    =
    \frac{\w'(\alpha)\k'(\alpha)-2\,\w(\alpha)\k''(\alpha)}{2\sqrt{\w(\alpha)}\,\k'(\alpha)^2}
    < 0,
    \end{aligned}
    \qquad\forall \alpha\in(0,1).
    \label{eq:slopesfromkw-c2}
\end{equation}
Under Hypothesis~\ref{hyp:kw}, the model exhibits \emph{strain hardening} (\textsf{SH}) in the homogeneous response and \emph{stress softening} (\textsf{SS}) in the equivalent cohesive law, as will be shown in Sections~\ref{sec:homogeneousSolutions} and~\ref{sec:localizedSolutions}, respectively. The reader can refer to~\autocite{pham2011GradientDamageModels} for the definition of analogous conditions in the context of gradient damage models.
In the twice-differentiable case, under~\eqref{eq:kwassumptions}, \textsf{(SS)} holds whenever $\k$ is convex ($\k''\geq 0$); if in addition $\w$ is convex, with $\w''$ and $\k''$ not simultaneously vanishing, \textsf{(SH)} holds as well. Convexity is however not necessary.

\section{Analysis of a simple shear problem}
\label{sec:modelProblem}
To investigate the behavior of the model, we consider a simple shear model problem consisting of a rectangular domain $\Omega = (-L/2,L/2)\times(-H/2,H/2)$ clamped on the left side and subjected to a prescribed displacement $\bar{u} $ on the right side, with free boundary conditions on the other sides and an initial undamaged state $\alpha_0 = 0$.

This problem can be interpreted as a shear test on a rectangular specimen of height $H$ and length $L$.
It is the antiplane analogue of the traction test considered in~\autocite{pham2011GradientDamageModels} and related works.

In this setting, we look for families of solutions ($u_t$,$\p_t$,$\alpha_t$) parametrized by $t>0$\footnote{Symmetric solutions would be obtained by considering $t<0$.}, with the imposed displacement  $\bar{u}=t\,L$.
To this end, we compute solutions of the first-order optimality conditions for the following static problem:
\begin{equation}
    (u_t, \p_t, \alpha_t) \in \argmin_{\substack{u \in \mathcal{C}_t,\, \alpha \in \mathcal{D}_0},\, \p \in \mathcal{P}} \mathcal{E}_\ell(u, \p, \alpha),
\end{equation}
with
\begin{eqnarray}
    \mathcal{C}_t &=& \{u \in BV(\Omega): u(-{L}/{2},y) = 0  \text{ and } u({L}/{2},y) = t\,L,\quad\forall y\in(-H/2,H/2)\}, \\
    \mathcal{D}_0 &=& \{\alpha \in H^1(\Omega): 0 \le \alpha \le 1 \text{ in } \Omega\}.
\end{eqnarray}
In particular, to allow for solutions with a non-zero homogeneous damage field, we do not impose Dirichlet boundary conditions on the damage.
Moreover, we only impose $\alpha\geq 0$ in the construction of the solutions.
We will then check the irreversibility condition $\alpha_t \ge \alpha_s$ for $t \ge s$ a posteriori.

We will look for two classes of solutions: (i) \emph{homogeneous solutions}, where the fields are constant in space, and (ii) \emph{localized solutions}, where the nonlinear deformation and damage localize along a band across the specimen.
The homogeneous solutions will provide insight into the equivalent local material response and the \emph{strength}, while the localized solutions will shed light on crack nucleation and the equivalent \emph{toughness}.

In all that follows, we look for fields that are invariant along the $y$-direction.
With an abuse of notation, we write $u(x)=u(x,0)$ and $\alpha(x)=\alpha(x,0)$, and denote by $\varepsilon(x) = u'(x)$ the strain field, where the prime denotes the derivative with respect to $x$.
The stress and nonlinear deformation fields are of the form $\t(x,0)=\tau(x)\,\underline{e}_1$ and $\p(x,0)=p(x)\,\underline{e}_1$, which defines the one-dimensional fields $\tau(x)$ and $p(x)$.

\subsection{Homogeneous solutions: material response}
\label{sec:homogeneousSolutions}
We first investigate homogeneous solutions, \emph{i.e.}~solutions where the deformations, the stress and the damage fields are constant in space, which is equivalent to focusing on  a material point subjected to a prescribed strain.

For the given geometry and boundary conditions, the displacement field must be of the form $u(x) = \bar{u} \,x/L$, so that the strain field is $\veps = \bar{u}/L=t>0$.
Mechanical equilibrium~\eqref{eq:equilibrium} and constitutive relation~\eqref{eq:constitutiveRelationAntiplane} together with the flow rule then give
\begin{subequations}
\begin{equation}
    p=\max\left(\veps - \k(\alpha)\dfrac{\tau_c}{\mu},\,0\right),
    \qquad
    \tau = \mu(\veps-p)=\min\left(\mu\,\veps,\;\k(\alpha)\,\tau_c\right).
\label{eq:homogeneous-flow-rule}
\end{equation}
The damage criterion~\eqref{eq:alpha-optimality-bulk}  reduces to the following pointwise condition:
\begin{equation}
    \label{eq:optFellConstant}
    \tau_c\,\k'(\alpha)\,\vert p\vert+ \dfrac{\Gc\,\w'(\alpha)}{4\cw\ell}
    \begin{cases}
        \ge 0 & \text{if } \alpha = 0, \\
        = 0 & \text{if } 0 < \alpha <1, \\
        \le 0 & \text{if } \alpha = 1.
    \end{cases}
\end{equation}
\end{subequations}
Since  $\k'(0)< 0$ and $\w'(0)\geq 0$,  we get that $\alpha = 0$ as long as
\begin{equation}
    \label{eq:elasticDomain}
   \vert p\vert \le p_c:=-\dfrac{\Gc}{\ell\,\tau_c}\dfrac{\w'(0)}{4\cw\k'(0)}
   \quad\Leftrightarrow\quad
   t \le \veps_c := \dfrac{\tau_c}{\mu} - \dfrac{\Gc}{\ell\,\tau_c}\dfrac{\w'(0)}{4\cw \k'(0)},
\end{equation}
and $p = 0$ as long as $t \le \veps_e := \frac{\tau_c}{\mu} \le \veps_c$.

For $t > \veps_e$, undamaged solutions are no longer admissible and $\alpha\in(0,1)$ must satisfy the following condition obtained by combining~\eqref{eq:optFellConstant} and~\eqref{eq:homogeneous-flow-rule}:
\begin{equation}
    \label{eq:elasticDomain2}
   t = \dfrac{\tau_c}{\mu}\k(\alpha) - \dfrac{\Gc}{\tau_c\ell}\dfrac{\w'(\alpha)}{4\cw\k'(\alpha)}.
\end{equation}
Summarizing the above, the homogeneous response is composed of four regimes:
\begin{equation}
    \label{eq:homogeneousRegimes}
    \begin{cases}
        \text{Elastic} & : t\in[0, \veps_e ), \quad \alpha = 0, \; p=0, \; \tau = \mu\veps \smallskip\\
        \text{Constant-stress} &: t\in[\veps_e, \veps_c) , \quad \alpha = 0, \; p \leq p_c, \;\tau = \tau_c \smallskip\\
        \text{Strength softening} &:t\in[\veps_c,\veps_u), \quad
             \alpha\text{ solves } \eqref{eq:elasticDomain2},\;
                p = \veps - \k(\alpha)\dfrac{\tau_c}{\mu}, \;
                \tau = \k(\alpha)\tau_c,\smallskip\\
        \text{Fully damaged} & :t\geq \veps_u,\quad \alpha = 1, \quad p = \veps, \quad \tau = 0,
    \end{cases}
\end{equation}
where the critical strains are given by
\begin{equation}
    \veps_e := \dfrac{\tau_c}{\mu},\quad
    \veps_c := \dfrac{\tau_c}{\mu} - \dfrac{\Gc}{\ell\,\tau_c}\dfrac{\w'(0)}{4\cw \k'(0)},\quad
    \veps_u :=
    \begin{cases}
        -\dfrac{\Gc}{\tau_c\ell} \dfrac{\w'(1)}{4\cw\k'(1)},\quad &\text{if }\k'(1)<0,\\
        +\infty, &\text{if }\k'(1)=0,
    \end{cases}
\end{equation}
using the fact that $\k(1)=0$.

    Note that~\eqref{eq:elasticDomain2} admits a unique solution $\alpha(t)$ satisfying the irreversibility condition if and only if its right-hand side is an increasing function of $\alpha$, and that a snap-back occurs otherwise.
    Assuming that $\w'/\k'$ is differentiable, this implies that
    \[
    \frac{\tau_c}{\mu} \k'(\alpha) - \frac{\Gc}{\tau_c \ell} \frac{1}{4\cw}\frac{\mathrm{d}}{\mathrm {d} \alpha}\left(\frac{\w'(\alpha)}{\k'(\alpha)}\right) > 0
    \]
    almost everywhere on $(0,1)$.
We rewrite this condition as
    \begin{equation}
    \label{eq:homogeneous-no-snapback-ellch}
    \frac{\ell}{\lch}
    <
    \frac{1}{4\cw\k'(\alpha)}
    \frac{\mathrm d}{\mathrm d\alpha}\left(\frac{\w'(\alpha)}{\k'(\alpha)}\right),
\end{equation}
with $\lch := \frac{\mu \Gc}{\tau^2_c}$ denoting the \emph{elasto-cohesive length} of the material, the antiplane analogue of Hillerborg's characteristic length~\autocite{hillerborg1976AnalysisCrackFormation}.
A necessary (but not sufficient) condition is that $\frac{\mathrm d}{\mathrm d\alpha}\left(\frac{\w'(\alpha)}{\k'(\alpha)}\right) < 0$, which is hypothesis \textnormal{\textsf{(SH)}} in~\eqref{eq:slopesfromkw}.

\begin{remark}[Snap-back and convexity of the bulk energy density at given strain]
\label{rem:homogeneous-snapback-convexity}

    In the strength-softening regime, we have $\e - p = \k(\alpha)\frac{\tau_c}{\mu}$ so that the total energy density is
    \begin{equation*}
    W(\veps,\alpha):=\psi(\veps,\alpha)+\frac{\Gc}{4\cw\ell}\w(\alpha)
    =\k(\alpha)\,\tau_c\,\veps-\frac{\k(\alpha)^2\tau_c^2}{2\mu}+\frac{\Gc}{4\cw\ell}\w(\alpha).
\end{equation*}
Differentiating with respect to $\alpha$ for a fixed $\e$ and noticing that $\frac{\partial p}{\partial \alpha} = -\k'(\alpha) \frac{\tau_c}{\mu}$, we get that
    \[
        \frac{\partial W}{\partial \alpha}(\e,\alpha)  = \k'(\alpha) \tau_c p + \frac{\Gc}{4\cw \ell}\w'(\alpha),
    \]
    and
    \[
        \frac{\partial^2W }{\partial \alpha^2}(\e,\alpha)  = \k''(\alpha) \tau_c p - \frac{\k'(\alpha)^2\tau_c^2}{\mu} + \frac{\Gc}{4\cw \ell}\w''(\alpha).
    \]
    Substituting $\w'' = \k''\w'/\k'+ \k'\,\tfrac{\mathrm{d}}{\mathrm{d}\alpha}(\w'/\k')$, and using the fact that since $ 0 < \alpha < 1$, optimality with respect to $\alpha$ implies that $\frac{\partial W}{\partial \alpha}(\e,\alpha) = 0$, we can obtain that

    \begin{align}
        \label{eq:SHvsEScompetition}
        \frac{\partial^2 W(\e,\alpha)}{\partial \alpha^2}  & = \frac{\Gc}{4\cw\ell} \k'(\alpha) \frac{\mathrm d}{\mathrm d\alpha}\left(\frac{\w'(\alpha)}{\k'(\alpha)}\right) - \frac{\tau_c^2\k'(\alpha)^2}{\mu}\\
        \notag & = \frac{\k'(\alpha)^2 \Gc}{\ell} \left( \frac{1}{4\cw\k'(\alpha)}\frac{\mathrm d}{\mathrm d\alpha}\left(\frac{\w'(\alpha)}{\k'(\alpha)}\right) - \frac{\ell}{\ell_\mathrm{ch}}\right)
    \end{align}
    so that condition~\eqref{eq:homogeneous-no-snapback-ellch} is equivalent to the strict convexity of the total energy with respect to $\alpha$ in the strength softening phase.
    It is also equivalent to mandating a negative tangential stiffness, $\frac{\mathrm {d}\tau}{\mathrm{d}t} < 0$.

    Note that~\eqref{eq:SHvsEScompetition} can be seen as the outcome of the competition between elastic softening ${\tau_c^2\k'(\alpha)^2/\mu}$, which is always destabilizing, and strain hardening     $\frac{\Gc}{4\cw\ell} \k'(\alpha) \frac{\mathrm d}{\mathrm d\alpha}\left(\frac{\w'(\alpha)}{\k'(\alpha)}\right)$.

  \end{remark}

\subsection{Localized solutions: equivalent cohesive law}
\label{sec:localizedSolutions}

Assume now that the damage field $\alpha(x)$ is not constant in space and attains a maximum $\alpha^*$ at a single point $x^* \in (-L/2,L/2)$.
In this setting, mechanical equilibrium implies that the shear stress $\tau(x)$ must be constant in space:
\begin{equation}
    \label{eq:1d-mechanical-equilibrium}
    \tau = \mu\left(\veps(x) - p(x)\right) = \text{constant}.
\end{equation}
The \emph{strength criterion} with $\k'(\alpha)<0$ and $\alpha^*>0$ implies that
\begin{equation}
    \label{eq:1d-strength-criterion}
    \tau \leq \k(\alpha^*)\tau_c<\tau_c,\qquad
    \tau \leq \k(\alpha^*)\tau_c<\k(\alpha(x))\tau_c,\quad \forall x\neq x^*,
\end{equation}
since $\k$ is decreasing and $\alpha(x)<\alpha^*$ for $x\neq x^*$.
Because of the flow rule~\eqref{eq:flow-rule},  the nonlinear strain $p(x)$ can be non-zero only at $x=x^*$ where $\tau\leq\k(\alpha^*)\tau_c$, while $\tau<\k(\alpha(x))\tau_c$ everywhere else.
Moreover, the displacement field must verify the \emph{loading condition}:
\begin{equation}
   t=
    \frac{u(L/2) - u(-L/2)}{L}
   =\frac{1}{L}\int_{-L/2}^{L/2} \veps(x)\,\mathrm{d}x = \frac{1}{L}\int_{-L/2}^{L/2} \left( \frac{\tau}{\mu} + p(x) \right) \,\mathrm{d}x= \frac{\tau}{\mu} + \frac{\jump{u}}{L},
\end{equation}
where we used the fact that $p=0$ for all $x\neq x^*$ and that~\eqref{eq:singular-strain-parts} mandates that $\jump{u}(x^*) = p^S(x^*) \not = 0$, \emph{i.e.} that the regular part of $p$ is identically zero and
\begin{equation}
    \label{eq:plastic-strain-decomposition}
    p(x) = p^S(x) = \jump{u}\,\delta_{x^*}(x),
\end{equation}
where $\delta_{x^*}$ is the Dirac delta distribution centered at $x=x^*$.
Normalizing the load by $\veps_e$, the loading condition above reads $t/\veps_e=\tau/\tau_c+\left(\jump{u}\,\tau_c/\Gc\right)\,(\lch/L)$, so that the global response of the bar depends on the brittleness ratio $L/\lch$ alone.

In the \emph{regular regions} $x\in(-L/2,x^*)\cup(x^*,L/2)$, where $\alpha(x)<\alpha^*\leq 1$ is smooth and $p(x)=0$, the {damage criterion} reduces to
\begin{equation}
    \label{eq:1d-damage-criterion-regular}
    \alpha(x)\geq0 ,\quad \w'(\alpha(x))- 2\ell^2\,\alpha''(x)\geq 0,\quad \left(\w'(\alpha(x))- 2\ell^2\,\alpha''(x)\right)\alpha(x)=0.
\end{equation}
The solution of this problem is classical in phase-field fracture models~\autocite[see \emph{e.g.}][]{pham2011GradientDamageModels}.
Since by regularity, either $\w'(\alpha(x))= 2\ell^2\,\alpha''(x)$ or $\alpha(x)=\alpha'(x)=0$ for $x\neq x^*$,
$$\left(\w'(\alpha(x))- 2\ell^2\,\alpha''(x)\right)\alpha'(x)=0$$ holds everywhere in the regular regions, which yields the first integral
\begin{equation}
    \ell^2\,\alpha'(x)^2  = {\w(\alpha(x))} - c_0,\quad \forall x\in(-L/2,x^*)\cup(x^*,L/2),
     \label{eq:firstIntegral}
\end{equation}
which is nothing but the \emph{optimal profile} problem for the Ambrosio--Tortorelli surface energy term (see~\autocite{Marigo-Maurini-EtAl-2016a} for instance).
We look for solutions where the damage is non-zero only in an interval of width $D<L$ centered at $x=x^*$, with $\alpha(x)=0$ for $\vert x - x^*\vert \geq D/2$.
For such solutions, the constant $c_0$ can be computed in the region where $\alpha=0$, giving $c_0=0$.
Hence, the integration of~\eqref{eq:firstIntegral} provides the profile of $\alpha(x)$ in the regular regions and the width of the damaged interval:
\begin{equation}
    \label{eq:1d-damage-profile-regular}
   \vert x - x^*\vert = \ell\int^{\alpha^*}_{\alpha(x)} \frac{\,\mathrm{d}\alpha}{\sqrt{ \w(\alpha)}},\quad
    D(\alpha^*) = 2\ell\int^{\alpha^*}_{0} \frac{\,\mathrm{d}\alpha}{\sqrt{ \w(\alpha)}}.
\end{equation}
The width $D(\alpha^*)$ is finite if and only if $\int_0 \mathrm{d}\alpha/\sqrt{\w(\alpha)}<\infty$, \emph{i.e.} if $\w'(0)>0$.
When $\w'(0)=0$, as for the model $\mathsf{M1}$ with $\zeta=1$, the optimal profile has unbounded support with exponential tails of width of order $\ell$, and $c_0=0$ holds only in the limit $L/\ell\to\infty$; the closed-form expressions below remain valid up to corrections of order $e^{-L/\ell}$, negligible for the values of $L/\ell$ used in Section~\ref{sec:numericalSimulations}.

At the displacement jump $x=x^*$, where $\alpha(x^*)=\alpha^*>0$ by hypothesis, the damage criterion reads as
\begin{equation}
    \label{eq:1d-damage-criterion-jump-set}
     \k'(\alpha^*)\tau_c\,\jump{u} - \frac{\Gc\ell}{2\cw} \jump{\alpha'} =0
\end{equation}
while the first integral~\eqref{eq:firstIntegral} gives
\begin{equation}
    \ell\,\jump{\alpha'}  =-2\sqrt{ \w(\alpha^*)}.
     \label{eq:firstIntegralJump}
\end{equation}
Eliminating $\jump{\alpha'}$ from~\eqref{eq:1d-damage-criterion-jump-set}--\eqref{eq:firstIntegralJump}, the flow rule at $x=x^*$ gives the following \emph{cohesive law} relating the stress $\tau$ to the displacement jump $\jump{u}$ at $x=x^*$, parametrized via the maximal damage value $\alpha^*$:
\begin{equation}
    \label{eq:1d-cohesive-law}
    \tau = \k(\alpha^*)\tau_c,\quad
    \text{with}\quad
     {\jump{u}}  = -\frac{\Gc}{\cw\tau_c}\frac{\sqrt{\w(\alpha^*) }}{\k'(\alpha^*)}.
\end{equation}
The global force-displacement response of the bar can then be obtained by combining the cohesive law~\eqref{eq:1d-cohesive-law} with the loading condition:
\begin{equation}
    \label{eq:1d-force-displacement}
    t = \frac{\tau}{\mu} + \frac{\jump{u} }{L}
    = \frac{\k(\alpha^*)\tau_c}{\mu} - \frac{1}{L}\frac{\Gc}{\cw\tau_c}\frac{\sqrt{\w(\alpha^*) }}{\k'(\alpha^*)}.
\end{equation}

Again, \eqref{eq:1d-force-displacement} admits a unique solution satisfying the irreversibility condition if and only if its right-hand side is an increasing function of $\alpha^*$.
Assuming again that $\sqrt{\w}/\k'$ is differentiable, this implies that
\[
    \frac{\k'(\alpha^*)\,\tau_c}{\mu} - \frac{1}{L}\frac{\Gc}{\cw \tau_c} \frac{\mathrm d}{{\mathrm d} \alpha^*}\frac{\sqrt{\w(\alpha^*)}}{\k'(\alpha^*)} > 0
\]
or equivalently that
\begin{equation}
    \label{eq:localized-no-snapback-ellch}
    \frac{L}{\lch}
    <
    \frac{1}{\cw\k'(\alpha^*)}
    \frac{\mathrm d}{\mathrm d\alpha^*}\left(\frac{\sqrt{\w(\alpha^*)}}{\k'(\alpha^*)}\right).
\end{equation}
As in the homogeneous response case, a necessary (but not sufficient) condition for \eqref{eq:localized-no-snapback-ellch} is that $    \frac{\mathrm d}{\mathrm d\alpha^*}\left(\frac{\sqrt{\w(\alpha^*)}}{\k'(\alpha^*)}\right) <0$, which is the stress softening hypothesis (\textsf{SS}).

\subsubsection*{Cohesive surface energy}
Because the nonlinear deformation is concentrated at $x^*$ and the damage profile is determined by $\alpha^*$ through~\eqref{eq:1d-damage-profile-regular}, the energy of the localized solution splits, per unit length in the $y$-direction, into one term coming from the {bulk} and two {surface} contributions:
\begin{equation}
    \label{eq:1d-energies}
    \mathcal{E}(\jump{u},\alpha^*)=\frac{\mu\,L}{2}\left(t-\frac{\jump{u}}{L}\right)^2+
    \phi(\jump{u},\alpha^*),
\end{equation}
the first term being the elastic energy of the bar, whose elastic strain $\veps-p=t-\jump{u}/L=\tau/\mu$ is uniform.
Writing $\delta:=\jump{u}$ for the opening, we further decompose the surface term in an elastic and a dissipated part, as in~\eqref{eq:surface-energy-densities-limit-model}:
\begin{equation}
    \label{eq:cohesive-energy-density}
    \phi(\delta,\alpha^*)=
    \underbrace{\k(\alpha^*)\,\tau_c\,\delta}_{\textstyle\phi_{\mathrm{coh}}}
    +\underbrace{\Gc\,\hat\alpha(\alpha^*)}_{\textstyle\phi_{\mathrm{dmg}}},\qquad{\text{with}}\qquad
    \hat\alpha(\alpha^*)=\dfrac{1}{\cw}\int_0^{\alpha^*}\sqrt{\w(\beta)}\,\mathrm{d}\beta.
\end{equation}
This is precisely the surface energy density~\eqref{eq:surface-energy-densities-limit-model} of the conjectured limit cohesive model, here parametrized by $\alpha^*$ rather than by $\hat\alpha=\hat\alpha(\alpha^*)$; in particular, it depends neither on the regularization length $\ell$ nor on the shear modulus $\mu$.

\subsubsection*{Qualitative properties of the surface energy and of the traction law}

The \emph{effective surface energy} $\Phi(\delta)$ of a cohesive crack and the associated \emph{traction law} $\tau(\delta)$ are obtained by minimizing the surface energy density~\eqref{eq:cohesive-energy-density} with respect to $\alpha^*$ at fixed opening $\delta$:
\begin{equation}
    \label{eq:effective-surface-energy}
    \Phi(\delta):=\min_{\alpha^*\in[0,1]}\phi(\delta,\alpha^*)
    =\min_{\hat\alpha\in[0,1]}\Big(\hat{\mathsf{k}}(\hat\alpha)\,\tau_c\,\delta+\Gc\,\hat\alpha\Big),
    \qquad
    \tau(\delta):=\k\big(\alpha^*(\delta)\big)\,\tau_c,
\end{equation}
where $\alpha^*(\delta)$ denotes the minimizer.
The stationarity condition of $\phi$ with respect to $\alpha^*$ at fixed $\delta$,
\begin{equation}
    \label{eq:cohesive-stationarity}
    \frac{\partial\phi}{\partial\alpha}(\delta, \alpha^*)
    =\k'(\alpha^*)\,\tau_c\,\delta+\frac{\Gc}{\cw}\sqrt{\w(\alpha^*)}=0,
\end{equation}
is precisely the damage criterion~\eqref{eq:1d-damage-criterion-jump-set} on the jump set combined with the first integral~\eqref{eq:firstIntegralJump}, and its solution $\alpha^*=\alpha^*(\delta)$ is the cohesive law~\eqref{eq:1d-cohesive-law}.
Under \textnormal{\textsf{(SS)}} the map $\hat\alpha\mapsto\phi(\delta,\hat\alpha)$ is convex, so the minimizer in~\eqref{eq:effective-surface-energy} is $\hat\alpha=0$ for $\delta\leq\delta_0$, the interior stationary point~\eqref{eq:cohesive-stationarity} for $\delta_0<\delta<\delta_u$, and $\hat\alpha=1$ for $\delta\geq\delta_u$, with
\begin{equation}
    \label{eq:cohesive-branch-endpoints}
    \delta_0:=-\frac{\Gc}{\tau_c\,\hat{\mathsf{k}}'(0^+)}\geq0,
    \qquad
    \delta_u:=-\frac{\Gc}{\tau_c\,\hat{\mathsf{k}}'(1^-)}.
\end{equation}
Since $\hat{\mathsf{k}}'(\hat\alpha)=\cw\,\k'(\alpha)/\sqrt{\w(\alpha)}$, one has $\delta_0=0$ whenever $\sqrt{\w(\alpha)}/\vert\k'(\alpha)\vert\to0$ as $\alpha\to0^+$, and the cohesive branch then emanates from the origin.
A finite $\hat{\mathsf{k}}'(0^+)$, which Hypothesis~\ref{hyp:kw} does not exclude, gives instead $\delta_0>0$ and an initial Dugdale-like plateau $\tau\equiv\tau_c$ on $[0,\delta_0]$.
Three properties follow, for every admissible pair $(\k,\w)$.

\emph{(i) $\Phi$ is concave.} It is the infimum of a family of affine functions of $\delta$. Consequently $\Phi$ depends on $\hat{\mathsf{k}}$ only through its convex envelope, and \textnormal{\textsf{(SS)}}, which is exactly the convexity of $\hat{\mathsf{k}}$, see Remark~\ref{rem:SS-convexity}, is the condition under which the entire branch parametrized by $\alpha^*$ is attained, so that the cohesive law is single-valued.

\emph{(ii) The traction law is the derivative of the surface energy.} By the envelope theorem, $\Phi'(\delta)=\partial\phi/\partial\delta=\k(\alpha^*(\delta))\,\tau_c=\tau(\delta)$, whence
\begin{equation}
    \label{eq:1d-surface-energy-area}
    \Phi(\delta)=\int_0^{\delta}\tau(s)\,\mathrm{d}s,
    \qquad
    \frac{\mathrm{d}\tau}{\mathrm{d}\delta}=\Phi''(\delta)
    =\frac{\tau_c^2\,\hat{\mathsf{k}}'(\hat\alpha)^3}{\Gc\,\hat{\mathsf{k}}''(\hat\alpha)}\le0
\end{equation}
on the softening branch, the last identity following from the second derivative~\eqref{eq:cohesive-second-derivative} of $\phi$ computed in Remark~\ref{rem:SS-convexity}:
the surface energy is the area under the traction--separation law, and concavity of $\Phi$ is the softening of $\tau$.

\emph{(iii) Strength and toughness are the only material constants.} Since $\k(0)=1$, the traction starts at $\tau(0^+)=\tau_c$; complete decohesion is reached at $\alpha^*=1$, that is at the opening $\delta_u$ of~\eqref{eq:cohesive-branch-endpoints}, which in terms of $(\k,\w)$ reads
\begin{equation}
    \label{eq:1d-critical-opening}
    \delta_u=-\frac{\Gc}{\cw\,\tau_c\,\k'(1)}\in(0,+\infty],
\end{equation}
using $\w(1)=1$; it is finite if $\k'(1)<0$ and infinite if $\k'(1)=0$. There $\tau(\delta_u)=0$ and $\Phi(\delta_u)=\Gc$; beyond it $\tau\equiv0$ and $\Phi\equiv\Gc$.

Hence $\Phi$ is a Barenblatt-type cohesive energy: increasing, concave, vanishing at the origin, with initial slope the \emph{strength} $\tau_c$ and plateau the \emph{toughness} $\Gc$,
\begin{equation}
    \label{eq:cohesive-law-properties}
    \Phi(0)=0,\qquad \Phi'(0^+)=\tau_c,\qquad
    \Phi(\delta)\le\min\left(\tau_c\,\delta,\;\Gc\right),\qquad
    \int_0^{\delta_u}\tau(s)\,\mathrm{d}s=\Gc,
\end{equation}
the last two bounds following from~\eqref{eq:effective-surface-energy} by testing with $\hat\alpha=0$ and $\hat\alpha=1$.
Both constants are inherited from the regularized model independently of $\ell$ and $\mu$, while the shape of $\tau(\delta)$ in between is set by the pair $(\k,\w)$; the Griffith model~\autocite{griffith1921PhenomenaRuptureFlow} is recovered in the limit $\delta_u\to0$.

\begin{remark}
    A stronger condition ensuring that~\eqref{eq:localized-no-snapback-ellch} is satisfied is
    \begin{equation}
    \label{eq:no-snapback-critical-length}
    L < L_c := \frac{\ell_{\mathrm{ch}}}{\cw}
    \inf_{\alpha\in(0,1)}
    \left(
        \frac{1}{|\k'(\alpha)|}
        \frac{\mathrm{d}}{\mathrm{d}\alpha}\frac{\sqrt{\w(\alpha)}}{|\k'(\alpha)|}
    \right).
\end{equation}
\end{remark}
\begin{remark}[Elastic and dissipated part of the surface energy]
\label{rem:surface-energy-split}
The two contributions to the surface energy~\eqref{eq:cohesive-energy-density} are of different nature.
The damage term $\phi_{\mathrm{dmg}}=\Gc\,\hat\alpha(\alpha^*)$ is dissipated, damage being irreversible.
The cohesive term $\phi_{\mathrm{coh}}=\tau\,\delta$ is instead a state function of $(\delta,\alpha^*)$: since the present formulation enforces no irreversibility condition on the nonlinear deformation (Remark~\ref{rem:irreversibility}), $\p$ is reversible and $\phi_{\mathrm{coh}}$ is entirely recovered upon unloading at frozen damage, so that only $\phi_{\mathrm{dmg}}$ is dissipated.
If instead $\p$ is assumed irreversible, the work spent in the band is dissipated incrementally, $\int\k(\alpha^*)\,\tau_c\,\mathrm{d}\bar p$, and not through the state function $\tau\,\delta$; along the cohesive branch it accumulates to $\int_0^{\delta}\tau(s)\,\mathrm{d}s=\phi_{\mathrm{coh}}+\phi_{\mathrm{dmg}}$ by~\eqref{eq:1d-surface-energy-area}, so that the whole surface energy $\phi$ is dissipated, $\phi_{\mathrm{dmg}}$ being already included in it.
The two conventions differ only away from complete decohesion: as $\alpha^*\to1$ the traction vanishes, $\phi_{\mathrm{coh}}\to0$, and both give $\phi\to\Gc$, in agreement with~\eqref{eq:cohesive-law-properties}.
In the notation of Section~\ref{sec:numericalSimulations} the bulk term of~\eqref{eq:1d-energies} is $E_{\mathrm{el}}$, while $\phi_{\mathrm{coh}}=E_{\mathrm{pl}}$ and $\phi_{\mathrm{dmg}}=E_{\mathrm{dmg}}$; the energy plots of the bar report the bulk term as \emph{elastic} and the whole surface energy $\phi$ as \emph{dissipated}.
\end{remark}

\begin{remark}[Stress softening as convexity of the surface energy]
\label{rem:SS-convexity}
Differentiating~\eqref{eq:cohesive-stationarity} once more at fixed opening and evaluating the result
\emph{along the cohesive branch}, \emph{i.e.}~substituting
$\tau_c\,\delta=-\tfrac{\Gc}{\cw}\sqrt{\w(\alpha^*)}/\k'(\alpha^*)$ into
$\partial^2_{\alpha^*\alpha^*}\phi=\k''(\alpha^*)\,\tau_c\,\delta
+\tfrac{\Gc}{\cw}\tfrac{\w'(\alpha^*)}{2\sqrt{\w(\alpha^*)}}$, gives
\begin{equation}
    \label{eq:cohesive-second-derivative}
    \frac{\partial^2\phi}{\partial{\alpha^*}^2}\bigg|_{\delta}
    =\frac{\Gc}{\cw}\,\k'(\alpha^*)\,
    \frac{\mathrm{d}}{\mathrm{d}\alpha^*}\left(\frac{\sqrt{\w(\alpha^*)}}{\k'(\alpha^*)}\right)>0
    \quad\Longleftrightarrow\quad\textnormal{\textsf{(SS)}},
\end{equation}
since $\k'<0$. The statement is sharper in the intrinsic damage variable $\hat\alpha$, in which the
dissipation is linear, $\phi=\hat{\mathsf{k}}(\hat\alpha)\,\tau_c\,\delta+\Gc\,\hat\alpha$, so that
$\partial^2_{\hat\alpha\hat\alpha}\phi\big|_{\delta}=\hat{\mathsf{k}}''(\hat\alpha)\,\tau_c\,\delta$:
condition \textnormal{\textsf{(SS)}} is exactly the convexity of $\hat{\mathsf{k}}$, hence the strict
convexity of $\phi(\delta,\cdot)$ in $\hat\alpha$ for \emph{every} $\delta>0$. (In the variable
$\alpha^*$, instead, \eqref{eq:cohesive-second-derivative} holds only along the branch, and $\phi(\delta,\cdot)$ need not be convex.)
By the envelope relation
$\mathrm{d}\tau/\mathrm{d}\jump{u}=-\left(\k'(\alpha^*)\,\tau_c\right)^2\big/\,
\partial^2_{\alpha^*\alpha^*}\phi$, this is in turn equivalent to a strictly softening cohesive law,
which motivates the name of the condition.

In contrast with the homogeneous case~\eqref{eq:SHvsEScompetition}, no destabilizing elastic term
appears and the equivalence holds for every $\mu$ and $\ell$: at the jump point the opening fixes the
singular part of $\p$ directly, cf.~\eqref{eq:plastic-strain-decomposition}, whereas in the bulk the
split $p=\veps-\k(\alpha)\tau_c/\mu$ shifts with $\alpha$ and produces the elastic softening term
$-\tau_c^2\k'(\alpha)^2/\mu$. The elastic energy stored in the bulk enters only through the structural
coupling with the rest of the bar: eliminating $\delta$ at fixed $t$ in the total energy~\eqref{eq:1d-energies},
which adds to $\phi$ the bulk contribution $\tfrac{\mu}{2L}(tL-\delta)^2$, one obtains, along the branch,
\begin{equation}
    \label{eq:SSvsEScompetition}
    \frac{\mathrm{d}^2\mathcal{E}}
         {\mathrm{d}{\alpha^*}^2}\bigg|_{t}
    =\k'(\alpha^*)^2\,\Gc\left(
    \frac{1}{\cw\,\k'(\alpha^*)}
    \frac{\mathrm{d}}{\mathrm{d}\alpha^*}\left(\frac{\sqrt{\w(\alpha^*)}}{\k'(\alpha^*)}\right)
    -\frac{L}{\ell_{\mathrm{ch}}}\right),
\end{equation}
whose positivity is exactly the no-snap-back condition~\eqref{eq:localized-no-snapback-ellch}, of
which~\eqref{eq:no-snapback-critical-length} is the $\alpha^*$-uniform sufficient version. The
destabilizing elastic term is now proportional to the specimen length $L$ rather than to $\ell$ as
in~\eqref{eq:SHvsEScompetition}: this is the origin of the size effect, and \textnormal{\textsf{(SS)}}
alone rules out snap-back only in the rigid limit $L/\ell_{\mathrm{ch}}\to0$.
\end{remark}

\begin{remark}
    The homogeneous condition involves the regularization length $\ell$, while the localized one involves the structural size $L$: for sufficiently small $\ell$ the homogeneous branch is always free of snap-back, whereas the brittleness of the global response is controlled by the ratio $L/\lch$ only.
\end{remark}

\begin{remark}[Selection between the homogeneous and the localized branch]
\label{rem:branch-selection}
The homogeneous and localized branches are two competing solutions of the same evolution problem.
The localized branch emanates from the end of the elastic phase: as $\alpha^*\to0$, $\jump{u}\to0$
and $t\to\k(0)\tau_c/\mu=\veps_e$ in~\eqref{eq:1d-force-displacement}.
Differentiating the energy along either branch, the terms proportional to
$\mathrm{d}\alpha/\mathrm{d}t$ cancel --- by~\eqref{eq:optFellConstant} on the homogeneous branch,
by the cohesive law~\eqref{eq:1d-cohesive-law} on the localized one --- leaving the energy balance
of the hard device, $\mathrm{d}\mathcal{E}/\mathrm{d}t=L\,\tau(t)$. Since both branches leave the
same elastic state at $t=\veps_e$, and since on the constant-stress plateau
$\tau_{\mathrm{loc}}=\k(\alpha^*)\tau_c<\tau_c=\tau_{\mathrm{hom}}$,
\begin{equation}
    \label{eq:1d-branch-energy-gap}
    \mathcal{E}_{\mathrm{loc}}(t)-\mathcal{E}_{\mathrm{hom}}(t)
    =L\int_{\veps_e}^{t}\left(\tau_{\mathrm{loc}}(s)-\tau_{\mathrm{hom}}(s)\right)\mathrm{d}s<0,
    \qquad t>\veps_e:
\end{equation}
the plateau is never a global minimizer. It is nonetheless a genuine stationary point: with
$p=t-\veps_e$ uniform, the damage criterion~\eqref{eq:optFellConstant} at $\alpha=0$ reads
$\Gc\,\w'(0)/(4\cw\ell)-\tau_c\vert\k'(0)\vert\,p\ge0$ and is \emph{strictly} satisfied for
$t<\veps_c$, so that an evolution started from the unperturbed homogeneous state remains on it up
to $\veps_c$. Its stability is however degenerate: since $\k(0)=1$, the energy of the plateau state
depends on $p$ only through the integral $P=\int p\,\mathrm{d}x$, and not on the distribution of $p$
along the bar,
so that concentrating the nonlinear deformation costs nothing, and growing damage on the resulting
jump is then favourable at first order in its amplitude. The plateau state is thus joined to states
of strictly lower energy by a path along which the energy never increases, and is not a local
minimizer. Any imperfection breaks the degeneracy strictly: with a non-uniform initial damage
$\alpha_0$ the term $\int\k(\alpha_0)\,\tau_c\,p\,\mathrm{d}x$ is minimized by concentrating $p$
where $\alpha_0$ is largest --- so that the load at which localization sets in is
imperfection-sensitive whereas the peak stress $\tau_c$ is not, as observed numerically in
Section~\ref{sec:numericsSimpleShear}.
\end{remark}

\begin{figure}[tp]
    \centering
    \includegraphics[width=.9\textwidth]{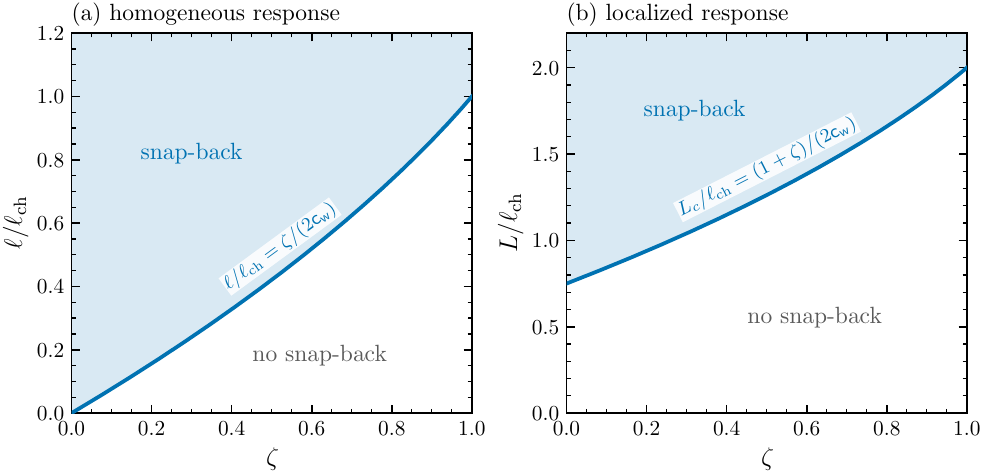}
    \caption{Snap-back phase diagrams for the family $\mathsf{M1}$ in the plane $(\zeta,\ell/\lch)$ for the homogeneous response (a) and $(\zeta,L/\lch)$ for the localized response (b). The solid lines are the critical values $\ell/\lch=\zeta/(2\cw)$ of~\eqref{eq:no-snapback-M1-homogeneous} and $L_c/\lch=(1+\zeta)/(2\cw)$ of~\eqref{eq:no-snapback-M1-localized}, with $\cw=\cw(\zeta)$; at $\zeta=1$ they take the values $\ell/\lch=1$ and $L_c/\lch=2$. The shaded regions correspond to snap-back of the corresponding branch under displacement control.}
    \label{fig:snapback_phase_diagram}
\end{figure}

\section{A one-parameter family of models}
\label{sec:M1}
The following family of constitutive laws $\mathsf{M1}$, depending on a scalar parameter $\zeta$, was introduced in~\autocite{bmmz25}.
It is defined by
\begin{equation}
\k(\alpha) = 1 - \alpha, \qquad
\w(\alpha) = (1 - \zeta)\alpha + \zeta \alpha^2, \qquad \zeta \in (0, 1].
\label{eq:model-1}
\end{equation}
The normalization constant $\cw=\int_0^1\sqrt{\w(s)}\,\mathrm{d}s$ is
\[
    \cw=\frac{4\,\zeta^{3/2} + 2\,\zeta^{1/2} \left(1 - \zeta\right) +  \left(1-\zeta\right)^{2} \log{\left(\dfrac{1 - \sqrt{\zeta}}{1 + \sqrt{\zeta}} \right)}}{8\,\zeta^{3/2}},
\]
a decreasing function of $\zeta$, ranging from $\cw=2/3$ as $\zeta\to0$ to $\cw=1/2$ at $\zeta=1$.
Hypotheses \textsf{(SS)} and \textsf{(SH)} hold for $0 < \zeta \le 1$; as $\zeta\to0$ the hardening function $\w'/\k'$ becomes constant, so that the inequality in \textsf{(SH)} degenerates into a non-strict one.
We nevertheless report the limit case $\zeta=0$ below and in the figures, as it admits particularly simple closed forms.

For $\zeta = 1$,
\begin{equation}
   \k(\alpha) = 1 - \alpha, \quad
   \w(\alpha) = \alpha^2, \quad
   \cw = \frac{1}{2},
   \label{eq:model-LS-zeta-1}
\end{equation}
the last term of~\eqref{eq:model} is the damage energy of the classical \textsf{AT2} phase-field model.

\begin{figure}[htbp]
    \centering
    \includegraphics[width=.9\textwidth]{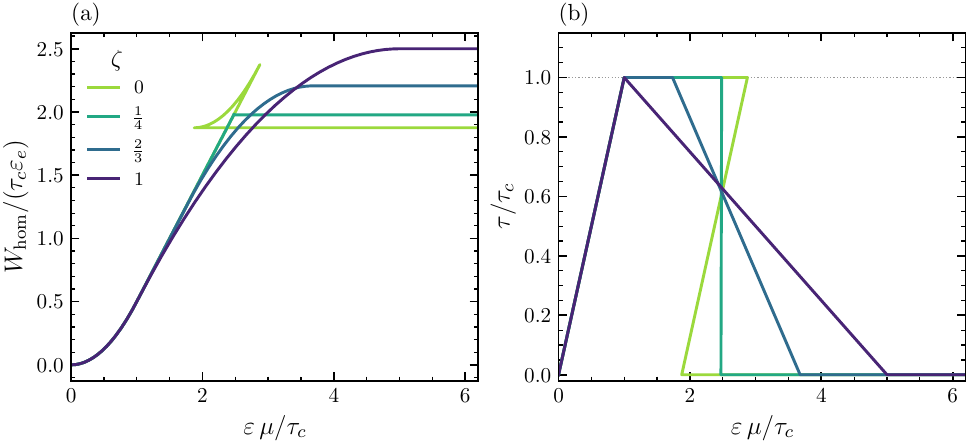}
    \caption{%
        \emph{Homogeneous (material-point) response} for the family $\mathsf{M1}$, for
        several values of $\zeta$ and length-scale ratio $\lch/\ell=5$:
        (a) normalized total energy density $W_{\rm hom}/(\tau_c\veps_e)$
        of~\eqref{eq:homogeneous-M1-energy} and
        (b) normalized stress $\tau/\tau_c$, \emph{versus} the applied strain
        $\varepsilon$ in units of the elastic-limit strain $\veps_e=\tau_c/\mu$.
        The constant-stress plateau $\veps_e\le\varepsilon<\veps_c$ is present only
        for $\zeta<1$, and the softening branch snaps back for $\zeta=0$ and
        $\zeta=1/4$.
    }
    \label{fig:homogeneous_response}
\end{figure}

The homogeneous response of Section~\ref{sec:homogeneousSolutions} is characterized by
\begin{equation}
    \label{eq:criticalStrainsM1}
    \veps_e = \dfrac{\tau_c}{\mu},\qquad
    \veps_c = \dfrac{\tau_c}{\mu} + \dfrac{\Gc}{\ell\tau_c}\dfrac{1-\zeta}{4\cw},\qquad
    \veps_u = \dfrac{\Gc}{\ell\tau_c}\dfrac{1+\zeta}{4\cw},
\end{equation}
so that the constant-stress regime $[\veps_e,\veps_c)$ is present for every $\zeta<1$ and closes at $\zeta=1$.
Condition~\eqref{eq:homogeneous-no-snapback-ellch} for the absence of snap-back is independent of $\alpha$ and becomes
\begin{equation}
    \label{eq:no-snapback-M1-homogeneous}
    \frac{\ell}{\lch}
    <
    \frac{\zeta}{2\cw},
\end{equation}
which is represented in the $(\zeta,\ell/\lch)$ plane in Figure~\ref{fig:snapback_phase_diagram}(a).
The threshold $\zeta/(2\cw)$ increases monotonically from $0$ as $\zeta\to0$ to $1$ at $\zeta=1$: the smaller $\zeta$, the more prone to snap-back the homogeneous branch, the limit $\zeta\to0$ snapping back for every $\ell>0$.
The corresponding stress-strain response is plotted in Figure~\ref{fig:homogeneous_response} for $\lch/\ell=5$, a value for which the threshold is crossed at $\zeta\simeq0.25$: the softening branch snaps back for $\zeta=0$ and, marginally, for $\zeta=1/4$, since $\ell/\lch=0.2$ is just above $\zeta/(2\cw)\simeq0.198$; it does not for $\zeta=2/3$ and $\zeta=1$.

In the strength-softening regime, the loading parameter, the damage and the stress are affine functions
\begin{equation}
    \label{eq:homogeneous-M1-softening}
    t = \veps_c + \alpha\,(\veps_u-\veps_c),
    \qquad
    \alpha = \frac{t - \veps_c}{\veps_u - \veps_c},
    \qquad
    \tau = \frac{\veps_u - t}{\veps_u - \veps_c}\,\tau_c,
\end{equation}
the branch being described by $\alpha\in[0,1]$ and snapping back whenever $\veps_u<\veps_c$, which is precisely the failure of~\eqref{eq:no-snapback-M1-homogeneous}.
The corresponding total energy density $W$ of Remark~\ref{rem:homogeneous-snapback-convexity} is
\begin{equation}
    \label{eq:homogeneous-M1-energy}
    W_{\rm hom} =\tau_c
    \frac{\veps_u - t}{\veps_u - \veps_c}
    \left(t - \frac{\veps_e}{2}\frac{\veps_u - t}{\veps_u - \veps_c}\right)
    +
    \frac{\Gc}{4\cw\ell}
     \frac{t - \veps_c}{\veps_u - \veps_c}
    \left((1-\zeta) + \zeta\,\frac{t - \veps_c}{\veps_u - \veps_c}\right),
\end{equation}
the two terms being respectively the strain energy density $\psi$ of~\eqref{eq:elastic-energy-density-antiplane} and the damage energy $\Gc\,\w(\alpha)/(4\cw\ell)$.
Since $\partial W/\partial\alpha=0$ along the branch, $W_{\rm hom}$ is also the work $\int_0^t\tau\,\mathrm{d}\veps$ of the applied load.

Turning to the localized solutions, the relation $\jump{u} = \Gc\sqrt{\w(\alpha^*)}/(\cw\,\tau_c)$ of~\eqref{eq:1d-cohesive-law} is a quadratic equation for $\alpha^*$ and is inverted explicitly, so that the cohesive law $\tau=\k(\alpha^*)\tau_c=(1-\alpha^*)\tau_c$ reads as
\begin{equation}
    \label{eq:1d-cohesive-law-M1-jump}
        \tau =\tau_c
        \begin{cases}
            \displaystyle
            1 - \dfrac{4\tau_c^2}{9\Gc^2}\jump{u}^{2}, & \zeta\to0,\\[10pt]
             \displaystyle
        1 - \frac{1}{2\zeta}\Bigg( \sqrt{(1-\zeta)^2 + 4\zeta\Big(\dfrac{\cw\tau_c}{\Gc}\jump{u}\Big)^2}- (1-\zeta) \Bigg), & 0<\zeta<1,\\[10pt]
        \displaystyle
        1 - \frac{\tau_c}{2\Gc}\jump{u}, & \zeta=1,
        \end{cases}
\end{equation}
where in each line the quantity subtracted from unity is the maximal damage $\alpha^*(\jump{u})$, and where $\cw=2/3$ in the first line and $\cw=1/2$ in the third.
The corresponding cohesive surface energy $\Phi=\k(\alpha^*)\,\tau_c\jump{u}+\Gc\,\hat\alpha(\alpha^*)$ of~\eqref{eq:cohesive-energy-density} reads as
\begin{equation}
    \label{eq:1d-cohesive-energy-M1}
    \Phi(\jump{u}) =
    \begin{cases}
         \tau_c\jump{u} - \dfrac{4\tau_c^3}{27\Gc^2}\jump{u}^3,
        & \zeta\to0,\\[6pt]
        \dfrac{\Gc}{\cw}\!\left[(1-\alpha^*)\sqrt{(1-\zeta)\alpha^*+\zeta{\alpha^*}^2}
        + \displaystyle\int_0^{\alpha^*}\!\sqrt{(1-\zeta)\beta+\zeta\beta^2}\,\mathrm{d}\beta\right],
        & 0<\zeta<1,\\[6pt]
       \tau_c\jump{u} - \dfrac{\tau_c^2}{4\Gc}\jump{u}^2,
        & \zeta=1,
    \end{cases}
\end{equation}
the middle line being expressed through the same $\alpha^*(\jump{u})$ as in~\eqref{eq:1d-cohesive-law-M1-jump}.
In all cases $\Phi$ is concave, with $\Phi'(0^+)=\tau_c$ and $\Phi(\delta_u)=\Gc$ at the critical opening $\delta_u=\Gc/(\cw\tau_c)$ of~\eqref{eq:1d-critical-opening}, in agreement with~\eqref{eq:cohesive-law-properties}.

The infimum in the critical length~\eqref{eq:no-snapback-critical-length} is attained at $\alpha^*=1$, since $\w'/\sqrt{\w}$ is non-increasing, so that the condition $L<L_c$ is here necessary as well as sufficient for the localized branch to be free of snap-back, with
\begin{equation}
    \label{eq:no-snapback-M1-localized}
    L_c = \frac{1+\zeta}{2\,\cw}\,\lch.
\end{equation}
As shown in Figure~\ref{fig:snapback_phase_diagram}(b), $L_c$ increases monotonically with $\zeta$, from $L_c = 3\lch/4$ in the limit $\zeta\to0$ to $L_c=2\lch$ at $\zeta=1$.

\begin{figure}[htbp]
    \centering

    \includegraphics[width=\textwidth]{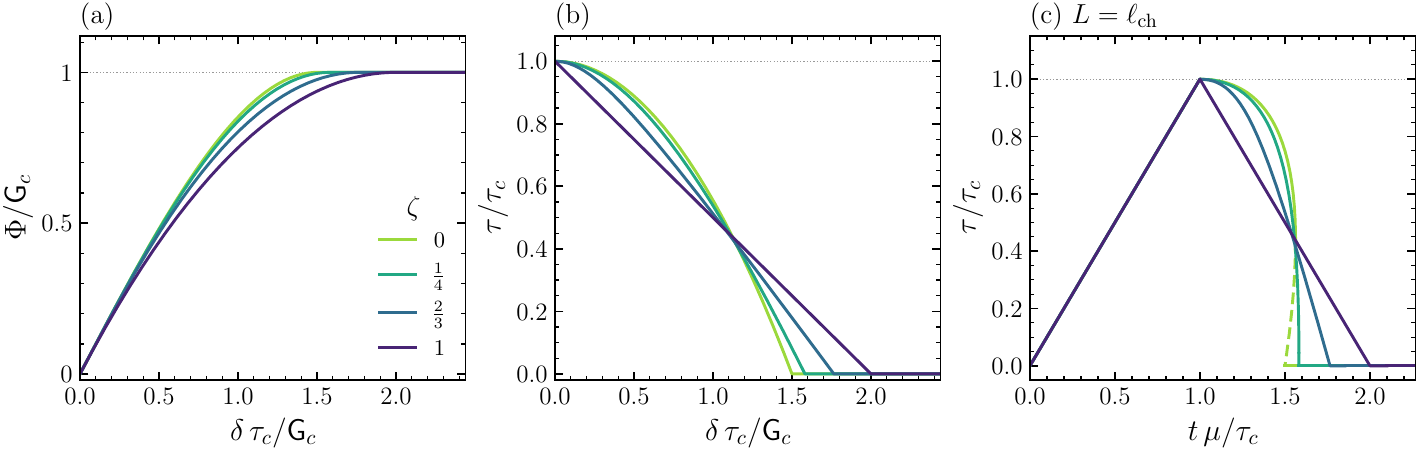}
    \caption{%
      \emph{Localized solution} for the family $\mathsf{M1}$, for the same values of the
      shape parameter $\zeta$ as in Figure~\ref{fig:homogeneous_response}:
      (a) equivalent cohesive surface energy $\Phi/\Gc$ and
      (b) cohesive traction $\tau/\tau_c$, both as functions of the normalized crack
      opening $\delta\,\tau_c/\Gc$, with $\delta=\jump{u}$;
      (c) stress $\tau/\tau_c$ \emph{versus} the normalized end load
      $t\,\mu/\tau_c$, with $t=u(L)/L$, for a bar of length $L=\lch$, combining the elastic loading branch with
      the localized softening branch~\eqref{eq:1d-force-displacement}, drawn dashed across
      the snap-back; the constant-stress plateau ($\zeta<1$) is deliberately
      not drawn.
    }
    \label{fig:cohesive_law}
\end{figure}

Figure~\ref{fig:cohesive_law} summarizes the properties of the localized solution for the same values of $\zeta$ as the homogeneous response of Figure~\ref{fig:homogeneous_response}.
It presents the cohesive surface energy $\Phi(\jump{u})$, the cohesive traction $\tau(\jump{u})$, and the resulting global response of a bar of length $L=\lch$.
Note the snap-back for the two smallest values of $\zeta$, for which $L>L_c$, see~\eqref{eq:no-snapback-M1-localized}.

\section{Numerical implementation}
\label{sec:numericalImplementation}
The numerical implementation is based on the minimization of the energy functional $\mathcal{E}_\ell(u,\p,\alpha)$ at each time step, under a prescribed time-dependent displacement on $\partial_u\Omega$ and the irreversibility condition on the damage $\alpha$.
Before presenting the numerical results, we first introduce the dimensionless formulation of the model and then describe the numerical solution strategy.

\subsection{Dimensionless formulation}
\label{sec:dimensionlessFormulation}
Besides the constitutive functions $\k$ and $\w$, the energy functional depends on the shear modulus $\mu$, the shear strength $\tau_c$, the fracture energy $\Gc$, and the regularization length scale $\ell$; we denote by $L$ a characteristic length of the domain $\Omega$.
The number of independent parameters can be reduced by rewriting the energy functional~\eqref{eq:energyFunctionalAntiplaneSR} in a dimensionless form.
Setting
\begin{equation}
    x=Lx^*,\quad
    u = u_0 u^*,\quad
    \nabla =\frac{1}{L} \nabla^*,\quad
    \p = \frac{u_0}{L}\p^*,\quad
    \dA = L^2\dA^* ,\quad
    \mathcal{E}_\ell=\Gc L \mathcal{E}_\ell^*,
\end{equation}
where $L$ and $u_0$ are characteristic length and displacement scales, we obtain
\begin{equation}
    \mathcal{E}_\ell^*(u^*,\p^*,\alpha)=
    \int_{\Omega^*}
    \frac{\mu u_0^2}{2\Gc L}  \Vert\nabla^* u^*-\p^*\Vert^2 + \frac{\tau_c u_0}{\Gc}\mathsf{k}(\alpha)\Vert\p^*\Vert
    + \frac{1}{4\cw}\left(\frac{L}{\ell}\mathsf{w}(\alpha)+\frac{\ell}{L}\Vert\nabla^*\alpha\Vert^2\right) \dA^*.
\end{equation}
Setting without loss of generality $u_0 := {\Gc}/{\tau_c}$, we get the following non-dimensional expression of the energy
\begin{equation}
    \mathcal{E}_\ell^*(u^*,\p^*,\alpha)=
    \int_{\Omega^*}
    \frac{\lch}{2L} \Vert\nabla^* u^*-\p^*\Vert^2
    +\mathsf{k}(\alpha)\Vert\p^*\Vert
    + \frac{1}{4\cw}\left(\frac{L}{\ell}\mathsf{w}(\alpha)+\frac{\ell}{L}\Vert\nabla^*\alpha\Vert^2\right)
    \dA^*.
\end{equation}
When written in this form, it is clear that the problem depends on the two dimensionless numbers
\begin{equation}
\text{{brittleness parameter}}:\quad \frac{L}{\lch}=\frac{\tau_c^2L}{\mu \Gc},\qquad
        \text{{regularization parameter}}:\quad \frac{\ell}{L}.
        \label{eq:dimensionless-parameter}
\end{equation}

The brittleness parameter is the antiplane form of the dimensionless brittleness number that governs size effects in quasi-brittle fracture~\autocite{carpinteri1982NotchSensitivityFracture,bazant1998FractureSizeEffect}.
In the numerical simulations below, each test is specified by the two dimensionless groups~\eqref{eq:dimensionless-parameter}, together with the constitutive parameter $\zeta$ and the numerical parameters ($h/\ell$, $\alpha_0$, $N$), and all results are reported in dimensionless form: stresses are normalized by the strength $\tau_c$, displacements by $u_0=\Gc/\tau_c$, and strains --- including the load $t$ --- by $\veps_e=\tau_c/\mu$, except in the rigid limit $\mu\to\infty$, where $\veps_e\to0$ and the load is reported as $t\,L/u_0=\jump{u}\,\tau_c/\Gc$. For the sake of readability, we drop the superscript $*$ from now on.

\subsection{Discretization and solvers}

At each loading step $i$, corresponding to the load parameter $t_i$, we solve the incremental minimization problem~\eqref{eq:incremental-minimization}, in which the lower bound $\alpha_{i-1}$ entering $\mathcal{D}_i$, the damage field at the previous loading step, enforces the irreversibility condition.
The energy is separately convex with respect to the pair $(u,\p)$ at fixed $\alpha$ and with respect to $\alpha$ at fixed $(u,\p)$, but not jointly convex.
We exploit this structure through an alternate minimization scheme~\autocite{BFM00}, a fixed-point algorithm where we iteratively solve the two subproblems until convergence:
\begin{enumerate}
    \item  Minimize the energy with respect to $(u,\p)$ at fixed $\alpha$.

    \item  Minimize the energy with respect to $\alpha$ at fixed $(u,\p)$ and under the bound constraints $\alpha_{i-1} \leq \alpha \leq 1$.
\end{enumerate}

We discretize the fields using finite elements with triangular cells.
For the displacement $u$ and the damage $\alpha$ we use linear simplicial Lagrange elements.
The nonlinear deformation $\p$ is discretized using a quadrature space that retains as the degrees of freedom only the values $\p^g=\p(x_g)$ of $\p$ at the quadrature points; in this paper, we use a simple one-point Gauss integration rule.
We denote by $\mathsf{P}$ the $n_g$-dimensional vector collecting the values of the nonlinear deformation at the quadrature points and by $\mathsf{U}$ and $\mathsf{D}$ the $n_n$-dimensional vectors collecting the degrees of freedom for the displacement and damage fields, such that $u(x) = \sum_{k=1}^{n_n} \mathsf{U}_k \chi_k(x)$ and $\alpha(x) = \sum_{k=1}^{n_n} \mathsf{D}_k \chi_k(x)$, where the $\chi_k$'s are the finite-element basis functions and $n_n$ and $n_g$ are the number of nodes and quadrature points in the mesh.
We use the finite element framework FEniCSx/DOLFINx for the discretization and the data management~\autocite{BarattaEtal2023}.

Both subproblems are convex and are solved to high accuracy with the conic interior-point optimizer of MOSEK~\autocite{aps2024mosek}: the first through the second-order cone programming (SOCP) reformulation of Appendix~\ref{sec:up-socp}, the second through the bound-constrained conic reformulation of Appendix~\ref{sec:alpha-qp}.
The alternate minimization loop is stopped when the increment of the damage field between two successive iterations is smaller, in the infinity norm, than a tolerance $\mathrm{tol}_{\mathrm{AM}}$, set to $10^{-3}$ unless otherwise stated.
The scheme extends without modification to the variant in which the nonlinear deformation is subject to an irreversibility condition (Remark~\ref{rem:irreversibility}), used in Section~\ref{sec:numericsSurfing}: it suffices to replace $\Vert\p\Vert$ by $\Vert\p-\p_{i-1}\Vert$ in the cone constraint of the $(u,\p)$ subproblem, and by the cumulated deformation $\bar p_i=\bar p_{i-1}+\Vert\p_i-\p_{i-1}\Vert$ in the damage subproblem.
Similarly, the rigid limit $\mu\to\infty$ used in Section~\ref{sec:numericsSimpleShear} is obtained by dropping the elastic term from the objective and enforcing $\nabla u=\p$ pointwise as an equality constraint, the stress being recovered as the associated Lagrange multiplier (Appendix~\ref{sec:up-socp}).

\begin{algorithm}[htbp]
\caption{Incremental alternate minimization scheme.}
\label{alg:alternate-minimization}
\begin{algorithmic}[1]
\State $\mathsf{D}_{0} \gets$ nodal values of the initial damage field $\alpha_0$
\For{$i=1,\ldots,N$} \Comment{loading steps}
    \State update the imposed displacement $\bar u(t_i)$; set $\mathsf{D}^{(0)}\gets\mathsf{D}_{i-1}$
    \For{$j=1,2,\ldots$, up to \textrm{maxiter}} \Comment{alternate minimization}
        \State $(\mathsf{U}^{(j)},\mathsf{P}^{(j)}) \gets$ solution of the SOCP~\eqref{eq:full-socp} with $\mathsf{D}=\mathsf{D}^{(j-1)}$
        \State $\mathsf{D}^{(j)} \gets$ solution of the SOCP~\eqref{eq:alpha-socp} with $\mathsf{P}=\mathsf{P}^{(j)}$ and bounds $\mathsf{D}_{i-1}\leq\mathsf{D}\leq 1$
        \If{$\max \vert \mathsf{D}^{(j)}-\mathsf{D}^{(j-1)}\vert \leq \mathrm{tol}_{\mathrm{AM}}$} \textbf{break}
        \EndIf
    \EndFor
    \State $(\mathsf{U}_{i},\mathsf{P}_{i},\mathsf{D}_{i})\gets(\mathsf{U}^{(j)},\mathsf{P}^{(j)},\mathsf{D}^{(j)})$
\EndFor
\end{algorithmic}
\end{algorithm}

\section{Numerical simulations}
\label{sec:numericalSimulations}
All the simulations of this section use the constitutive family $\mathsf{M1}$ of~\eqref{eq:model-1}:
$$\k(\alpha) = 1-\alpha,\qquad
\w(\alpha) = (1-\zeta)\,\alpha+\zeta\,\alpha^2,$$
with $\zeta\in(0,1]$ controlling the shape of the cohesive law and of the homogeneous material response, as shown in Sections~\ref{sec:homogeneousSolutions}--\ref{sec:localizedSolutions} and Figures~\ref{fig:homogeneous_response} and~\ref{fig:cohesive_law}.
We adopt the dimensionless setting of Section~\ref{sec:dimensionlessFormulation}, so that $\ell_\mathrm{ch}=1$ and $\veps_e=1$, except in Figure~\ref{fig:notch-cl}, where $\lch$ is varied.
Each test is then characterized by the brittleness ratio $L/\ell_\mathrm{ch}$, the regularization ratio $\ell/\ell_\mathrm{ch}$, and the constitutive parameter $\zeta$.
In all the figures of this section, markers denote numerical results and lines the corresponding analytical predictions or reference curves.

\subsection{Simple shear}
\label{sec:numericsSimpleShear}

We begin by focusing on the simple shear problem introduced in Section~\ref{sec:modelProblem}, for which the homogeneous and localized solutions constructed in Sections~\ref{sec:homogeneousSolutions} and~\ref{sec:localizedSolutions} provide closed-form references.
The aim is twofold: to verify that the alternate minimization scheme of Section~\ref{sec:numericalImplementation} correctly captures the localized solutions and the associated equivalent cohesive law, and to illustrate the structural responses predicted by the model when varying the dimensionless parameters identified in Section~\ref{sec:dimensionlessFormulation}.

We impose $\alpha=0$ on the loaded end points, so that solutions with localization at the boundary or non-zero homogeneous damage are not admissible.
Unless otherwise stated, the localization point is selected by a small initial damage field $\alpha_0\cos^2(\pi x/L)$, imposed as an irreversible lower bound.
All runs use unstructured triangular meshes of size $h<\ell$, generated by the mesh generator Gmsh~\autocite{Geuzaine-Remacle-2009a}.
The boundary displacements are applied in 50 to 100 time-increments.
The discretization and the solver settings are those of Section~\ref{sec:numericalImplementation}.
The values specific to each test ($L/\ell_\mathrm{ch}$, $\ell/\ell_\mathrm{ch}$, $\zeta$, $\alpha_0$, and $N$) are reported in the figure captions.
In all the figures, the analytical reference curves are the homogeneous response~\eqref{eq:homogeneousRegimes} and the localized response obtained by combining the equivalent cohesive law~\eqref{eq:1d-cohesive-law} with the loading condition~\eqref{eq:1d-force-displacement}.

We first focus on the case $\zeta=1$ in which the constant-stress regime of the homogeneous solution disappears ($\veps_c=\veps_e$, see~\eqref{eq:criticalStrainsM1}) and the equivalent cohesive law~\eqref{eq:1d-cohesive-law-M1-jump} is linear with the stress vanishing at the ultimate opening $
    \delta_u^f:=\jump{u}\big\vert_{\alpha^*=1}={2\Gc}/{\tau_c}$.

\begin{figure}[htbp]
    \centering
    \includegraphics[width=\textwidth]{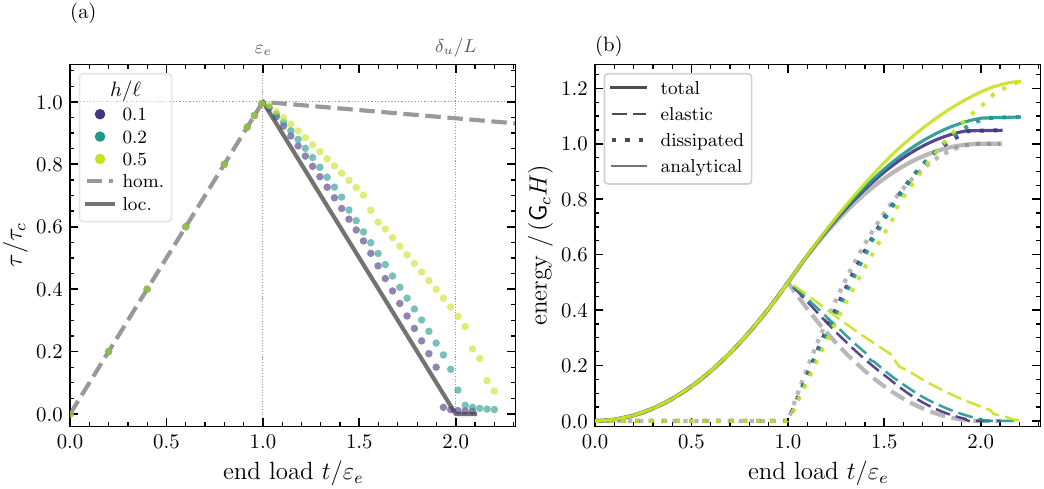}
    \caption{Simple shear, reference case: $L/\lch=1$, $\zeta=1$, $\ell/\lch=0.05$, $\alpha_0=10^{-3}$, $N=80$ load steps, for uniform meshes $h/\ell\in\{0.1,0.2,0.5\}$. (a) Stress $\tau/\tau_c$ versus end load $t/\veps_e$; (b) total, elastic and dissipated energies normalized by $\Gc H$, against the exact solution~\eqref{eq:1d-energies}.}
    \label{fig:barReferenceForceEnergy}
\end{figure}

Figure~\ref{fig:barReferenceForceEnergy}(a) shows the global response of a bar of length $L=\ell_\mathrm{ch}$ with $\ell/\ell_\mathrm{ch}=0.05$ for decreasing discretization sizes $h = \ell/2$, $\ell/5$, and $\ell/10$.
The evolution of the total, elastic (the bulk term $\int_\Omega\frac{\mu}{2}\Vert\nabla u-\p\Vert^2\dA$), and dissipated (the plastic and damage terms, see Remark~\ref{rem:surface-energy-split}) energies are shown in Figure~\ref{fig:barReferenceForceEnergy}(b), together with the exact solution from Section~\ref{sec:modelProblem}.
We observe that in all cases the numerical solution bifurcates from the  homogeneous response at the onset of the strength-softening branch $t = \veps_c = \veps_e = 1$, and the stresses vanish at $t \simeq \delta_u^f/L$, the final failure load inheriting the mesh-induced toughening discussed below (visible for $h=\ell/2$).
As expected, the energy dissipated along the crack is overestimated by a factor $1+ {h}/{(4\cw\ell)}$, which could be counteracted by using a numerical fracture toughness $\Gc^{\mathrm{num}} = \Gc / \left(1+ {h}/({4\cw\ell})\right)$ accounting for this mesh-induced toughening~\autocite{bourdin2008VariationalApproach}.

These observations are made quantitative in Table~\ref{tab:barReferenceCritical}. The peak strength $\tau_{\max}$, which sets the critical load for crack nucleation ($t=\veps_c=\veps_e=1$), is matched within $1\%$ and is insensitive to the mesh; the final failure load follows $\delta_u^f/L=2$ amplified by the same mesh-toughening factor as the dissipated energy, since $\delta_u^f$ is proportional to $\Gc$; the residual gap on $\tau_{\max}$ merely reflects the finite load increment $\Delta t\simeq0.02$ and the initial imperfection $\alpha_0=10^{-3}$, the last step before nucleation being still elastic. The normalized dissipated energy $\mathcal{E}_{\mathrm{diss}}/(\Gc H)$ is overestimated by an amount that follows the analytical mesh-toughening factor $1+h/(4\cw\ell)$ to better than $1\%$ for $h\le\ell/5$, and to within $2\%$ at the coarsest mesh $h=\ell/2$, where the bar is not yet fully broken at the last computed load.

\begin{table}[htbp]
    \centering
    \begin{tabular}{c cc c}
        \toprule
        & \multicolumn{2}{c}{Dissipated energy $\mathcal{E}_{\mathrm{diss}}/(\Gc H)$}
        & Strength $\tau_{\max}/\tau_c$\\
        \cmidrule(lr){2-3}        \cmidrule(lr){4-4}

        $h/\ell$ & numerical & $1+\tfrac{h}{4\cw\ell}$ & numerical\\
        \midrule
        $0.1$ & $1.049$ & $1.050$ & $0.994$ \\
        $0.2$ & $1.096$ & $1.100$ & $0.994$ \\
        $0.5$ & $1.222$ & $1.250$ & $0.996$ \\
        \bottomrule
    \end{tabular}
    \caption{Dissipated fracture energy and strength as a function of the mesh size for the simple-shear test of Figure~\ref{fig:barReferenceForceEnergy}.}
    \label{tab:barReferenceCritical}
\end{table}

\begin{figure}[htbp]
    \centering
    \includegraphics[width=\textwidth]{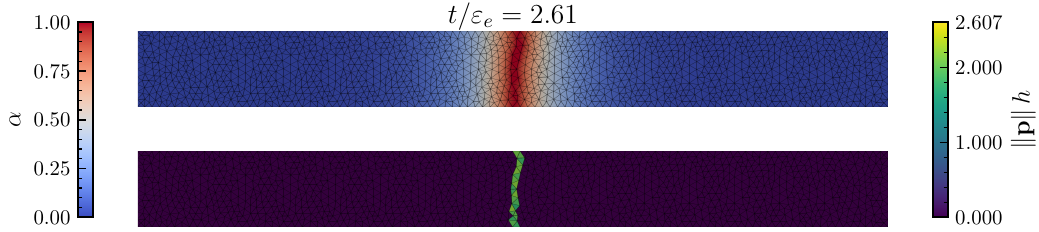}
    \caption{Damage $\alpha$ (top) and nonlinear deformation scaled by the mesh size, $\Vert\p\Vert h$ (bottom), at the end of loading for the test of Figure~\ref{fig:barReferenceForceEnergy} with $H/L=0.1$ and a deliberately coarse mesh, $h/\ell=0.2$, to make the triangulation visible.}
    \label{fig:barReferenceFields}
\end{figure}

The damage and nonlinear deformation fields $\alpha$ and $\Vert\p\Vert h$ after failure are shown in Figure~\ref{fig:barReferenceFields}, using a coarse discretization in order to make the mesh visible.
As expected, they are invariant in the $y$ direction, \emph{up to the discretization size}.
The crack can be identified as the one-element wide band along which $\p$ localizes, and $\alpha = 1$.

\begin{figure}[htbp]
    \centering
    \includegraphics[width=\textwidth]{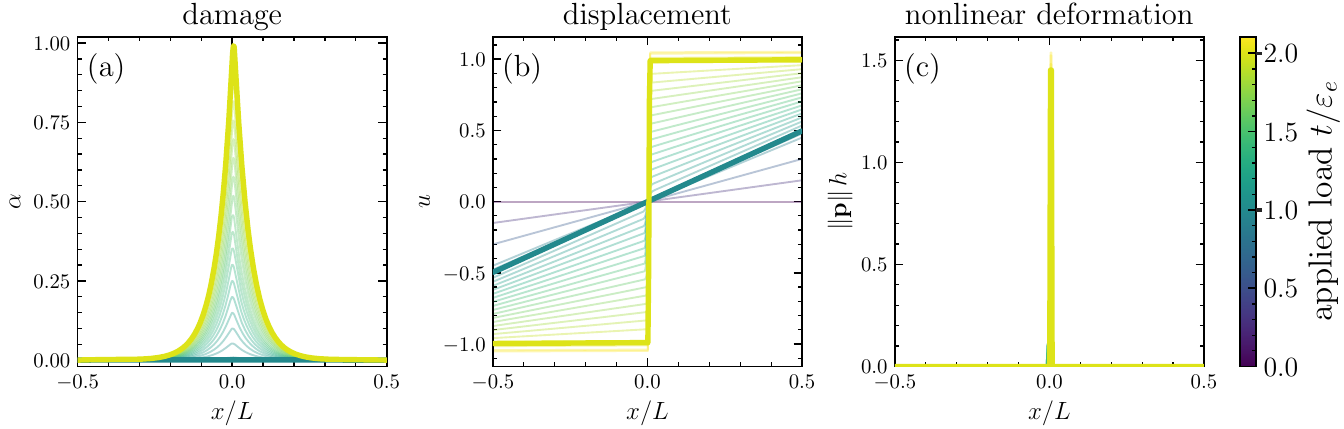}
    \caption{Reference test of Figure~\ref{fig:barReferenceForceEnergy} at $h/\ell=0.1$: profiles along the mid-line $y=0$ of (a) the damage $\alpha$, (b) the displacement $u$ and (c) the scaled nonlinear deformation $\Vert\p\Vert h$, versus $x/L$ and colored by the load $t/\veps_e$. Thick lines: onset of localization $t=\veps_c=1$ and ultimate failure $t=\delta_u^f/L=2$.}
    \label{fig:barReferenceProfiles}
\end{figure}

Figure~\ref{fig:barReferenceProfiles} shows snapshots in time of one-dimensional profiles of $u$, $\alpha$, and $\p$ along the symmetry axis $y = 0$.
We observe the elastic phase ($\alpha = 0$, linear displacement field), followed by the nucleation of a cohesive crack indicated by a non-zero damage in a strip of width of order $\ell$ centered at 0 along which $\Vert\p\Vert>0$ and the displacement field is discontinuous.
Unlike classical damage models based on stiffness degradation, where the displacement can be discontinuous only when $\alpha=1$, it jumps even when $\alpha<1$: the jump is the singular part $\p^{\mathrm{S}}=\jump{u}\,\mathbf{n}\,\delta_{J_u}$ of~\eqref{eq:singular-strain-parts}, obtained by concentration of the nonlinear deformation on the localization line.
Since $\p$ concentrates on a single band of elements, where it scales as $1/h$, we report the scaled field $\Vert\p\Vert h$, which for the one-point quadrature discretization is the discrete transverse integral of $\Vert\p\Vert$ across the band and converges to $\left|\!\jump{u}\!\right|$ as $h\to0$, rather than the mesh-dependent $\Vert\p\Vert$.
Residual forces along the crack faces result in non-constant displacement in each ligament and vanish after the ultimate failure load $t = \delta_u^f/L$.

\begin{figure}[htbp]
    \centering
    \includegraphics[width=\textwidth]{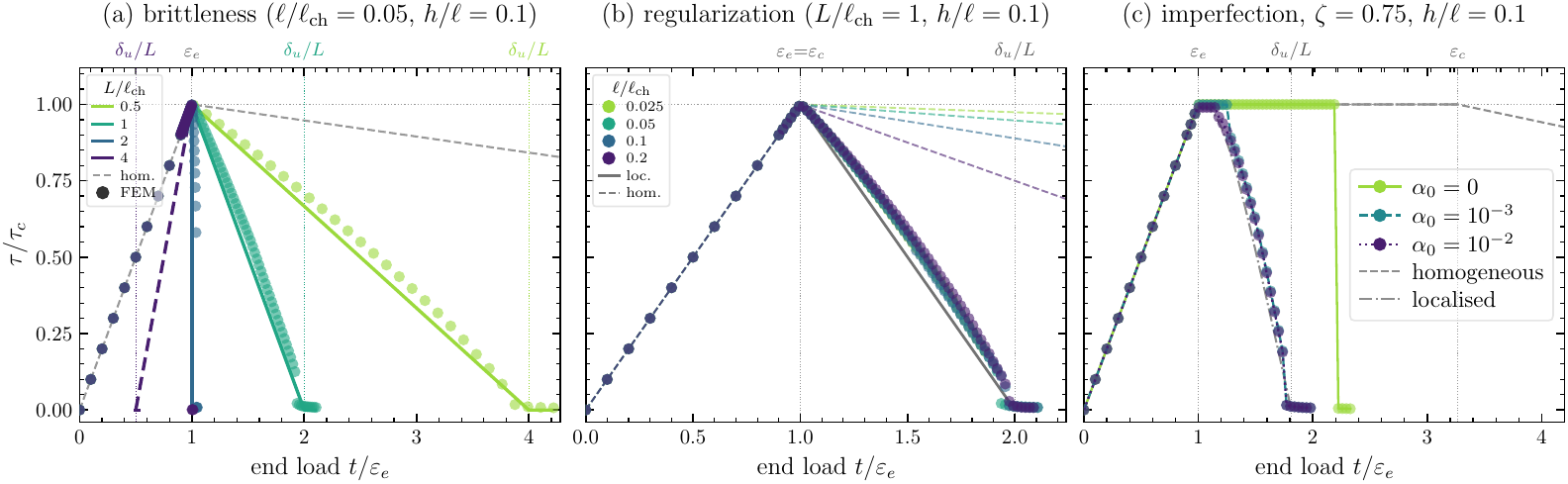}
    \caption{Stress--strain response of the simple shear test: influence of (a) the brittleness $L/\lch\in\{0.5,1,2,4\}$ at $\ell/\lch=0.05$, the localized branch dashed across its snap-back; (b) the regularization length $\ell/\lch\in\{0.025,0.05,0.1,0.2\}$ at $L/\lch=1$; (c) the imperfection $\alpha_0\in\{0,10^{-3},10^{-2}\}$ at $\zeta=0.75$, $N=100$. Other parameters as in Figure~\ref{fig:barReferenceForceEnergy}, with $h/\ell=0.1$.}
    \label{fig:barBrittlenessEll}
\end{figure}

In Figure~\ref{fig:barBrittlenessEll}(a), we  explore the effect of the brittleness parameter $L/\ell_\mathrm{ch}$ on the post-peak global response at fixed $\ell/\ell_\mathrm{ch}=0.05$.
The remaining parameters are $\zeta = 1$, $\alpha_0=10^{-3}$, $h/\ell=0.1$, and $N=80$.
The numerical results match the analytical expression~\eqref{eq:1d-force-displacement}: after an elastic regime for $0<t<1$, we observe the creation of a cohesive branch for $L<L_c=2\,\ell_\mathrm{ch}$ (see~\eqref{eq:no-snapback-M1-localized} and Figure~\ref{fig:snapback_phase_diagram}(b)). For $L>L_c$, since one cannot follow the snap-back in a displacement-controlled setting, we observe the sudden nucleation of a crack with full release of the elastic energy, \emph{i.e.} a brittle crack.

This is consistent with the structural size effects of cohesive models: short bars fail gradually, long bars fail in a brittle manner, the transition being governed by the ratio $L/\ell_\mathrm{ch}$.

Figure~\ref{fig:barBrittlenessEll}(b) highlights the influence of the regularization length $\ell/\ell_\mathrm{ch}$ when the ratio $L/\ell_\mathrm{ch} = 1$ is fixed.
All other parameters are similar to those of the previous case.
As $\ell/\ell_\mathrm{ch}$ decreases from $0.2$ down to $0.025$, we observe that the overall response is virtually unaffected by changes in $\ell$ and matches the localized branch from Section~\ref{sec:localizedSolutions}.
Both behaviors are consistent with~\autocite{zolesi2024StabilityCrackNucleation} for standard phase-field models with stiffness degradation.
This test confirms numerically a key property of the model: in contrast with the standard phase-field models, where the equivalent strength scales as $\sqrt{\mu\Gc/\ell}$, here both the strength and the equivalent cohesive law are $\ell$-independent, and $\ell$ acts as a pure numerical regularization parameter.

In Figure~\ref{fig:barBrittlenessEll}(c), we consider the case $\zeta = 0.75$, where the homogeneous response exhibits a constant-stress phase, shown in  Figure~\ref{fig:homogeneous_response}(b), and study the impact of the magnitude of the initial perturbation $\alpha_0 \in\{0,10^{-3},10^{-2}\}$.
The other parameters are $L/\ell_\mathrm{ch}=1$, $\ell/\ell_\mathrm{ch}=0.05$, $N=100$ load steps, and the alternate-minimization tolerance was tightened to $10^{-5}$.
In this situation, the critical load at which we observe the bifurcation from the constant-stress to the localized regime depends on the magnitude of the small perturbation $\alpha_0$, but the peak stress remains unaffected.
This is the numerical counterpart of Remark~\ref{rem:branch-selection} and is again consistent with~\autocite{bmmz25}: since $\k(\alpha)\,\tau_c\Vert\p\Vert$ is homogeneous of degree one, the second variation of the energy is degenerate along redistributions of $\p$ on the constant-stress branch, so that its stability is decided by higher-order terms and the load at which localization sets in by the imperfection.

\begin{figure}[htbp]
    \centering
    \includegraphics[width=\textwidth]{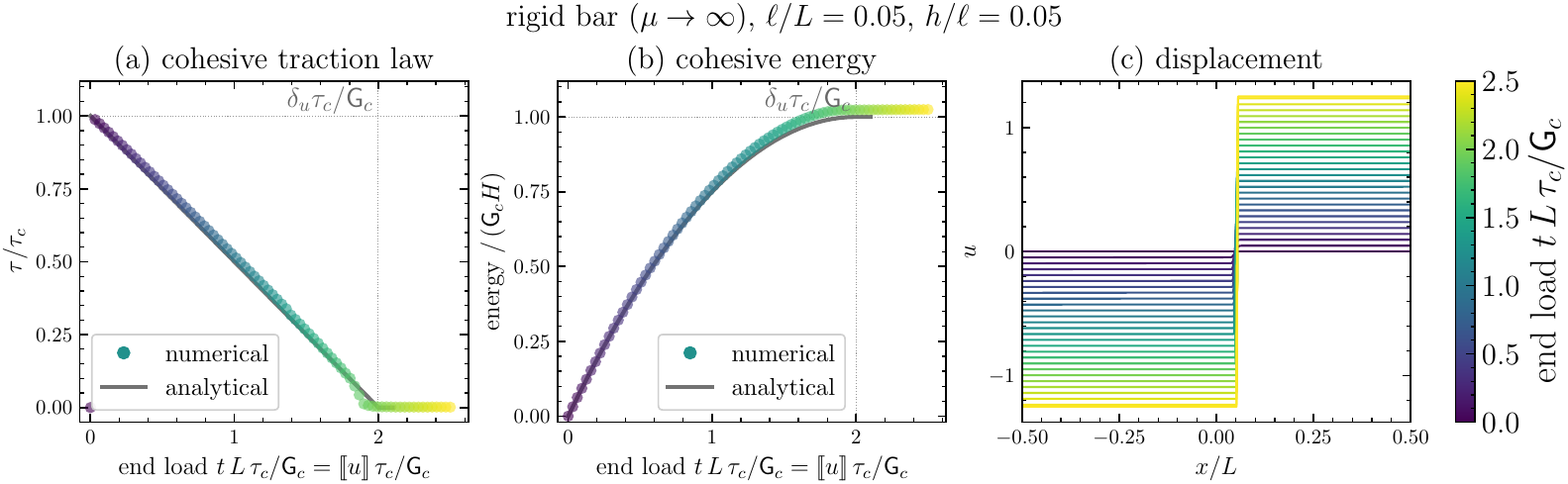}
    \caption{Simple shear in the rigid limit $\mu\to\infty$ ($\zeta=1$, $L\,\tau_c/\Gc=1$, $\ell/L=0.05$, $h/\ell=0.05$, $\alpha_0=0$, $N=80$ load steps), for which the imposed load is the opening, $t\,L=\jump{u}$: (a) equivalent cohesive law~\eqref{eq:1d-cohesive-law-M1-jump}, (b) dissipated energy normalized by $\Gc H$, (c) displacement $u$ along the bar; color scale: the load $\jump{u}\,\tau_c/\Gc$.}
    \label{fig:barRigid}
\end{figure}

Finally, Figure~\ref{fig:barRigid} deals with the rigid limit $\mu\to\infty$.
In this case, minimizers of the total energy satisfy $\nabla u=\p$ and the elastic energy is always 0.
The $(u,\p)$ subproblem reduces to the minimization of $\int_\Omega \k(\alpha)\,\tau_c\Vert\nabla u\Vert\dA$.
Since the bar carries no elastic strain, the imposed displacement coincides with the displacement jump, $t\,L=\jump{u}$, and the measured force--displacement response is the equivalent cohesive law itself.
This test isolates the cohesive response and provides the most direct numerical verification of the equivalence between the phase-field model and the cohesive law of Section~\ref{sec:localizedSolutions}.

\subsection{Surfing problem}
\label{sec:numericsSurfing}
\subsubsection{Geometry and loading}

\begin{figure}[t]
\centering
  \includegraphics[width=\linewidth]{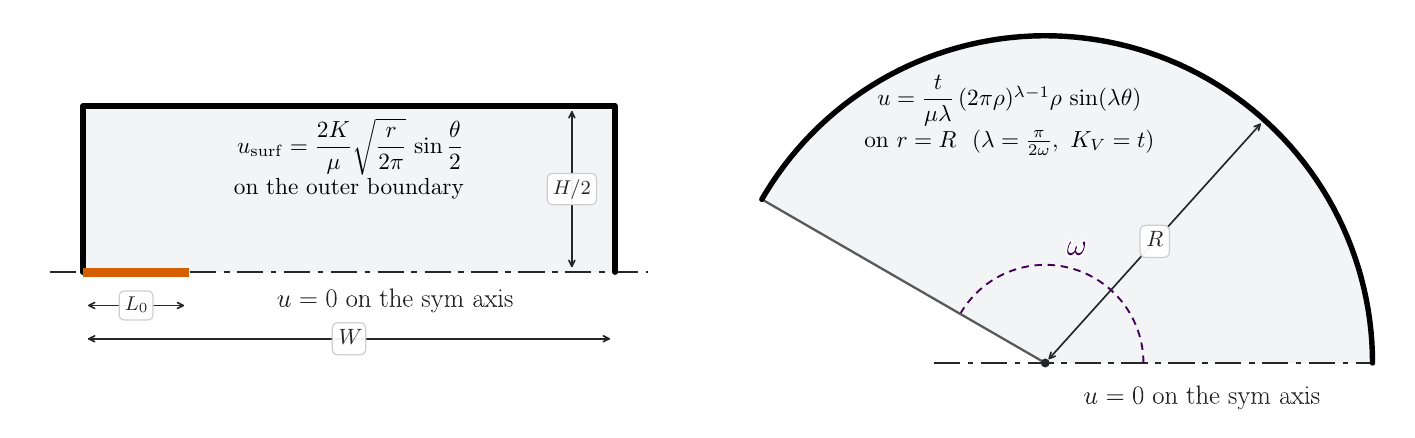}
  \caption{Geometry and loading of the two antiplane test cases. Half-domain for the (left) Surfing, (right) Pac-Man simulations.}
  \label{fig:geometry-bc}
\end{figure}

Our second set of numerical simulations is based on the ``surfing problem'' from~\autocite{Hossain-Hsueh-EtAl-2014a}.
We consider a rectangular domain $(0,W)\times(-H/2,H/2)$ with an initial crack $(0, L_0]\times \{0\}$ (see Figure~\ref{fig:geometry-bc}).
Along the outer boundary of the domain, we prescribe a displacement of the form
\begin{equation}
  u_{\mathrm{surf}}^{K,x_c}(x,y)=\frac{2K(t)}{\mu}\sqrt{\frac{r}{2\pi}}\,\sin\frac{\theta}{2},
  \quad r=\sqrt{(x-x_c(t))^2+y^2},\quad \theta=\operatorname{atan2}(y,\,x-x_c(t)),
  \label{eq:tear-surf}
\end{equation}
with
\begin{equation}
  K(t) = \begin{cases}
    \frac{t}{t_\mathrm{ramp}}K_\infty & \text{ if } t < t_\mathrm{ramp},\\
    K_\infty & \text{ otherwise}
  \end{cases},\qquad
  \label{eq:surfingXc}
  x_c(t) = \begin{cases}
    L_0  & \text{ if } t < t_\mathrm{ramp},\\
    L_0+(t-t_{\rm ramp}) & \text{ otherwise}
  \end{cases}
\end{equation}
while the pre-existing crack edges are left stress free.
Past an initial stage for $t < t_{\rm ramp}$, this boundary displacement is a translation of the mode-III asymptotic field associated with a stress intensity factor $K_\infty$ along the $x$-axis at unit speed and is fully characterized by the dimensionless parameter $G/\Gc=K_\infty^2/(2\mu \Gc)$.
Owing to the symmetry of the problem, one expects a crack growing along the $x$-axis, and we performed our simulations on a half domain.

While $\p$ was introduced as a nonlinear deformation, it is equally natural to view it as a plastic strain and subject it to an irreversibility constraint, as in the theory of perfect plasticity (Remark~\ref{rem:irreversibility}).
We therefore performed two series of simulations with the \textsf{M1} model at $\zeta=1$, without and with this constraint --- referred to in the sequel as the \emph{reversible} and \emph{irreversible} cases --- the latter requiring only a minor modification of the two convex subproblems (Section~\ref{sec:numericalImplementation}).
As we shall see, plastic irreversibility barely affects the damage field but changes the crack-tip fields and the energy balance appreciably: the work spent in the cohesive band, entirely recoverable when $\p$ is reversible, is then dissipated (Remark~\ref{rem:surface-energy-split}), and a residual plastic wake is left behind, and dragged along with, the advancing tip.
The two settings thus realize two distinct fracture phenomenologies: with $\p$ reversible, dissipation is carried by damage alone and the macroscopic toughness reduces to the prescribed $\Gc$, as for brittle cohesive fracture; with $\p$ irreversible, the plastic work accumulated in the wake adds to it and the effective toughness exceeds $\Gc$, the mechanism by which plastic dissipation inflates the measured fracture energy in ductile failure.

In the field computations below (Figures~\ref{fig:tear-tipzoom}--\ref{fig:tear-energy}), the domain size is $W=8$, $H=5$ and the initial crack length is $L_0=1/4$, in units of $\lch$; the load is applied in $121$ increments, the first $15$ forming the ramp, with $G/\Gc=1.2$ in the reversible case and $G/\Gc=1.55$ in the irreversible one --- the latter must exceed the effective toughness $\Gceff\simeq1.3\,\Gc$ measured below for the crack to propagate.
The regularization and mesh sizes ($\ell/\lch$, $h/\ell$) are reported in the figure captions.

\subsubsection{Simulation results}

\begin{figure}
    \centering
  \includegraphics[width=0.72\linewidth]{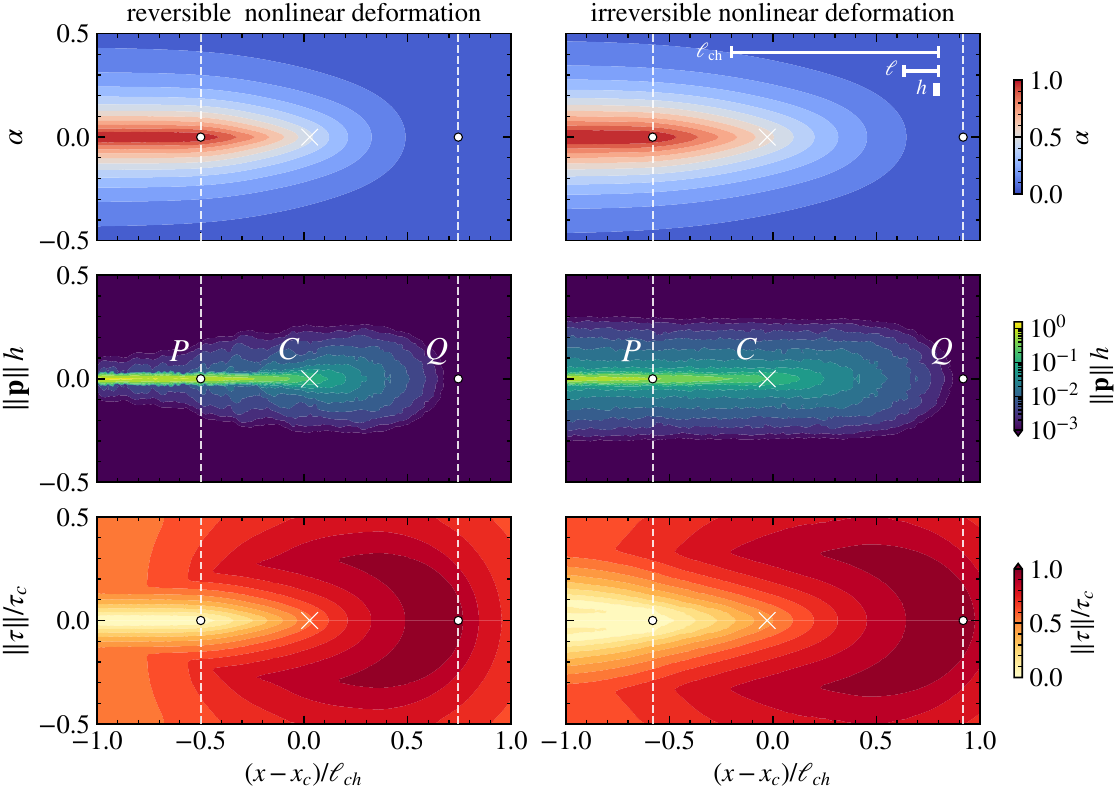}
  \caption{Crack-tip fields at the finest mesh ($\ell/\lch=1/6$, $h/\ell=0.1$): damage $\alpha$
  (top), plastic slip $\|\p\|h$ (middle), normalized stress $\|\t\|/\tau_c$
  (bottom), for the reversible (left) and irreversible (right) run. Note the residual plastic wake behind the irreversible tip.}
  \label{fig:tear-tipzoom}
\end{figure}

\begin{figure}[t]
\centering
  \includegraphics[width=0.78\linewidth]{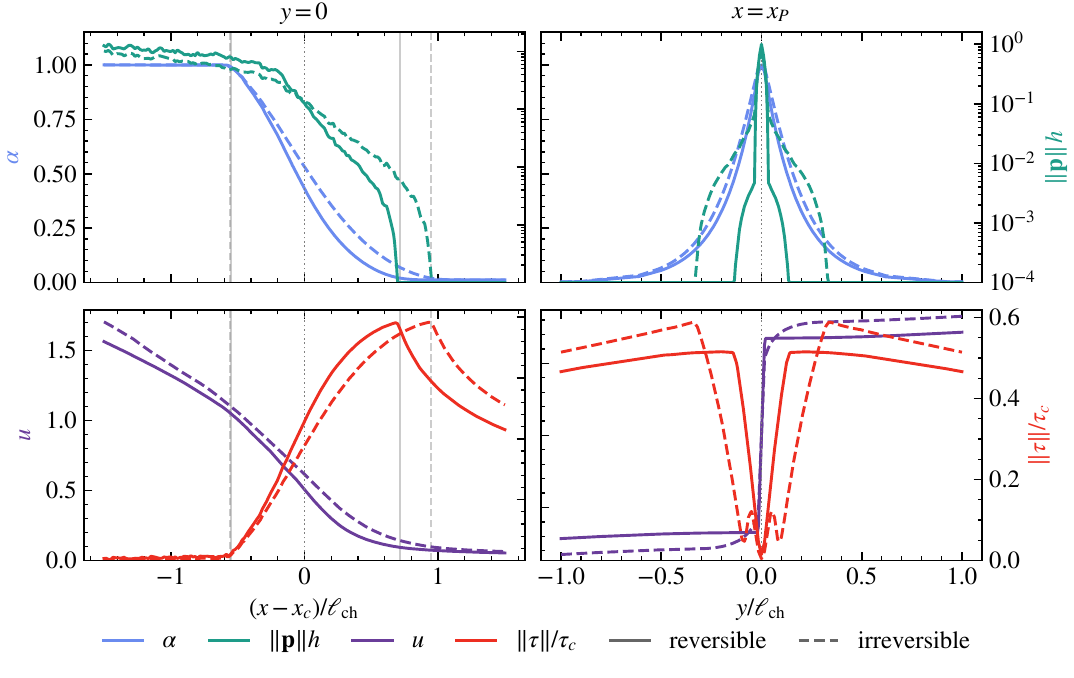}
  \caption{1-D field profiles along $y=0$ (left) and the tip normal $x=x_P$ (right). Top: damage
  $\alpha$, plastic slip $\|\p\|h$ (log); bottom: displacement $u$, stress
  $\|\t\|/\tau_c$. Solid: reversible, dashed: irreversible.}
  \label{fig:tear-profile}
\end{figure}

After the loading ramp phase, we  observed progressive crack propagation associated with self-similar profiles for $\alpha$, $\p$ and $\t$, translating along the $x$-axis at unit speed.
False-color plots of these fields in the reversible and irreversible cases are shown in Figure~\ref{fig:tear-tipzoom}, while Figure~\ref{fig:tear-profile} represents their profiles along the $x$-axis and along a vertical line through the fully developed crack.
Since we solved the problem on a half-domain and $u$ is pinned to 0 along the $x$-axis, in Figure~\ref{fig:tear-profile} (bottom left), $u$ is sampled at $y = \ell_{\mathrm ch}/10$.
The propagation appears to be smooth in time, as observed in~\autocite{rodella2026SharpInterface} for sharp-interface cohesive models and in contrast to  the  ``jerky'' motion, where a crack grows intermittently, which has been observed and analyzed in~\autocite{alessi2014GradientDamageModels,Brach-Tanne-EtAl-2019a,maso2020NumericalStudyJerky,dalmaso2020JerkyCrackGrowth}.

\begin{figure}[t]
\centering
  \includegraphics[width=\linewidth]{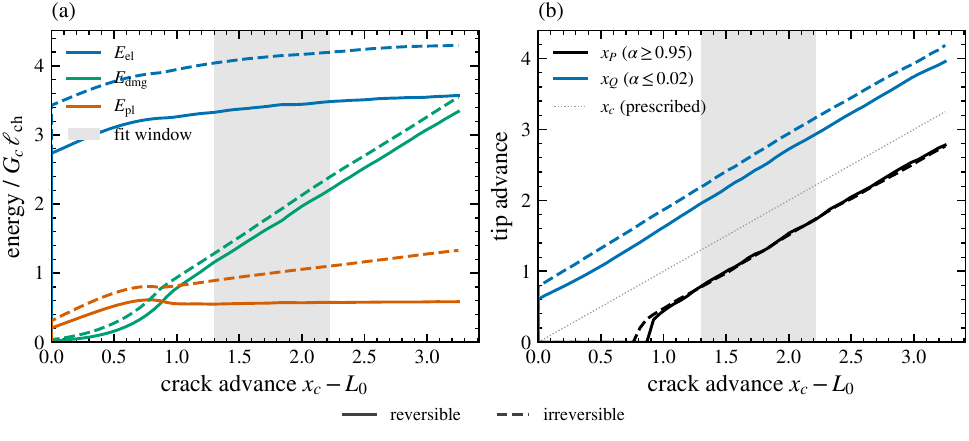}
  \caption{Energy and crack-tip evolution vs.\ crack advance $x_c-L_0$ (solid: reversible,
  dashed: irreversible). (left) Energy components.
  (right) Position of the brittle and cohesive crack tips $P$ ($\alpha\ge0.95$) and $Q$
  ($\alpha\le0.02$). }
  \label{fig:tear-energy}
\end{figure}

As expected, we observe that the damage field $\alpha$ is similar in both cases (Figure~\ref{fig:tear-tipzoom} (top)).
Along vertical cross sections in the crack wake, the  localization width is of order $\mathcal{O}(\ell)$ and the profile consistent with the exponential optimal profile for the AT2 model (Figure~\ref{fig:tear-profile} (top right)).
Ahead of the crack however, we observe an elongated area with $\alpha > 0$ (Figure~\ref{fig:tear-profile} (top left)).
In both figures, we quantify this region by marking three points on the $x$-axis: $C = (x_c,0)$, as given by~\eqref{eq:surfingXc} and $P=(x_P,0)$ and $Q=(x_Q,0)$ corresponding  to the region in which $\alpha$ transitions from 1 to 0 along the $x$-axis, so that $Q$ is also the maximum extent of the region $\|\p\| > 0$.
As shown in Figure~\ref{fig:tear-energy} (right), after a transition region corresponding to the loading ramp, $P$ and $Q$ move along with $C$.
In Figure~\ref{fig:tear-profile} (bottom left) we see that in both cases, the crack edges are stress-free while the stress attains its maximum near $Q$.
We thus think of $P$ as the ``brittle crack tip'' and of $Q$ as the ``cohesive crack tip''.
The length of the cohesive crack $|PQ|$ is slightly larger in the irreversible case with $|PQ| \simeq \lch$.
The displacement profile along the crack edges (Figure~\ref{fig:tear-profile} (bottom left)) transitions from the parabolic opening of a traction-free brittle crack to a convex profile decaying to zero within the cohesive crack.

The main difference between the reversible and irreversible cases consists in the presence of a residual plastic wake along the brittle crack edges in the irreversible case (Figure~\ref{fig:tear-tipzoom} (middle right)), while the inelastic strain $\p$ returns to 0 outside of a strip of width 1 element along which it localizes (Figure~\ref{fig:tear-tipzoom} (middle left)).
This has a significant impact on the balance between stored elastic energy and that dissipated during crack growth.

Figure~\ref{fig:tear-energy} reports the evolution of the elastic energy $E_\mathrm{el}$, the plastic work $E_\mathrm{pl}$, and the damage dissipation $E_\mathrm{dmg}$,
\[
E_\mathrm{el} =\int_\Omega \frac{\mu}{2}\|\nabla u - \p\|^2\dA
\qquad
E_\mathrm{pl} = \int_\Omega \k(\alpha)\,\tau_c\, \bar p\,\dA
\qquad
E_\mathrm{dmg} = \frac{\Gc}{4\cw}\int_\Omega \left(\frac{\w(\alpha)}{\ell} + \ell \|\nabla \alpha\|^2\right)\dA
\]
where $\bar p$ is the cumulated nonlinear deformation of Remark~\ref{rem:irreversibility}, reducing to $\|\p\|$ in the reversible case; the plastic work is recoverable in the reversible case and dissipated when $\p$ is irreversible (Remark~\ref{rem:surface-energy-split}).
Past the loading ramp, in the reversible case the damage dissipation grows at a constant rate while the plastic work remains constant.
In the irreversible case, however, the plastic work grows linearly as well, which is consistent with the presence of a residual plastic wake behind a crack tip translating at constant speed.

\begin{figure}[t]
\centering
  \includegraphics[width=\linewidth]{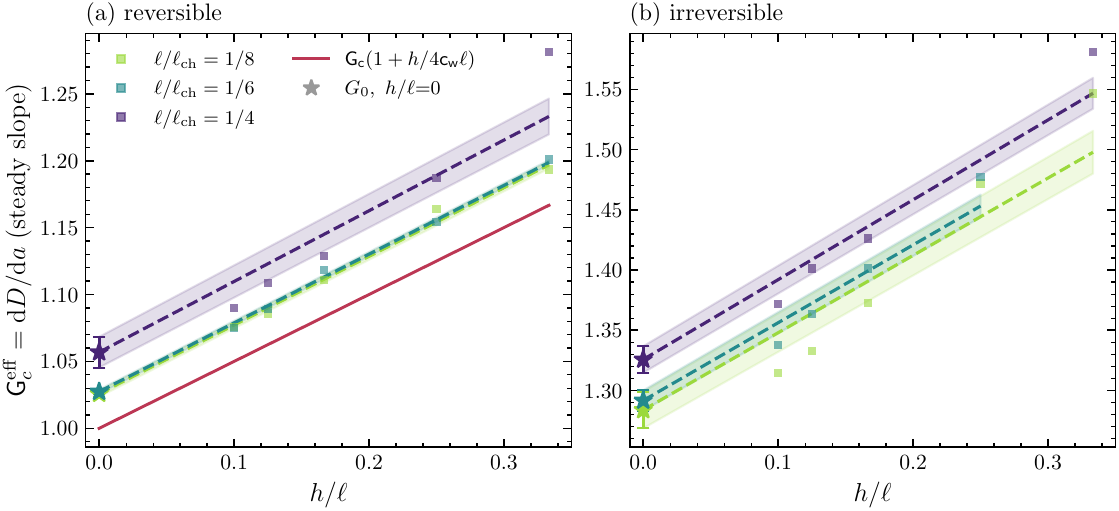}
  \caption{Effective toughness $\Gceff=\dDda$ versus $h/\ell$ (steady-state slope, one series per $\ell/\lch$; domain length $W=4\,\lch$, mesh refined in a band of half-width $\lch/2$ around the crack path) for reversible (left) and irreversible (right) plasticity. Dashed: fits $G_0\,(1+h/(4\cw\ell))$ with $\cw=\tfrac12$, the value of $\mathsf{M1}$ at $\zeta=1$; red: the same law with $G_0=\Gc$, \emph{i.e.}~the classical mesh-induced toughening. Bands and error bars: one standard deviation.}
  \label{fig:tear-conv}
\end{figure}

Figure~\ref{fig:tear-conv} shows the rate of energy dissipated per unit of crack length
\[
    \Gceff := \dDda = \frac{\partial (E_{\rm dmg}+E_{\rm pl})}{\partial a}
\]
as the crack grows through the region shown in gray in Figure~\ref{fig:tear-energy} for multiple values of the ratios $\ell/\lch$ and $h/\ell$ in order to compensate for mesh size-induced toughening in the reversible (left) and irreversible (right) cases.
The average and standard deviation of $\Gceff$ while the crack propagates through this area are denoted by colored markers and shaded areas respectively.
In the reversible case $\Gceff$ matches the classical correction $\Gc\left(1+h/(4\cw\ell)\right)$ and approaches $\Gc$ as $h/\ell \to 0$, independently of $\lch$, so that $\Gc$ can be interpreted as the rate at which energy is dissipated during the fracture growth process; this is not the case in the irreversible one.
Instead, we observe that in the limit of $h \to 0$, $\Gceff \simeq 1.3 \Gc$.
This is consistent with the fact that in order to grow a crack by an increment of length $\mathrm{d}a$, one needs to propagate the damage profile by the same amount (at a cost of $\Gc\,\mathrm{d}a$, up to discretization effects) and also the plastic zone, at a cost that remains to be identified analytically.
We argue that in this case, the  material's fracture toughness is $\Gceff$ and not $\Gc$, which only accounts for the damage dissipation, and would need to be evaluated numerically by estimating ${\mathrm{d}E_{\rm pl}}/{\mathrm{d} a}$, perhaps using this surfing experiment.

\subsection{V-notch}
\label{sec:numericsVnotch}
\subsubsection{Geometry and loading}

Our last set of numerical experiments focuses on crack growth from a V-notch and, in doing so, bridges two classical  descriptions of failure at a stress concentrator: the small-scale yielding (SSY) solution of {perfect plasticity} at the notch tip~\autocite{hult1956ElasticPlastic,rice1966ContainedPlastic} and the {cohesive crack} of Dugdale~\autocite{dugdale1960YieldingSteelSheets} and Barenblatt~\autocite{barenblatt1962MathematicalTheoryEquilibrium}.

We consider the ``Pac-Man'' geometry from~\autocite{tanne2018CrackNucleation}, consisting of a disk of radius $R$ with a re-entrant V-notch with complementary half-angle $\pi/2 \le \omega \le \pi$ occupied by an isotropic homogeneous material (see Figure~\ref{fig:geometry-bc}, right).

The dominant terms in a power series expansion of the antiplane mode-III elastic displacement and stress fields are given in~\autocite{rice1967StressesSharpNotch} by
\begin{equation}
  \label{eq:notchAsymptoticFields}
u(\rho,\theta)=\frac{t}{\mu\,\lambda}\,(2\pi\rho)^{\lambda-1}\,\rho\,\sin(\lambda\theta),\qquad
\t(\rho,\theta)=t\,(2\pi\rho)^{\lambda-1}\bigl(\sin(\lambda\theta)\,\underline{e}_\rho+\cos(\lambda\theta)\,\underline{e}_\theta\bigr),
\end{equation}
where $(\rho, \theta)$ denotes polar coordinates emanating from the notch tip with $\theta = 0$ corresponding to the ligament and $\theta = \pm \omega$ the notch faces.
The loading magnitude is $t$, and  $\lambda=\frac{\pi}{2\omega}\in\bigl(\tfrac12,1\bigr)$ is the exponent of the displacement singularity at the notch tip.
The magnitude of the stress  field near $\rho = 0$ is therefore independent of $\theta$: $\|\t(\rho, \theta)\| = t (2\pi\rho)^{\lambda-1}$, so that $\t(\rho,0)=t\,(2\pi\rho)^{\lambda-1}\,\underline{e}_\theta$.
It is customary to define the generalized stress intensity factor (GSIF) of the notch:
\[
  K_V=\lim_{r\to0^+}(2\pi r)^{\,1-\lambda}\,\tau(r,0)=t,
\]
which reduces to the classical mode-III stress-intensity factor in the limit $\omega\to \pi$, corresponding to a crack, with $\lambda=\tfrac12$.

On the circular part of the boundary of the domain, we prescribe the boundary displacement given by~\eqref{eq:notchAsymptoticFields} so that $K_V=t$ \emph{i.e.} the magnitude of the prescribed displacement is the V-notch GSIF, while the notch edges are left stress-free.
In this setting, the critical loading parameter $t_c=K_V^*$ plays the role of a notch toughness whose physical dimension $\tau_c\,\ell^{\,1-\lambda}$ varies with $\omega$, and which interpolates, as in the plane-elasticity coupled criterion of~\autocite{leguillon2002StrengthToughness,cornetti2006FiniteFractureMechanics}, between the strength $\tau_c$ in the limit ($\omega\to\pi/2$, $\lambda\to1$) of a straight edge and the Griffith toughness $K_c=\sqrt{2\mu\Gc}$ when the notch degenerates into a crack ($\omega=\pi$, $\lambda=\tfrac12$).
Again, we assume  symmetry (or anti-symmetry) of the solution with respect to the $x$-axis and perform our computations on a half-domain.

While the elastic solution admits a singularity at the notch tip, in the small-scale yielding  regime, this singularity gives way to a process zone where the stress constraint $\|\t\| = \tau_c$ is saturated.
Following the construction of~\autocite{rice1966ContainedPlastic,rice1967StressesSharpNotch}, it is a lobe of the curve defined by
\begin{equation}
  r(\theta)=a_0\cos(m\theta),\quad |\theta|\le\frac{\pi}{2m},\qquad
  m=\frac{\lambda}{1-\lambda},\quad a_0=\frac1\pi\left(\frac{|t|}{\tau_c}\right)^{1/(1-\lambda)},
  \label{eq:notch-petal}
\end{equation}
and depends only on the notch angle $\omega$ and the loading parameter $t$.
Note that in the limit of a crack ($\omega=\pi$, $\lambda=\tfrac12$, $m=1$) it degenerates to the Hult--McClintock circle through the tip~\autocite{hult1956ElasticPlastic} while as the wedge blunts ($\omega\to\pi/2$) the lobe narrows into a needle along the ligament.
The amplitude $a_0$ is the forward reach of this lobe (the value of $r$ at $\theta=0$), and $b_0=2a_0\max_{|\theta|\le\pi/2m}\cos(m\theta)\sin\theta$ its full transverse width, the small-scale yielding counterparts of the measured process-zone extents $a$ and $b$. For a crack the lobe is circular and $a_0=b_0$, while the aspect ratio $b_0/a_0$ falls to $0.71$ at $\omega=5\pi/6$ and vanishes as $\omega\to\pi/2$.

\subsubsection{Simulation results}
As in the rest of this section we adopt the $\mathsf{M1}$ family with $\zeta=1$, for which the equivalent cohesive law is linear and the homogeneous response has no constant-stress plateau, and we keep $\mu=\tau_c=1$, so that the elasto-cohesive length $\lch =\mu\Gc/\tau_c^2=\Gc$ is set by the toughness alone.
The nonlinear deformation $\p$ is taken reversible, as in the first series of simulations of Section~\ref{sec:numericsSurfing}.
The domain radius is $R = 10$ and we set the regularization length $\ell = 0.4$.
The mesh is refined along the expected crack path, with $h\simeq0.02$ ($h/\ell=0.05$) in a band of half-width $1.5\ell$ around the ligament, coarsening to $h=2$ at $\rho=R$; the loading is applied in $151$ increments and the alternate-minimization tolerance is $10^{-3}$.
Two critical loads are reported below: the peak-force load $t_c\equiv K_V^*$, and the load $t^*$ at which localization is first detected ($\max_\Omega\alpha\ge0.95$).
They are separated by $1.5$ load increments ($\Delta t\simeq0.014\,\tau_c$), \emph{i.e.} they coincide to within the load discretization.

\begin{figure}[tp]
  \centering
  \includegraphics[width=0.85\linewidth]{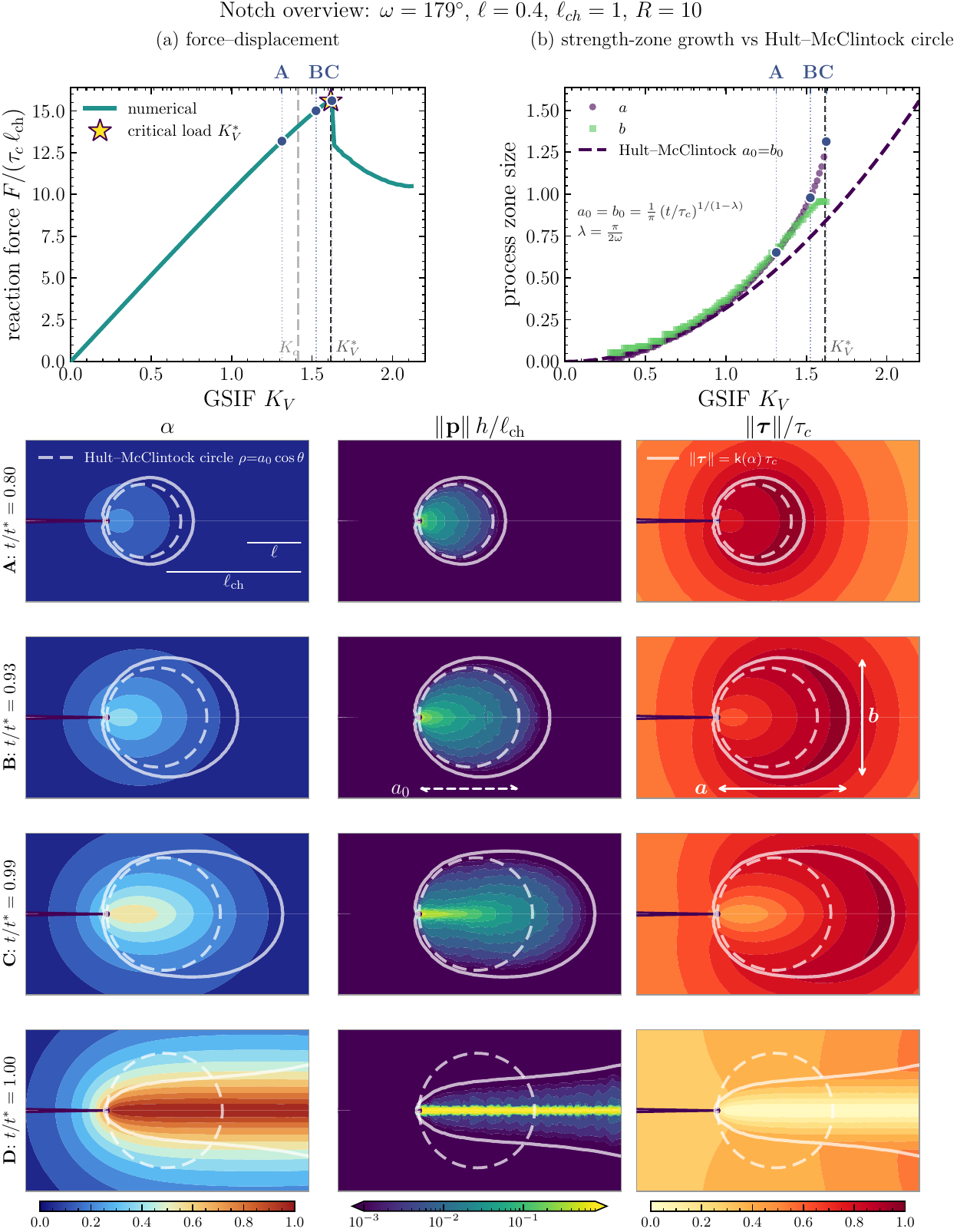}
  \caption{Near-crack overview, $\omega=179^\circ$ ($\ell=0.4$, $\lch=1$, $R=10$; $t_c=1.616$, $t^*=1.637$). Top: (a)~force $F/(\tau_c\lch)$ versus the GSIF $t=K_V$, with $t_c=K_V^*$ (dashed), the Griffith load $K_c$ (gray) and the snapshots $A$--$C$; (b)~extents $a,b$ of the strength zone against the SSY $a_0=b_0$~\eqref{eq:notch-petal}, up to nucleation. Bottom: tip-zoom maps of $\alpha$, $\Vert\p\Vert\,h/\lch$ (log) and $\Vert\t\Vert/\tau_c$ at $A$--$D$ (rows), $D$ the first localized state.}
  \label{fig:notch-ov179}
\end{figure}

We focus first on the degenerate case of a crack ($\omega = \frac{179\pi}{180}\simeq \pi$).
Figure~\ref{fig:notch-ov179} (top row) shows the evolution of the reaction force and of the strength zone $\{\Vert\t\Vert=\k(\alpha)\tau_c\}$ as a function of the loading parameter, while the bottom rows show snapshots in time of the  damage variable, inelastic strain, and magnitude of the stress.
In all the field maps that follow, the white solid line is the boundary $\Vert\t\Vert=\k(\alpha)\,\tau_c$ of the strength zone and the white dashed line the small-scale-yielding lobe~\eqref{eq:notch-petal}.
The snapshots $A$ and $B$ are taken at $t=0.80\,t^*$ and $t=0.93\,t^*$, $C$ at the last converged step before localization, and $D$ at the first localized one.

We identify three regimes.
In a first phase, until a point labeled as $A$ in Figure~\ref{fig:notch-ov179}, we observe a circular plastic zone closely matching the Hult--McClintock circle.
We observe a small damaged region near the crack tip.
In this region, we have $\Vert\t\Vert = \k(\alpha)\,\tau_c< \tau_c$ which explains why the magnitude of the stress is not  constant.

From $A$ onward, the strength zone becomes progressively elongated, with a forward reach $a$ and a transverse width $b$ such that $a>b$, $b$ exceeding $b_0$ by $20$--$25\%$.
Damage near the notch tip increases significantly.
This is the precursor to crack growth, where the solution progressively departs from the classical elastic / perfectly plastic one.
The snapshots at $t=B$ and $t=C$ confirm this trend: a significant increase of the damage near the notch and a plastic zone becoming more elongated.
The stresses do not vanish ahead of the pre-existing crack, so that we interpret this phase as the progressive growth of a cohesive crack.

Finally, at $t = D$ we observe a sudden bifurcation in the solution with the nucleation of a brittle crack, characterized by the classical phase-field crack profile for $\alpha$, the localization of $\p$ on a strip along the $x$-axis in which stresses vanish, and the sudden fall of the reaction force.
We view this instant as the sudden nucleation of a brittle crack.

Note that the onset of both cohesive and brittle cracks exceeds the threshold at which a brittle crack would nucleate in a classical phase-field model, even when using undamaged boundary conditions applied to crack edges, as seen in~\autocite{tanne2018CrackNucleation}.

Note that throughout the evolution, the size $a$ of the process zone remains small compared to the domain size, but grows to be of the order of $\lch$ when the brittle crack nucleates, reaching $a/R\simeq0.13$: the hypotheses of the small-scale yielding theory are therefore only marginally satisfied at nucleation.

\begin{figure}[htbp]
  \centering
  \includegraphics[width=0.85\linewidth]{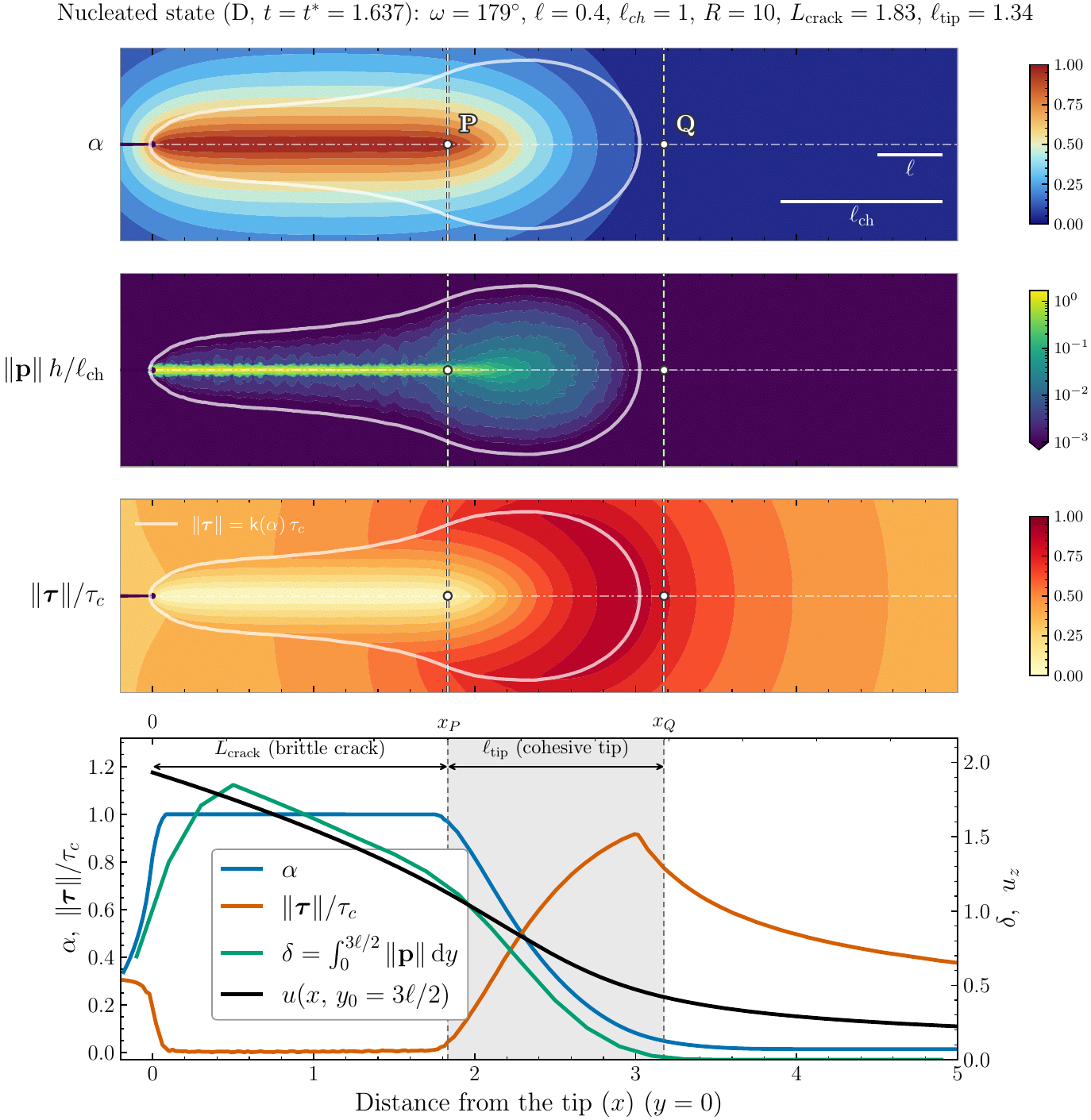}
  \caption{Nucleated state at $\omega=\tfrac{179\pi}{180}$ ($179^\circ$; $t=t^*=1.637$; crack length $L_\mathrm{crack}=1.83$, tip process-zone length $\ell_\mathrm{tip}=1.34$). Top: tip-zoom maps as in Figure~\ref{fig:notch-ov179}, with the crack tip $P$ and the process-zone tip $Q$ at $x_P=L_\mathrm{crack}$ and $x_Q=x_P+\ell_\mathrm{tip}$. Bottom: profiles along $y=0$ of $\alpha$ and $\Vert\t\Vert/\tau_c$ (left axis) and of the opening $\delta=\int\Vert\p\Vert\,\mathrm{d}y$ (right axis); the gray band is the process zone $PQ$.}
  \label{fig:notch-nuc}
\end{figure}

Figure~\ref{fig:notch-nuc} focuses on the last phase of the evolution.
Just as in the  surfing computations, we observe a brittle crack ending at point $x_P$ followed by a cohesive tip extending to $x = x_Q$, clearly indicated by the slow transition of $\alpha$ from 1 to 0 along the $x$-axis and the elongated plastic zone originating from $x_P$ ($x_P$ is the last point on $y=0$ where $\Vert\t\Vert<0.05\,\tau_c$, and $x_Q$ the first one where $\alpha<0.05$).
Stresses vanish along the edges of the brittle crack but not of the cohesive one, and reach a maximum of the order of $\tau_c$ near its tip (see Figure~\ref{fig:notch-ov179} (bottom)).
Half the crack opening is estimated by integrating $\|\p\|$ over $0<y<3\ell/2$ on the computed half-domain, so that $\delta=\jump{u}/2$.
We see that along the length of the brittle crack, this measure compares to the value of $u$ taken at $y=3\ell/2$, which is consistent with a stress-free crack, and decays down to 0 within the cohesive crack, forming a convex curve, a key feature of a cohesive crack as compared to the parabolic profile of a brittle crack.

\begin{figure}[tp]
  \centering
  \includegraphics[width=0.85\linewidth]{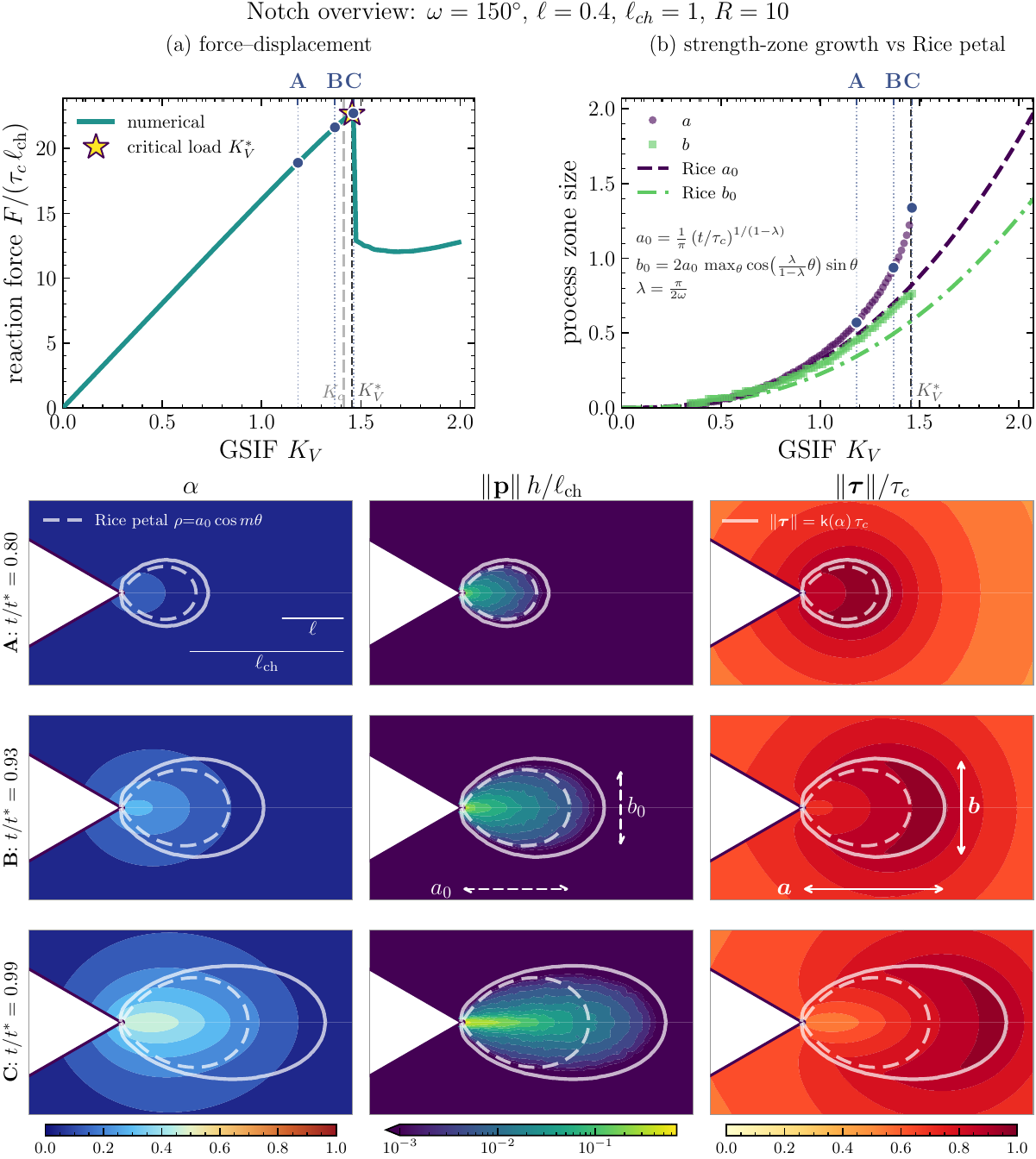}
  \caption{Blunt-wedge overview, $\omega=\tfrac{5\pi}{6}$ ($150^\circ$; $\ell=0.4$, $\lch=1$, $R=10$; $t_c=1.456$, $t^*=1.475$). Same panels, fields and colormaps as Figure~\ref{fig:notch-ov179}, with the Rice petal~\eqref{eq:notch-petal} ($a_0\neq b_0$) in place of the Hult--McClintock circle; here the three snapshots $A$--$C$ all precede nucleation, so there is no row $D$.}
  \label{fig:notch-ov150}
\end{figure}

As the notch angle decreases, we observe a qualitatively similar scenario (see Figure~\ref{fig:notch-ov150} for a notch angle of $150^\circ$).
Up to the load labeled $A$ in the figure, damage is limited and the forward reach $a$ of the strength zone follows the SSY prediction~\eqref{eq:notch-petal} to within $7\%$ up to $t\simeq\tau_c$, while its transverse width exceeds $b_0$ by about $30\%$ throughout.
From point $A$ onward, we observe the growth of a cohesive crack, while the plastic zone becomes increasingly elongated, up to the critical load $K_V = K^*_V$ at which a brittle crack is suddenly nucleated, as indicated by the sudden drop in reaction force in Figure~\ref{fig:notch-ov150} (top left) and the change in the geometry of the plastic zone (Figure~\ref{fig:notch-ov150} (top right)).
Note how the geometry of the plastic zone becomes similar to that observed at the crack tip in the surfing problem (Figure~\ref{fig:tear-tipzoom}).

\begin{figure}[htbp]
  \centering
  \includegraphics[width=\textwidth]{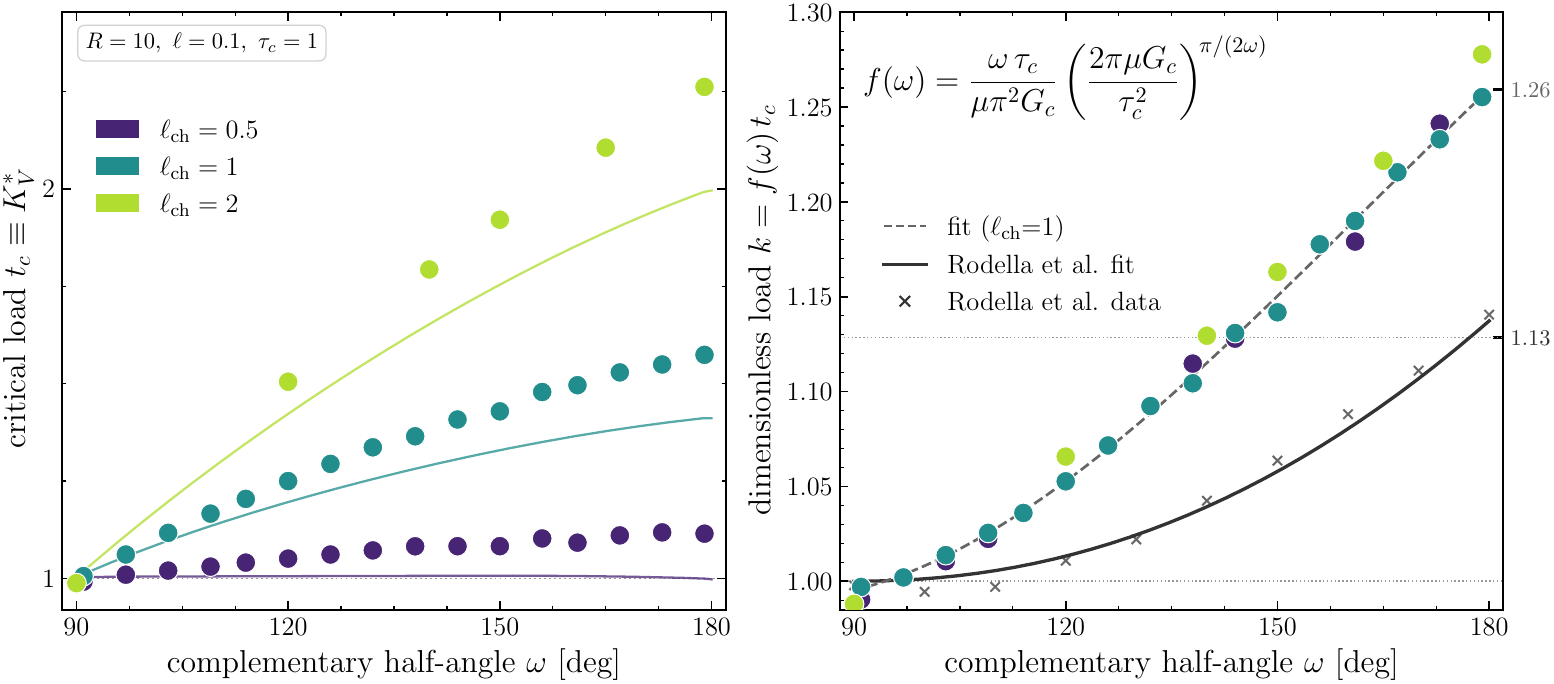}
  \caption{Critical load at the onset of the brittle crack, $\ell=0.1$, $\lch\in\{0.5,1,2\}$ (one color per series), against the sharp-interface model of~\autocite{rodella2026SharpInterface} (solid lines, crosses). (left) Peak-force load $t_c=K_V^*$ versus $\omega$. (right) Same data through the group $k$, which collapses the series (gray dashed: cubic fit through $\lch=1$); dotted: the strength and Griffith limits $k=1$, $2/\sqrt{\pi}\simeq1.13$.}
  \label{fig:notch-cl}
\end{figure}

Figure~\ref{fig:notch-cl} (left) shows the critical load $K^*_V$ as a function of the notch angle and of the elasto-cohesive length $\lch$, compared to that computed using the sharp-interface model of~\autocite{rodella2026SharpInterface}; these simulations use $\ell=0.1$, with a mesh size $h=\ell/5$ in the refined band, rather than the $\ell=0.4$ of the field computations above.
As $\omega \to 90^\circ$, we observe nucleation at $K_V^* = \tau_c$, which is consistent with our one-dimensional analysis and the tearing simulations.
As in Figure~\ref{fig:notch-ov179}, for sharp cracks, we observe that re-nucleating a crack from an existing notch requires $K_V^*>K_c = \sqrt{2\mu\Gc}$.
While this is consistent with the fact that some energy is dissipated creating a plastic zone near the crack tip before re-nucleation, we cannot rule out that our numerical scheme may overshoot the nucleation threshold.
However, based on the experience gained in~\autocite{tanne2018CrackNucleation}, we argue that this overshoot is typically small and would not account for the discrepancy observed here.

Figure~\ref{fig:notch-cl} (right) shows the same data through the dimensionless group
\[
  k=\frac{1}{\lambda}\left(2\pi\lch\right)^{\lambda-1}\frac{K_V^*}{\tau_c},
\]
normalized so that $k=1$ in the strength-dominated limit ($\omega=\pi/2$, $\lambda=1$, $K_V^*=\tau_c$) and $k=2/\sqrt{\pi}\simeq1.13$ in the Griffith limit ($\omega=\pi$, $\lambda=\tfrac12$, $K_V^*=K_c$).
All the simulations collapse on a single curve, independent of $\lch$, following the ``universal law'' suggested in~\autocite{marigo2023ModellingFractureCohesive}.
The collapsed curve reaches $k\simeq1.26$ in the crack limit, about $12\%$ above the sharp-interface value of~\autocite{rodella2026SharpInterface}; about half of this excess is accounted for by the mesh-induced toughening factor $(1+h/(4\cw\ell))^{1-\lambda}\simeq1.05$ at the discretization used here, the remaining $\simeq6\%$ being attributable to the plastic dissipation preceding re-nucleation.

\section{Conclusions and extensions}
\label{sec:conclusions}

We have presented a mathematical analysis and carefully tailored numerical simulations of the model~\eqref{eq:plastic-damage-energyIntro}, introduced in~\autocite{bmmz25}, in the antiplane setting. Its variational structure is the basis of our formulation and our numerical algorithm, which alternates the minimization of the total energy with respect to the displacement and the nonlinear deformation, and with respect to the damage variable, each subproblem being reformulated as a second-order cone program and solved to global optimality.

In the antiplane setting, one of the most delicate features of the full model disappears: since the nonlinear deformation is a vector, every localized deformation is a compatible jump, and the strength domain no longer restricts the admissible jump directions as it does in three dimensions.
This is precisely what makes this simplified setting precious: it allows us to study all the other properties of the model independently of the jump compatibility condition, and to highlight the most fundamental aspect of this family of models --- its ability to bridge the gap between limit analysis, perfect plasticity, cohesive fracture, and brittle fracture.
In particular, we have shown that the small-scale yielding crack-tip plastic zone can localize along a line, leading to the propagation of a cohesive crack which, as the opening becomes larger, degenerates into a Griffith-like brittle crack.
Our numerical results thus disclose the missing path for unifying small-scale-yielding analysis with cohesive crack-band approaches within a single consistent nonlinear model, able to predict both crack nucleation and propagation.

The numerical simulations substantiate these claims quantitatively.
We have focused our investigations on simple geometries of universal interest, to disclose the fundamental behavior of the proposed model.
On the simple shear problem, the computed evolutions reproduce the closed-form solutions of Sections~\ref{sec:modelProblem} and~\ref{sec:M1}: the equivalent cohesive law and the associated energy balance are recovered independently of $\ell$, the dissipated energy exceeds $\Gc$ only by the mesh-toughening factor $1+h/(4\cw\ell)$, matched to better than $1\%$ for $h\le\ell/5$, and the size effect --- the transition from progressive failure to snap-back --- is governed by the ratio $L/\lch$; in the linearly rigid limit, the measured force--displacement curve is the cohesive law itself.
The surfing simulations measure the energy dissipated per unit crack advance: the effective toughness $\Gceff$ coincides with the prescribed $\Gc$ when $\p$ is unconstrained, while an irreversibility constraint on $\p$ leaves a residual plastic wake behind the tip and raises $\Gceff$ to about $1.3\,\Gc$.
At the V-notch, a single monotonic loading drives the tip fields through three classical regimes --- the confined small-scale-yielding plastic zone of Hult and McClintock, a Barenblatt-type cohesive crack, and a stress-free brittle crack --- and the nucleation loads for all notch angles and material lengths collapse onto the ``universal law'' of~\autocite{marigo2023ModellingFractureCohesive}, interpolating between the strength-dominated limit $K_V^*=\tau_c$ and the toughness-dominated Griffith limit $K_V^*=\sqrt{2\mu\Gc}$, within about $6\%$ of the sharp-interface computations of~\autocite{rodella2026SharpInterface}, once the mesh-induced toughening of the regularized model is discounted.
Taken together, these results exhibit limit analysis, perfect plasticity, cohesive fracture, and brittle fracture as asymptotic regimes of a single variational model with one set of material data $(\mu,\tau_c,\Gc)$, the transitions between them selected by the loading and by the ratios $\ell/\lch$ and $L/\lch$, rather than by a change of constitutive description.

Unlike conventional phase-field models, originally introduced in~\autocite{BFM00} and the subject of countless extensions since then, our model accounts for a strength surface that is fully independent of the elastic properties and of the regularization parameter $\ell$.
As such, the parameter $\ell$, which is often interpreted as an internal length in classical phase-field models, can be chosen as a purely numerical regularization parameter with little impact on the computational results, provided that it is small compared to the dimensions of the structure and to the characteristic size of the patterns it develops.

Viewed as a regularized model of softening plasticity, the present model also achieves what, to our knowledge, other softening-plasticity approaches~\autocite{pijaudier-cabot1987NonlocalDamageTheory,muhlhaus1991VariationalPrincipleGradient,leblond1994BifurcationEffectsDuctile,bazant2002NonlocalIntegralFormulations,jirasek2003ComparisonIntegraltypeNonlocal,forest2004LocalizationPhenomenaRegularization,lorentz2008NumericalSimulationDuctile,bacquaert2025RegularizationSofteningPlasticity} have not reached: the identification of a clear underlying sharp-interface model, the cohesive energy~\eqref{eq:limit-energy}.

Finally, the model we studied and the numerical implementation we proposed extend naturally to two- and three-dimensional elasticity~\autocite{bmmz25}, where the choice of strength surface, constitutive relation, and cohesive law will lead to a broader gamut of behaviors that remain to be explored.
In this vectorial setting, with tensorial strains and stresses, the jump compatibility condition comes into play: a careful analysis of its practical implications is the subject of ongoing work and will be presented in a separate paper.
Similarly, minor modifications, such as the irreversibility constraint on $\p$ already explored in the surfing example of Section~\ref{sec:numericsSurfing}, open the way to more complex material behaviors, including ductile fracture or fatigue.
However, while the present model offers a promising starting point, it is still far from a complete theory of ductile fracture: this would require substantial long-term developments, including a proper account of permanent plastic deformations, the modeling of a hardening regime, and multiaxial strength surfaces --- a program to which our future work will be devoted.

\ifcras\else
  \section*{Acknowledgements}
  \ackstext

  \section*{Appendices}
\fi
\appendix

\section{Conic reformulations of the alternate-minimization subproblems}
\label{sec:numericalAppendix}
This appendix details the conic-programming reformulations of the two convex subproblems solved at each iteration of the alternate-minimization scheme of Section~\ref{sec:numericalImplementation} (Algorithm~\ref{alg:alternate-minimization}).

\subsection{Convex-cone reformulation of the $(u,\p)$ subproblem}
\label{sec:up-socp}

The first subproblem consists in minimizing the energy with respect to $(u,\p)$ at fixed $\alpha$:
\begin{equation}
    \min_{u,\p} \int_{\Omega} \varphi(\nabla u, \p, \alpha)\dA, \qquad
    \varphi(\nabla u, \p, \alpha) = \frac{\mu}{2} \|\nabla u - \p\|^2 + \mathsf{k}(\alpha)\,\tau_c\|\p\|.
    \label{eq:up-problem}
\end{equation}
The non-smooth term $\|\p\|$ limits the effectiveness of standard gradient-based optimization algorithms.
Following~\autocite{bleyer-convex-optim}, we reformulate~\eqref{eq:up-problem} as a Second-Order Cone Programming (SOCP) problem~\autocite{lobo1998ApplicationsSecondOrderCone}, which can be solved robustly and efficiently using the MOSEK software~\autocite{aps2024mosek}, without the need for smoothing.
This approach is also flexible and can be extended to vector-valued displacements and non-smooth multiaxial strength domains.

Introducing auxiliary scalar variables $s$ and $z$, the energy density can be equivalently expressed as the partial minimization
\begin{equation}
\begin{aligned}
    \varphi(\nabla u,\p,\alpha) &= \min_{s,z \geq 0}\; \mu\,s + \mathsf{k}(\alpha)\,\tau_c\,z\\
    &\quad\text{subject to}\\
    &\qquad 2s \geq \|\nabla u - \p\|^2,\\
    &\qquad z \geq \|\p\|,
\end{aligned}
\label{eq:local-socp}
\end{equation}
which turns the non-smooth objective into a linear one at the cost of introducing cone constraints.
These constraints are expressed using the second-order cone $\mathcal{Q}^{n+1}$ and the rotated second-order cone $\mathcal{Q}_{\mathrm{r}}^{n+2}$, standard types natively supported by MOSEK~\autocite{aps2024mosek}:
\begin{equation}
\begin{aligned}
\mathcal{Q}^{n+1} &:= \{(r,x)\in\mathbb{R}\times\mathbb{R}^n : r \geq \|x\|\},\\
\mathcal{Q}_{\mathrm{r}}^{n+2} &:= \{(r_1, r_2, x)\in\mathbb{R}^2\times\mathbb{R}^n : 2\,r_1 r_2 \geq \|x\|^2,\; r_1,r_2 \geq 0\}.
\end{aligned}
\label{eq:cones}
\end{equation}
Setting $e_i := u_{,i} - p_i$ for $i=1,2$, the constraint $z \geq \|\p\|$ reads $(z, p_1, p_2)\in\mathcal{Q}^{3}$ and the constraint $2s \geq \|\nabla u - \p\|^2$ reads $(s, 1, e_1, e_2)\in\mathcal{Q}_{\mathrm{r}}^{4}$.

In order to discretize~\eqref{eq:local-socp}, we introduce the discrete gradient matrix $\mathsf{B}$ of size $2n_g\times n_n$, defined by
\begin{equation}
    \mathsf{B}_{gk} := \nabla \chi_k(x_g), \qquad \text{so that} \qquad
    \mathsf{B}\mathsf{U} = \left\{\nabla u(x_g)\right\}_{g=1}^{n_g},
    \label{eq:gradient-matrix}
\end{equation}
mapping the nodal values of the discretization $\mathsf{U}$ of the displacement by $\mathbb{P}_1$ simplicial Lagrange finite elements to its gradient at all quadrature points.
Dirichlet boundary conditions are enforced as equality constraints on the nodal values:
\begin{equation}
    \label{eq:dirichlet-constraint}
    \mathsf{U}_k = \bar{u}(x_k), \quad \forall\, k\in\mathcal{I}_{\partial_u\Omega},
\end{equation}
where $\mathcal{I}_{\partial_u\Omega}$ denotes the set of indices of nodes on $\partial_u\Omega$.
Applying the local reformulation~\eqref{eq:local-socp} pointwise at each quadrature point $x_g$ and assembling over the mesh with the quadrature weights $\rho_g$ (equal to the cell areas for the one-point rule), the discrete elastoplastic subproblem becomes the SOCP:
\begin{equation}
\begin{aligned}
    \min_{\mathsf{U},\mathsf{P},\mathsf{s},\mathsf{z}} \quad & \sum_{g=1}^{n_g} \rho_g \left( \mu\, s^g + \mathsf{k}(\alpha^g)\,\tau_c\, z^g \right)\\
    \text{subject to} \quad
                & e_{i}^g = (\mathsf{B}\mathsf{U})_i^g - p_{i}^g, \quad i=1,2,\; g=1,\ldots,n_g,\\
                & (s^g,\, 1,\, e_1^g,\, e_2^g) \in \mathcal{Q}_{\mathrm{r}}^{4}, \quad g=1,\ldots,n_g,\\
                & (z^g,\, p_1^g,\, p_2^g) \in \mathcal{Q}^{3}, \quad g=1,\ldots,n_g,\\
                & \mathsf{U}_k = \bar{u}(x_k), \quad \forall\, k\in\mathcal{I}_{\partial_u\Omega}.
\end{aligned}
\label{eq:full-socp}
\end{equation}
In the implementation, the strength degradation function is replaced by $\k(\alpha)+\k_{\mathrm{res}}$ with a small residual strength ($\k_{\mathrm{res}}=10^{-6}$ for the simple-shear computations, $10^{-4}$ for the surfing and V-notch ones), so that the coefficient of $z^g$ in the objective never vanishes when $\alpha\to1$ and the conic problem remains well-posed.
The SOCP is solved with the interior-point conic optimizer of MOSEK~\autocite{aps2024mosek} with feasibility and relative-gap tolerances set to $10^{-8}$.

The formulation above extends directly to the rigid limit $\mu\to\infty$, used in Section~\ref{sec:numericsSimpleShear} to compute the response of a rigid bar.
In this case the elastic strain must vanish: the auxiliary variables $s^g$ and the rotated-cone constraints are dropped, and the elastic strain definition is replaced by the pointwise kinematic constraint
\begin{equation}
    \label{eq:rigidity-constraint}
    \rho_g\left((\mathsf{B}\mathsf{U})_i^g - p_{i}^g\right) = 0, \quad i=1,2,\; g=1,\ldots,n_g,
\end{equation}
so that the objective reduces to the plastic term $\sum_g \rho_g\, \mathsf{k}(\alpha^g)\,\tau_c\, z^g$.
The stress at the quadrature points is then recovered as the dual (Lagrange) multiplier of the equality constraints~\eqref{eq:rigidity-constraint}, directly provided by the conic optimizer.

\subsection{Minimization with respect to the damage field}
\label{sec:alpha-qp}
The second subproblem consists in minimizing the energy with respect to $\alpha$ at fixed $(u,\p)$, under the bound constraints given by irreversibility and the upper bound $\alpha\leq1$:
\begin{equation}
    \min_{\alpha} \int_{\Omega} \mathsf{k}(\alpha)\,\tau_c\|\p\| + \frac{\Gc}{4\cw}\left(\frac{\mathsf{w}(\alpha)}{\ell}+\ell\,\Vert\nabla\alpha\Vert^2\right) \dA, \qquad \text{subject to} \quad \alpha_{i-1} \leq \alpha \leq 1.
\end{equation}
Under Hypothesis~\ref{hyp:kw} with $\k$ convex and $\w$ convex, this is a convex optimization problem with simple pointwise bound constraints.
For the family $\mathsf{M1}$, where $\k$ is affine and $\w$ is quadratic in $\alpha$, we also discretize the damage field with $\mathbb{P}_1$ simplicial Lagrange finite elements and cast it directly in conic form.
Let $\mathsf{D}$ collect the nodal values of the damage field, so that the damage $\alpha(x_g)=\sum_k\mathsf{D}_k\chi_k(x_g)$ and its gradient $\nabla\alpha(x_g)$ are determined at the quadrature points.
The subproblem then reads
\begin{equation}
\begin{aligned}
    \min_{\mathsf{D},\,z} \quad & z + \mathsf{b}^T \mathsf{D}\\
    \text{subject to} \quad
                & (z,\,\tfrac12,\,\mathsf{L}\mathsf{D})\in\mathcal{Q}_\mathrm{r},\qquad
                 \mathsf{D}_{i-1,k} \leq \mathsf{D}_k \leq 1, \quad \forall\, k=1,\ldots,n_n,
\end{aligned}
\label{eq:alpha-socp}
\end{equation}
where the auxiliary scalar $z$ and the rotated cone $\mathcal{Q}_\mathrm{r}$ of~\eqref{eq:cones} enforce $z\geq\Vert\mathsf{L}\mathsf{D}\Vert^2$, turning the quadratic damage energy into a linear objective.
We do not factorize a global stiffness matrix: since with one-point quadrature the quadratic damage energy is already a sum of squares over the quadrature points, the vector $\mathsf{L}\mathsf{D}$ is built point by point by stacking the damage and its gradient at each $x_g$, each scaled by the square root of the corresponding (positive, scalar) energy weight,
\begin{equation}
    \mathsf{L}\mathsf{D} = \Big\{\,
        \sqrt{c_g}\;\alpha(x_g),\;\;
        \sqrt{d_g}\;\partial_1\alpha(x_g),\;\;
        \sqrt{d_g}\;\partial_2\alpha(x_g)
    \,\Big\}_{g=1}^{n_g},
    \qquad
    c_g=\frac{\Gc\,\zeta\,\rho_g}{4\cw\ell},\quad
    d_g=\frac{\Gc\,\ell\,\rho_g}{4\cw},
    \label{eq:alpha-socp-factor}
\end{equation}
with $\rho_g$ the quadrature weights (cell areas) of Appendix~\ref{sec:up-socp}.
By construction
$\Vert\mathsf{L}\mathsf{D}\Vert^2=\sum_g\big(c_g\,\alpha(x_g)^2+d_g\,\Vert\nabla\alpha(x_g)\Vert^2\big)=\frac{\Gc}{4\cw}\int_\Omega\big(\tfrac{\zeta}{\ell}\,\alpha^2+\ell\,\Vert\nabla\alpha\Vert^2\big)\dA$, \emph{i.e.}~the quadratic part of the damage energy.
The load vector
\begin{equation}
    \mathsf{b}_k = \frac{\Gc(1-\zeta)}{4\cw\ell}\int_\Omega \chi_k\dA
    - \tau_c\int_\Omega \Vert\p\Vert\,\chi_k\dA
    \label{eq:alpha-socp-load}
\end{equation}
collects the linear part of $\w(\alpha)$ and the non-smooth coupling term $\k(\alpha)\,\tau_c\|\p\|$ with $\k'(\alpha)=-1$, respectively.
The program is solved with the conic interior-point optimizer of MOSEK~\autocite{aps2024mosek}, with maximal tolerances set to $10^{-8}$.

\ifcras
  \printCOI
  \section*{License}
  \noindent\ccby\enspace This article is licensed under the Creative Commons
  Attribution 4.0 International License. To view a copy of this license, visit
  \CDRliclinkurl.
\fi

\printbibliography

\end{document}